\documentclass[acmsmall]{acmart}
\AtBeginDocument{%
  }

\setcopyright{cc}
\setcctype{by} 
\acmJournal{PACMMOD}
\acmYear{2026} \acmVolume{4}
\acmNumber{3} \acmArticle{178}
\acmMonth{6} \acmPrice{}
\acmDOI{10.1145/3802055}
\received{October 2025}
\received[revised]{January 2026}
\received[accepted]{February 2026}

\usepackage{array}
\usepackage{algorithm}
\usepackage{algpseudocode}
\usepackage{enumitem}
\usepackage{float}
\usepackage{graphicx}
\usepackage{multirow}
\usepackage{subcaption}
\usepackage{tikz}
\usetikzlibrary{automata,arrows}
\usetikzlibrary{backgrounds}
\usetikzlibrary{calc}

\usepackage{pifont}

\usepackage{listings}
\usepackage{color}
\definecolor{dkgreen}{rgb}{0,0.6,0}
\definecolor{gray}{rgb}{0.5,0.5,0.5}
\definecolor{mauve}{rgb}{0.58,0,0.82}
\usepackage{xcolor}

\usepackage{marginnote}
\usepackage{dblfloatfix}

\lstdefinestyle{promptstyle}{
	basicstyle=\ttfamily\small,
	numbers=left,
	numberstyle=\tiny\color{gray},
	numbersep=8pt,
	frame=single,
	rulecolor=\color{black!20},
	backgroundcolor=\color{black!02},
	showstringspaces=false,
	breaklines=true,
	breakatwhitespace=true,
	columns=fullflexible,
	upquote=true,
	tabsize=2,
	xleftmargin=1em,
	xrightmargin=1em,
}

\newcommand{\tab}[1]{Table~\ref{#1}}
\newcommand{\fig}[1]{Fig.~\ref{#1}}
\newcommand{\sect}[1]{\S\ref{#1}}

\newcommand{\alg}[1]{Alg.~\ref{#1}}
\newcommand{\eg}{\textit{e.g.}, } 
\newcommand{\ie}{\textit{i.e.}, }

\newcommand{\etal}{\textit{et al.} }

\begin{document}

\title{Factorized and Vectorized Execution: Optimizing Analytical and Semantic Queries over Relations}

\author{Sunny Yasser}
\affiliation{%
	\institution{Polytechnique Montréal}
	\city{Montréal, QC}
	\country{Canada}
}
\email{sunny.yasser@polymtl.ca}

\author{Anas Dorbani}
\affiliation{
	\institution{Polytechnique Montréal}
	\city{Montréal, QC}
	\country{Canada}
}
\email{anas.dorbani@polymtl.ca}

\author{Amine Mhedhbi}
\affiliation{
	\institution{Polytechnique Montréal}
	\city{Montréal, QC}
	\country{Canada}
}
\email{amine.mhedhbi@polymtl.ca}

\begin{abstract} 
Many-to-many joins are central to analytical and semantic workloads such as fraud detection, network analysis, and recommendation, where insights arise from relationships between entities. 
These workloads often suffer from an explosion of intermediate results, sometimes orders of magnitude larger than the inputs. 
Factorized representations address this problem by exploiting conditional independence among attributes to encode intermediates more compactly. In some cases, they can reduce the output size asymptotically below the worst-case output size. 
However, adopting factorization in modern vectorized query processors remains challenging: 
factorized representations are hierarchical, whereas vectorized execution is built around flat, block-oriented processing. 
Prior approaches either rely on full materialization or support only restricted factorization layouts, sacrificing much of the benefits of both factorization and vectorization. 

We present \textsc{FFX}, a novel engine for \underline{\textbf{F}}ast \underline{\textbf{F}}actorized e\underline{\textbf{X}}ecution. 
\textsc{FFX} is the first pipelined engine to support arbitrary factorization schemes while preserving full vectorization. 
The engine introduces packed factorized vectors and operators that maintain cache-friendly, contiguous layouts. 
Beyond analytics, \textsc{FFX} also co-optimizes semantic operators by serializing factorized intermediates into compact prompts for large language models (LLMs), substantially reducing token usage and inference cost while maintaining output quality and, in some cases, improving it. 
Together, these contributions enable efficient execution of join-heavy analytical queries, including queries augmented with semantic operators. 
\end{abstract}



\maketitle

\section{Introduction}

Many-to-many joins underpin a wide range of analytical workloads, 
in which insights arise from the structure of relationships between entities. 
A common analytical task is the enumeration of network substructures, such as common neighbourhoods and paths. 
For example, follower patterns inform social recommendations~\cite{DBLP:conf/www/GuptaGLSWZ13}, 
while path queries trace traffic flow in computer networks~\cite{DBLP:conf/nsdi/NarayanaTRW16}. 
Beyond analytics, these substructures are increasingly used as context for LLM inference in tasks such as recommendation~\cite{DBLP:conf/sigir/LiuY0ZGZLSG25} and network diagnosis~\cite{DBLP:conf/nsdi/WangALZGMWLZDJW24}. 

Most networks are stored as records in RDBMSs, 
where enumerating substructures translates into complex multi-way joins. 
However, join evaluation often becomes a scalability bottleneck because intermediate results can exceed input sizes by orders of magnitude. 
As a result, this large size limits the efficiency of both traditional RDBMSs and even specialized graph analytics frameworks~\cite{DBLP:journals/vldb/SahuMSLO20}. 
The problem is even more severe in emerging systems that interleave data analytics with semantic reasoning using LLMs~\cite{DBLP:conf/cidr/BiswalPJKLGGZ25,DBLP:journals/corr/abs-2407-11418,DBLP:conf/cidr/LiuRCCCCFKMSG25,DBLP:conf/sigmod/JoT24,DBLP:conf/sigmod/SatrianiVSRVP25}. 
In these systems, intermediate relations must be serialized as text and included in LLM prompts, 
making inference cost directly proportional to their size. 
This motivates our focus on join-heavy queries, including those augmented with LLM-powered semantic operators. 

A promising approach to mitigating the explosion in intermediate result size is to use factorized representations~\cite{DBLP:conf/icdt/OlteanuZ12}. 
They rewrite a multiset of flat tuples as 
a nested expression over unions and products by exploiting the distributivity of product over union, along with the commutativity of both operators. 
This representation factors tuples according to shared attribute-value prefixes: 
unions enumerate alternative values, while products combine subexpressions that vary independently under a shared prefix. 
For example, 
the tuples $(a_1,a_2,a_3)\in\{(0,1,4),(0,2,4),(0,3,4),(0,1,5),$ $(0,2,5),(0,3,5),(0,1,6),(0,2,6),(0,3,6)\}$ 
can be rewritten more succinctly as $(\cup_{a_1}\{0\} \times \cup_{a_2}\{1,2,3\} \times \cup_{a_3}\{4,5,6\})$. 
Factorization can yield representations whose size is asymptotically smaller than the maximum possible output size, known as the AGM bound~\cite{DBLP:conf/focs/AtseriasGM08}. 

Despite these theoretical advantages, 
adopting factorization in vectorized query engines remains challenging. 
Factorization organizes results hierarchically, grouping variable-length sets of values, 
whereas 
vectorized execution processes flat blocks of tuples for CPU-efficient execution over contiguous memory. 
A second challenge is maintaining state dynamically. 
Selections at one level may propagate to dependent values, causing \emph{cascading updates} across multiple levels of the hierarchy. 
This behaviour contrasts with vectorized execution, where tuple reduction is typically confined to a single selector. 
Existing systems such as FDB~\cite{DBLP:journals/pvldb/BakibayevOZ12,DBLP:journals/pvldb/BakibayevKOZ13} sidestep cascading updates by fully materializing 
fresh factorized results at each operator, at the cost of 
higher memory utilization and the loss of pipelining and cache-efficient processing. 
Other systems restrict factorization to simplify the engine, sacrificing much of the potential performance benefit~\cite{DBLP:journals/pvldb/GuptaMS21,DBLP:conf/cidr/JinFCLS23}. 

\subsection{Existing Vectorized Approaches}

Prior systems demonstrate the benefits of factorization, but remain limited in how they adopt it in vectorized pipelines. 
Kuzu~\cite{DBLP:conf/cidr/JinFCLS23} and Graphflow~\cite{DBLP:conf/sigmod/KankanamgeSMCS17} adopt list-based processing~(LBP)~\cite{DBLP:journals/pvldb/GuptaMS21}, which is pipelined but supports only restricted factorization layouts and yields sparsely populated vectors with high interpretation overhead. 
DuckPGQ~\cite{DBLP:conf/cidr/WoldeSSB23,DBLP:journals/pvldb/WoldeSB23} exploits factorization for specialized optimizations (\eg early aggregation), rather than as a general intermediate representation propagated across operators. 
What is missing is a unified execution model that treats intermediates as native factorized, packed vectors within the pipeline. 
Such a model preserves both compactness (smaller intermediates) and vectorization (cache-efficient CPU execution). 
\emph{Bridging these paradigms for join-heavy analytical queries with semantic operators is the central goal of this work}. 

Predicate-transfer  techniques~\cite{DBLP:journals/pvldb/BirlerKN24,DBLP:journals/sigmod/ZhaoSYYKZ25,DBLP:journals/pvldb/StoianZKNDKK25,DBLP:conf/cidr/YangZYK24} 
make Yannakakis-style processing practical~\cite{DBLP:conf/pvldb/Yannakakis81} by pruning dangling tuples early, thereby improving the robustness of query processing. 
However, they still enumerate large join results as flat tuples and do not propagate factorized intermediates between operators. 
Factorization and predicate-transfer are complementary: pruning reduces intermediates, while factorization compactly represents the output without flattening. 

We argue that \emph{passing factorized intermediates is essential for end-to-end co-optimization}. 
In this work, this approach optimizes semantic processing by allowing factorized join intermediates to flow directly into LLM-powered semantic operators. 
More broadly, the same idea could support other workloads, such as gradient boosting~\cite{DBLP:journals/pvldb/HuangSL023}, group-by-over-join analytics for decision trees and Rk-means~\cite{DBLP:journals/pvldb/SchleichO20}, and join-free feature engineering~\cite{DBLP:conf/sigmod/KumarNPZ16}.

\begin{table}[t!]
	\centering
	\small
	\setlength{\tabcolsep}{6pt}
	\begin{tabular}{lcccc}
		\toprule
		& \textbf{Flat} & \textbf{FDB} & \textbf{LBP} & \textbf{\textsc{FFX}} \\
		\midrule
		\textbf{Pipelined execution} 
		& \ding{51} & \ding{55} & \ding{51} & \ding{51} \\
		\textbf{Fully packed vectors} 
		& \ding{51} & \ding{55} & \ding{55}  & \ding{51} \\
		\textbf{Factorized intermediates} 
		& \ding{55} & \ding{51} & \ding{51} & \ding{51} \\
		\textbf{Supports arbitrary f-trees} 
		& \ding{55} & \ding{51} & \ding{55} & \ding{51} \\
		\bottomrule
	\end{tabular}
	\caption{Comparison of \textsc{FFX} with prior execution models. 
		We use \ding{51}/\ding{55} to denote support.}
	\label{tab:flat-listbased-ffx-comparison}
	\vspace*{-2.34em}
\end{table}

\subsection{Contributions}
\label{subsec:contributions}
We present \textsc{FFX}\footnote{https://github.com/dais-polymtl/ffx}, a novel query engine for \underline{\textbf{F}}ast \underline{\textbf{F}}actorized e\underline{\textbf{X}}ecution. 
\textsc{FFX} co-optimizes join-heavy analytics and LLM-powered semantic operators through the following contributions: 
\begin{itemize}
	\setlength{\itemsep}{0pt}
	\setlength{\parskip}{0pt}
	\setlength{\parsep}{0pt}
	\item \emph{Packed factorized vectors.} (\sect{sec:vector-representation}) 
	We propose a new representation for intermediates that supports arbitrary factorization while enabling fully packed, contiguous in-memory vectors. 
	\item \emph{Vectorization with graceful fallback.} (\sect{subsec:join-evaluation}) 
	We adapt operators to execute over packed factorized vectors in tight loops without pointer chasing. 
	When factorization offers limited benefit, \textsc{FFX} falls back to standard vectorized execution without overhead.
	\item \emph{Cascade update operator.} (\sect{sec:cascade-update}) 
	We propose a new operator to propagate updates, such as tuple reductions, across the factorized hierarchy without breaking vectorized execution. 
	\item \emph{Factorized semantic processing.} (\sect{sec:prompt-compression}) 
	We adapt the LLM-powered operators of Flock~\cite{DBLP:journals/pvldb/DorbaniYLM25} to consume serialized factorized vectors and materialize Cartesian combinations as part of their output. 
	This substantially reduces input token counts and, in turn, inference cost. 
\end{itemize}

\tab{tab:flat-listbased-ffx-comparison} compares \textsc{FFX} with prior work. 
Our experiments (\sect{sec:evaluation}) demonstrate that factorized and vectorized execution are complementary. 
\textsc{FFX} achieves 
major speedups on heavy-join workloads over state-of-the-art systems and maintains comparable or improved accuracy for semantic operators over flat representations while reducing inference cost. 

\section{Background}
\label{sec:background}

\textsc{FFX} builds on three foundations: vectorized execution, factorized query processing, and semantic query operators. We provide the necessary background on each below. 

\subsection{Vectorized Execution}
\label{sec:background-vectorized-execution}

Modern analytical DBMSs widely employ \emph{vectorized execution} for query processing, as implemented in systems such as Photon~\cite{DBLP:conf/sigmod/BehmPAACDGHJKLL22}, 
Apache DataFusion~\cite{DBLP:conf/sigmod/LambSHCKHS24}, 
Velox~\cite{DBLP:journals/pvldb/PedreiraEBWSPHC22}, 
DuckDB~\cite{DBLP:conf/sigmod/RaasveldtM19}, 
and DB2~\cite{DBLP:journals/pvldb/RamanABCKKLLLLMMPSSSSZ13}. 
Unlike traditional tuple-at-a-time processing in Volcano~\cite{DBLP:journals/tkde/Graefe94} or full materialization in MonetDB~\cite{monetdb-thesis}, 
vectorized engines follow a \emph{block-at-a-time} model in which operators process small, fixed-size batches of tuples in a columnar layout. 
Pioneered by MonetDB/X100~\cite{DBLP:conf/cidr/BonczZN05} and VectorWise~\cite{DBLP:conf/icde/ZukowskiWB12}, vectorized execution evaluates fixed-size contiguous vectors in tight loops. 
This design is well suited to modern superscalar CPUs: it yields cache-friendly execution, amortizes interpretation overhead, hides memory latency, reduces branching, and thereby improves instructions per cycle~\cite{DBLP:journals/pvldb/KerstenLKNPB18}. 

\subsection{Factorized Query Processing}
\label{sec:background-factorized-query-processing}

Factorized representations mitigate the intermediate-result explosion in many-to-many joins by compactly representing join outputs~\cite{DBLP:journals/tods/OlteanuZ15}. 
They exploit the conditional independence induced by joins to express results as algebraic combinations of products and unions or, equivalently, as vertical and horizontal partitions. 
By avoiding or delaying full Cartesian expansion, factorized representations eliminate redundancy and can reduce intermediate result size below the worst-case output bound, namely, the AGM bound. 
We introduce f-representations, the specific factorized representation used in this work, and then provide an overview of representative factorized query engines. 
We also note that more compact representations exist, namely d-representations. 
However, such representations require materialization, whereas our focus is on pipelined execution. 

\subsubsection{F-representations}
\label{subsec:background-f-representations}

A relation over $n$ attributes is typically represented as a multiset of flat $n$-ary tuples, \ie a union of tuples. 
An \emph{f-representation} compresses this form into a nested algebraic expression by exploiting the distributivity of product over union and the commutativity of both operators. 
Specifically, an f-representation is a nested expression over unions, products, and attribute values. 
The expression has a trie-like structure that factors tuples by their shared attribute-value prefixes. 
Unions enumerate alternative values of an attribute, 
while products over unions combine subexpressions that vary independently given the current prefix of values 
(\ie along any root-to-leaf path, values under the same union share an attribute-value prefix). 
Flattening distributes products over unions, producing the Cartesian combinations needed to recover flat tuples.  

We say that an f-representation is over an \emph{f-tree} $\mathcal{T}$. 
An f-tree is a rooted tree specifying how attributes are grouped: each node corresponds to one attribute, and each parent-child relationship, indicated by an edge $x - y$, means that values of attribute $y$ (the child) are stored grouped by a given value of attribute $x$ (the parent), together with all ancestors of $x$. 
Equivalently, every root-to-leaf path $a_{i_1} - a_{i_2} - \cdots - a_{i_k}$ defines a nesting order in which a union of $a_{i_{j+1}}$ values appears under each fixed prefix assignment $(a_{i_1},\dots,a_{i_j})$. 
Branching in f-trees captures conditional independence: if a node $x$ has children $y$ and $z$, then the subrelations rooted at $y$ and $z$ are 
represented independently for each fixed assignment to the attributes on the path from the root to $x$. 
Intuitively, branching avoids enumerating Cartesian products between sibling substructures, because their combinations are implied by the product in the factorized expression rather than listed as flat tuples. 

\begin{figure*}[t!]
	\centering
	\begin{subfigure}[b]{0.3\textwidth}
		\centering
		\captionsetup{justification=centering}
		\begin{tabular}{|>{\centering\arraybackslash}p{0.7cm}>{\centering\arraybackslash}p{0.7cm}| }
			\hline
			$src$ & $dst$ \\
			\hline
			0 & 2 \\
			1 & 2 \\
			1 & 3 \\
			1 & 4 \\
			2 & 5 \\
			2 & 6 \\
			2 & 7 \\
			3 & 2 \\
			3 & 4 \\
			4 & 8 \\
			\hline
		\end{tabular}
		\caption{Tabular format.}
		\label{subfig:input-relation-r}
	\end{subfigure}
	\begin{subfigure}[b]{0.3\textwidth}
		\centering
		\captionsetup{justification=centering}
		\begin{tikzpicture}[scale=0.4, transform shape,->,>=stealth', shorten >=1pt, auto,node distance=2.8cm, thick, main node/.style={circle,draw,font=\sffamily\Huge\bfseries,minimum size=13mm}, text node/.style={font=\sffamily\Huge}]
			\node[main node] (1) {$1$};
			\node[main node] (3) [right of=1]  {$3$};
			\node[main node] (2) [above of=3]  {$2$};
			\node[main node] (9) [left of=2]  {$0$};
			\node[main node] (10) [right of=2]  {$6$};
			\node[main node] (4) [below of=3]  {$4$};
			\node[main node] (5) [above right of=2]  {$5$};
			\node[main node] (6) [below right of=2]  {$7$};
			\node[main node] (7) [right of=4]  {$8$};
			\path[every node/.style={font=\sffamily\Huge}]
			(1) edge  (2) edge (3) edge (4)
			(2) edge (5) edge (6) edge (10)
			(3) edge (2) edge (4)
			(4) edge (7) 
			(9) edge (2)
			;
		\end{tikzpicture}
		\caption{Graph format.}
		\label{subfig:input-relation-as-graph}
	\end{subfigure}\vspace*{0.25em}
	
	\begin{subfigure}[b]{0.1\textwidth}
		\centering
		\captionsetup{justification=centering}
		\begin{tikzpicture}[
			level 1/.style={sibling distance=0.8cm,level distance=0.7cm},
			level 2/.style={sibling distance=0.8cm,level distance=0.7cm}]
			\node { $a_1$ }
			child { node { $a_2$ }
				child { node { $a_3$ } }
			}
			;
		\end{tikzpicture}
		\caption{$\mathcal{T}_1$.}
		\label{fig:t-1}
	\end{subfigure}
	\begin{subfigure}[b]{0.3\textwidth}
		\centering
		\captionsetup{justification=centering}
		\begin{tikzpicture}[every node/.style={circle,draw},
			level  1/.style={sibling distance=1.6cm,level distance=0.65cm},
			level  2/.style={sibling distance=0.86cm,level distance=0.65cm},
			level  3/.style={sibling distance=0.8cm,level distance=0.65cm},
			level  4/.style={sibling distance=0.6cm,level distance=0.65cm},
			level  7/.style={sibling distance=0.25cm,level distance=0.65cm}]
			\node [rectangle,draw=none] {$\cup_{a_1}$}
			child { node [rectangle,draw=none] {$0$} 
				child { node [rectangle,draw=none] {$\times$}
					child { node [rectangle,draw=none] {$\cup_{a_2}$}
						child { node [rectangle,draw=none] {$2$}
							child { node [rectangle,draw=none] {$\times$}
								child { node [rectangle,draw=none] {$\cup_{a_3}$}
									child { node [rectangle,draw=none] {$5$} }
									child { node [rectangle,draw=none] {$6$} }
									child { node [rectangle,draw=none] {$7$} }
								}
							}
						}
					}
				}
			}
			child { node [rectangle,draw=none] {$1$} 
				child { node [rectangle,draw=none] {$\times$}
					child { node [rectangle,draw=none] {$\cup_{a_2}$}
						child { node [rectangle,draw=none] {$2$}
							child { node [rectangle,draw=none] {$\times$}
								child { node [rectangle,draw=none] {$\cup_{a_3}$}
									child { node [rectangle,draw=none] {$5$} }
									child { node [rectangle,draw=none] {$6$} }
									child { node [rectangle,draw=none] {$7$} }
								}
							}
						}
						child { node [rectangle,draw=none] {$3$}
							child { node [rectangle,draw=none] {$\times$}
								child { node [rectangle,draw=none] {$\cup_{a_3}$}
									child { node [rectangle,draw=none] {$2$} }
									child { node [rectangle,draw=none] {$4$} }
								}
							}
						}
						child { node [rectangle,draw=none] {$4$}
							child { node [rectangle,draw=none] {$\times$}
								child { node [rectangle,draw=none] {$\cup_{a_3}$}
									child { node [rectangle,draw=none] {$8$} }
								}
							}
						}
					}
				}
			}
			child { node [rectangle,draw=none] {$3$} 
				child { node [rectangle,draw=none] {$\times$}
					child { node [rectangle,draw=none] {$\cup_{a_2}$}
						child { node [rectangle,draw=none] {$2$}
							child { node [rectangle,draw=none] {$\times$}
								child { node [rectangle,draw=none] {$\cup_{a_3}$}
									child { node [rectangle,draw=none] {$5$} }
									child { node [rectangle,draw=none] {$6$} }
									child { node [rectangle,draw=none] {$7$} }
								}
							}
						}
						child { node [rectangle,draw=none] {$4$}
							child { node [rectangle,draw=none] {$\times$}
								child { node [rectangle,draw=none] {$\cup_{a_3}$}
									child { node [rectangle,draw=none] {$8$} }
								}
							}
						}
					}
				}
			}
			;
		\end{tikzpicture}
		\caption{output over $\mathcal{T}_1$.}
		\label{subfig:t1-rep}
	\end{subfigure}
	\begin{subfigure}[b]{0.16\textwidth}
		\centering
		\captionsetup{justification=centering}
		\begin{tikzpicture}[
			level 1/.style={sibling distance=0.8cm,level distance=0.7cm},
			level 2/.style={sibling distance=0.8cm,level distance=0.7cm}]
			\node { $a_2$ }
			child { node { $a_1$ } }
			child { node { $a_3$ } }
			;
		\end{tikzpicture}
		\caption{$\mathcal{T}_2$.}
		\label{fig:t-2}
	\end{subfigure}
	\begin{subfigure}[b]{0.3\textwidth}
		\centering
		\captionsetup{justification=centering}
		\begin{tikzpicture}[every node/.style={circle,draw},
			level  1/.style={sibling distance=1.45cm,level distance=0.7cm},
			level  2/.style={sibling distance=0.8cm,level distance=0.7cm},
			level  3/.style={sibling distance=0.8cm,level distance=0.7cm},
			level  4/.style={sibling distance=0.25cm,level distance=0.7cm}]
			\node [rectangle,draw=none] {$\cup_{a_2}$}
			child { node [rectangle,draw=none] {$2$} 
				child { node [rectangle,draw=none] {$\times$}
					child { node [rectangle,draw=none] {$\cup_{a_1}$}
						child { node [rectangle,draw=none] {$0$} }
						child { node [rectangle,draw=none] {$1$} }
						child { node [rectangle,draw=none] {$3$} }
					}
					child { node [rectangle,draw=none] {$\cup_{a_3}$}
						child { node [rectangle,draw=none] {$5$} }
						child { node [rectangle,draw=none] {$6$} }
						child { node [rectangle,draw=none] {$7$} }
					}
				}
			}
			child { node [rectangle,draw=none] {$3$} 
				child { node [rectangle,draw=none] {$\times$}
					child { node [rectangle,draw=none] {$\cup_{a_1}$}
						child { node [rectangle,draw=none] {$1$} }
					}
					child { node [rectangle,draw=none] {$\cup_{a_3}$}
						child { node [rectangle,draw=none] {$2$} }
						child { node [rectangle,draw=none] {$4$} }
					}
				}
			}
			child { node [rectangle,draw=none] {$4$} 
				child { node [rectangle,draw=none] {$\times$}
					child { node [rectangle,draw=none] {$\cup_{a_1}$}
						child { node [rectangle,draw=none] {$1$} }
						child { node [rectangle,draw=none] {$3$} }
					}
					child { node [rectangle,draw=none] {$\cup_{a_3}$}
						child { node [rectangle,draw=none] {$8$} }
					}
				}
			}
			;
		\end{tikzpicture}
		\caption{output over $\mathcal{T}_2$.}
		\label{subfig:t2-rep}
	\end{subfigure}
	\caption{Example relation $R(src,dst)$ in tabular and graph formats, and factorized outputs of $R(a_1,a_2)\Join R(a_2,a_3)$ over f-trees $\mathcal{T}_1$ and $\mathcal{T}_2$.} 
	\label{fig:input-relation-example}
	\vspace*{-1em}
\end{figure*}
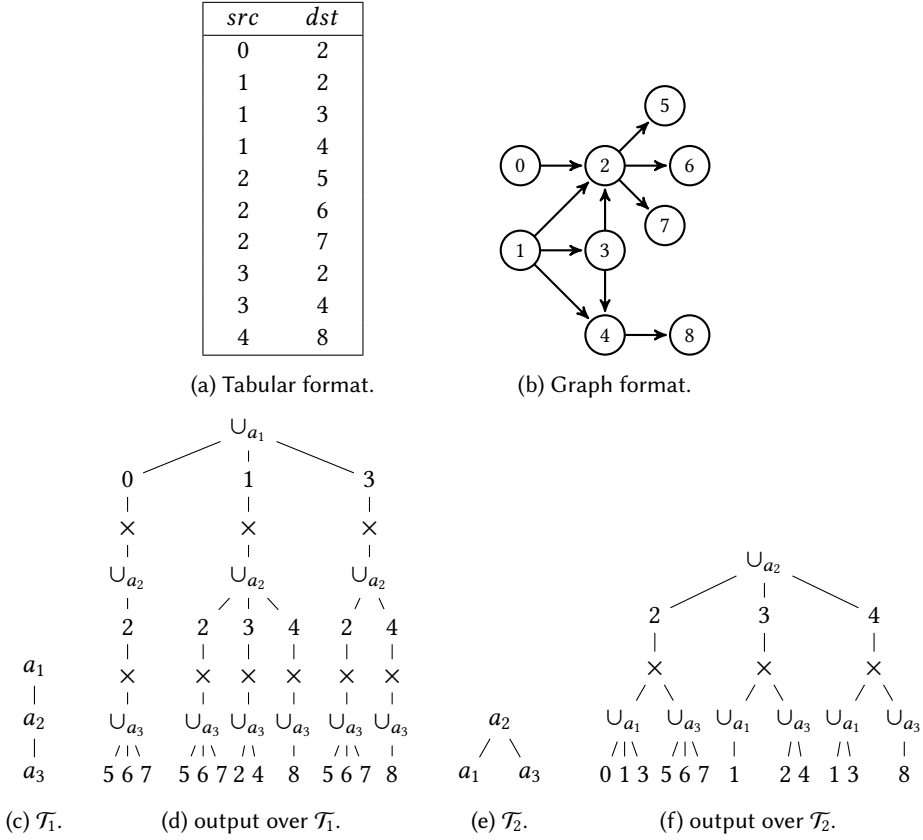

We illustrate f-representations and f-trees with a concrete example. For a formal treatment, we refer the reader to Olteanu \etal~\cite{DBLP:journals/tods/OlteanuZ15}. 
\setlist[enumerate]{leftmargin=*}
Consider the base relation $R(src,dst)$ in \fig{subfig:input-relation-r} and the query 
$Q_{2H} = R(a_1,a_2) \Join R(a_2,a_3)$. 
The output of $Q_{2H}$ contains $13$ flat tuples, for a total of $39$ atomic values, with substantial repetition.

Figures~\ref{subfig:t1-rep} and~\ref{subfig:t2-rep} show factorized outputs of $Q_{2H}$ over the \emph{f-trees} $\mathcal{T}_1$ and $\mathcal{T}_2$. 
An f-tree specifies the grouping layout. 
For instance, $\mathcal{T}_1$ groups values as $a_1 - a_2 - a_3$, mirroring flat tuple enumeration except that repeated prefixes are removed, resulting in $22$ atomic values.  
In contrast, $\mathcal{T}_2$ exploits conditional independence and groups values as $a_2 - a_1$ and $a_2 - a_3$, yielding a more compact intermediate encoding of $15$ values. 
In general, more branching and smaller height imply greater compactness. 
We use the arrow notation and the visual tree diagrams interchangeably. 

These are examples of data-independent factorization layouts: the same f-tree applies regardless of the specific input instance. 
Valid f-trees must satisfy the \emph{path constraint}~\cite{DBLP:journals/tods/OlteanuZ15}, which requires that any two dependent attributes, \ie attributes in the same relation or linked via a chain of join predicates with all join keys projected out, appear along the same root-to-leaf path.

\subsubsection{Factorized Engines --- FDB~\cite{DBLP:journals/pvldb/BakibayevOZ12,DBLP:journals/pvldb/BakibayevKOZ13}} 
\label{subsubsec:fdb} 
FDB was the first engine to evaluate queries over f-representations. 
Like MonetDB~\cite{monetdb-thesis}, it fully materializes operator outputs. 
Each operator produces a fresh factorized result, represented as a multi-level trie, for the next operator to consume. 
This design avoids complex in-place hierarchical updates during joins and filters, but requires reconstructing full factorized intermediates at every operator. 
While factorization reduces the size of intermediates, the full-materialization model forfeits cache locality and pipelined execution, leading to excessive memory traffic from iterating over, copying, and rebuilding factorized intermediates. 

Subsequent work further optimized FDB by refining the underlying data structures to reduce pointer chasing and improve locality through more contiguous memory layouts~\cite{Lombardo2016StorageLayerFactorized}. 
While these optimizations improve the efficiency of reading and writing factorized representations, the operators' execution model still relies on fully materializing intermediate results. 

\subsubsection{Factorized Engines --- List-based Processing (LBP)~\cite{DBLP:journals/pvldb/GuptaMS21}} 
\label{subsubsec:list-based-processing}

LBP improves on FDB by adopting factorization into pipelined execution. 
Graph DBMSs such as Graphflow~\cite{DBLP:conf/sigmod/KankanamgeSMCS17} and Kuzu~\cite{DBLP:conf/cidr/JinFCLS23} operate on \emph{factorized vectors}~\cite{DBLP:journals/ftdb/MhedhbiDS24} aligned to adjacency lists~\cite{DBLP:conf/icde/MhedhbiGKS21}. 

Existing implementations use worst-case-optimal join plans. 
Each plan is defined by a query attribute ordering. 
Query evaluation follows that ordering: it first enumerates candidate values for the first attribute, namely those common to the base relations that contain it, and then extends the bindings one attribute at a time by applying a binary or multi-way join over the relations involving the next attribute, conditioned on the bound values. 

\begin{figure}[t!]
	\centering
	\begin{subfigure}[b]{0.32\columnwidth}
		\centering
		\includegraphics[width=\textwidth]{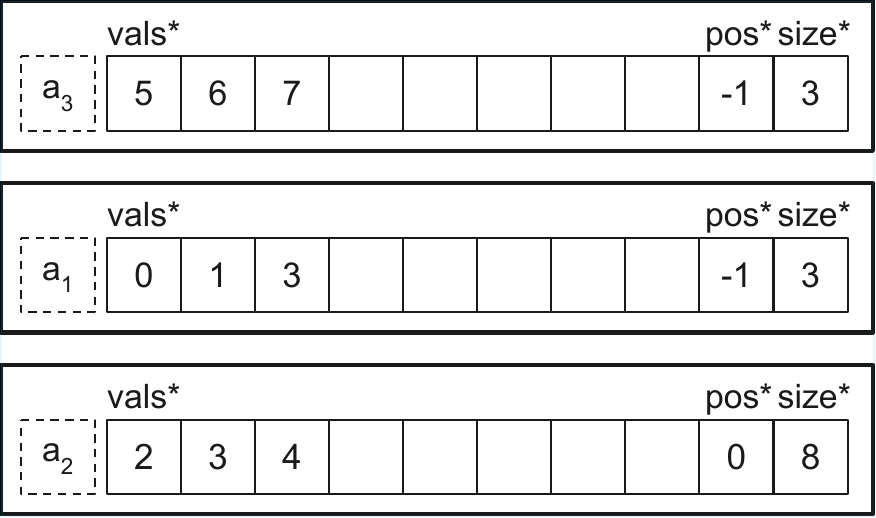}
	\end{subfigure} 
	\caption{First output vectors produced by list-based processing~\cite{DBLP:journals/pvldb/GuptaMS21} for $R(a_1,a_2)\Join R(a_2,a_3)$ 
		and ordering $(a_2,a_1,a_3)$ evaluated on the base relation $R$ in~\fig{subfig:input-relation-r}.}
	\label{fig:listbased-output-vectors}
\end{figure}

LBP allocates fixed-size vectors (\eg of size $2048$), and each vector has a \texttt{state} containing its \texttt{size}, a position field (\texttt{pos}), and a selection array (\texttt{sel}). 
We explain the vector state shortly through a concrete example. 
Consider $R(a_1,a_2)\bowtie R(a_2,a_3)$ and the ordering $(a_2,a_1,a_3)$, which yields the pipeline 
\texttt{Scan[$a_2$]$\rightarrow$Join[$R(a_1,a_2)$]$\rightarrow$Join[$R(a_2,a_3)$]}. 
After one call to \texttt{Scan} and each subsequent \texttt{Join}, a subset of the output relation, also called a \emph{chunk}, is stored in the output vectors shown in \fig{fig:listbased-output-vectors}. 
We next explain step by step how these vector states are obtained. 

The \texttt{Scan} produces a vector of $a_2$ values common to $R(a_1,a_2)$ and $R(a_2,a_3)$, namely $\{2,3,4\}$, 
with \texttt{pos} $=-1$, marking the vector as a list of values, and \texttt{size} $=3$. 
The output vector of a \texttt{Scan} thus represents a one-attribute relation, equivalent to an f-representation over a single-node f-tree rooted at $a_2$. 
When \texttt{pos} \(\ge 0\), the vector is single-valued and bound to the $a_2$ value at position \texttt{pos}. 

The first \texttt{Join[$R(a_1,a_2)$]} is many-to-many, 
\ie each $a_2$ value can match multiple tuples in $R(a_1,a_2)$. 
Thus, LBP binds $a_2$ to one value at a time by iterating 
\texttt{pos} $\in\{0,1,2\}$, \ie $a_2\in\{2,3,4\}$, until it finds non-empty matches. 
This iteration advances \texttt{pos} from $-1$ to $0$, thereby binding $a_2=2$, 
and produces a vector of $a_1$ matches, namely $\{0,1,3\}$, with \texttt{pos} $=-1$ and \texttt{size} $=3$. 
Logically, the intermediate relation represented by these two vectors is 
$(a_2,a_1)\in\{\cup_{a_2}\{2\} \times \cup_{a_1}\{0,1,3\}\}$, which is equivalent to the flat tuples $(a_2,a_1)\in\{(2,0),(2,1),(2,3)\}$. 
Since $a_1$ depends on $a_2$, the two output vectors are equivalent to an f-representation over an f-tree with $a_2$ as the root and $a_1$ as its child. 

The second \texttt{Join[$R(a_2,a_3)$]} proceeds similarly, 
but does not need to iterate because $a_2$ is already single-valued, bound to $2$ with \texttt{pos} $=0$. 
This produces a vector of $a_3$ matches, namely $\{5,6,7\}$, with \texttt{pos} $=-1$ and \texttt{size} $=3$. 
Logically, the intermediate relation represented by these three vectors is 
$(\cup_{a_2}\{2\}\times \cup_{a_1}\{0,1,3\}\times \cup_{a_3}\{5,6,7\})$, which is equivalent to the flat tuples 
$(a_2,a_1,a_3)\in\{(2,0,5),(2,0,6),(2,0,7),(2,1,5),$ $(2,1,6),(2,1,7),(2,3,5),(2,3,6),(2,3,7)\}$. 
Since $a_3$ depends on $a_2$, the three output vectors are equivalent to an f-representation over the f-tree $\mathcal{T}_2$ in~\fig{fig:t-2}. 
These tuples form the first output chunk of the relation. 
Execution then backtracks to \texttt{Join[$R(a_1,a_2)$]} and resumes by advancing \texttt{pos} from $0$ to $1$. 

As the walkthrough and \fig{fig:listbased-output-vectors} illustrate, 
LBP materializes each node of $\mathcal{T}_2$ (\ie each output attribute) as a separate factorized vector. 
These vectors maintain the f-tree's parent-child groupings implicitly through their state: 
internal nodes become single-valued (\texttt{pos}$\ge 0$) because join evaluation iterates over one parent binding at a time, while child nodes are produced as lists of values (\texttt{pos}$=-1$) conditioned on the current single-valued parent. 
As a result, child vectors are populated at the granularity of join matches, typically tens to hundreds of values in network datasets, rather than being filled to the fixed vector width. 
\fig{fig:listbased-output-vectors} shows most slots in the $8$-wide vectors empty, which is a major shortcoming: sparsely populated vectors incur high interpretation overhead and underutilize caches relative to fully packed flat vectors~\cite{DBLP:journals/ftdb/MhedhbiDS24}. 
Moreover, orderings equivalent to flat tuple evaluation can degenerate into a few-tuples-at-a-time execution. 
For example, the ordering $(a_1,a_2,a_3)$ yields an output over $\mathcal{T}_1$ in~\fig{fig:t-1}, 
which offers no factorization benefit and therefore performs worse than vectorized engines because of higher interpretation cost.

Finally, LBP supports only a restricted class of f-trees. 
Because parent-child groupings are maintained implicitly through the vector state, LBP supports only f-trees with limited branching, \ie at any node, at most one child branch can extend beyond a height of 1. 
For example, for the query $R(a_1,a_2)\Join R(a_2,a_3)\Join R(a_3,a_4)\Join R(a_4,a_5)$, the most compact f-tree $\mathcal{T}_3$ in~\fig{fig:t-3} yields f-representations with worst-case size $N^2$: 
branching at $a_3$ separates independent subqueries conditioned on $a_3$ values, 
so the representation size becomes the sum of the sizes of $R(a_1,a_2)\Join R(a_2,a_3)$ and $R(a_3,a_4)\Join R(a_4,a_5)$ rather than their Cartesian product. 
In contrast, LBP can realize only $\mathcal{T}_4$ or $\mathcal{T}_5$, shown in~\fig{fig:t-4} and~\fig{fig:t-5}, respectively, both of which have worst-case size $N^3$. 
Thus, because of LBP's intermediate layouts, substantial compression benefits remain unattainable.

\begin{figure}[t!]
	\centering
	\captionsetup{justification=centering}
	\begin{subfigure}[b]{0.15\textwidth}
		\centering
		\captionsetup{justification=centering}
		\begin{tikzpicture}[
			level 1/.style={sibling distance=0.8cm,level distance=0.7cm},
			level 2/.style={sibling distance=0.8cm,level distance=0.7cm}]
			\node { $a_3$ }
			child { node { $a_2$ }
				child { node { $a_1$ } }
			}
			child { node { $a_4$ }
				child { node { $a_5$ } }
			}
			;
		\end{tikzpicture}
		\caption{$\mathcal{T}_3$.}
		\label{fig:t-3}
	\end{subfigure}
	\begin{subfigure}[b]{0.15\textwidth}
		\centering
		\captionsetup{justification=centering}
		\begin{tikzpicture}[
			level 1/.style={sibling distance=0.8cm,level distance=0.7cm},
			level 2/.style={sibling distance=0.8cm,level distance=0.7cm}]
			\node { $a_3$ }
			child { node { $a_2$ }
				child { node { $a_1$ } }
				child { node { $a_4$ }
					child { node { $a_5$ } }
				}
			}
			;
		\end{tikzpicture}
		\caption{$\mathcal{T}_4$.}
		\label{fig:t-4}
	\end{subfigure}
	\begin{subfigure}[b]{0.15\textwidth}
		\centering
		\captionsetup{justification=centering}
		\begin{tikzpicture}[
			level 1/.style={sibling distance=0.8cm,level distance=0.7cm},
			level 2/.style={sibling distance=0.8cm,level distance=0.7cm}]
			\node { $a_3$ }
			child { node { $a_4$ }
				child { node { $a_2$ }
					child { node { $a_1$ } }
				}
				child { node { $a_5$ } }
			}
			;
		\end{tikzpicture}
		\caption{$\mathcal{T}_5$.}
		\label{fig:t-5}
	\end{subfigure}
	\vspace*{-0.5em}
	\caption{Valid f-trees for $R(a_1,a_2)$\hspace*{0.19em}$\bowtie$\hspace*{0.19em}$R(a_2,a_3)$\hspace*{0.19em}$\bowtie$\hspace*{0.19em}$R(a_3,a_4)$\hspace*{0.19em}$\bowtie$\hspace*{0.19em}$R(a_4,a_5)$.}
	\label{fig:ftrees-supported}
	\vspace*{-0.5em}
\end{figure}

\subsection{Semantic LLM Operators}
\label{subsec:semantic-llm-operators}

A new wave of analytical-semantic DBMSs (\textsc{Tag}~\cite{DBLP:conf/cidr/BiswalPJKLGGZ25}, \textsc{Palimpsest}~\cite{DBLP:conf/cidr/LiuRCCCCFKMSG25}, \textsc{Lotus}~\cite{DBLP:journals/corr/abs-2407-11418}, \textsc{ThalamusDB}~\cite{DBLP:conf/sigmod/JoT24}, and \textsc{Galois}~\cite{DBLP:conf/sigmod/SatrianiVSRVP25}) integrate LLMs into query execution via semantic operators over relations, such as summarization, classification, and extraction. 
In these systems, LLMs are invoked within scalar or aggregate functions applied in \texttt{Projection} operators over intermediate relations. 
However, LLMs consume text, so intermediates must be serialized as text and included in prompts. 

Consider the relations \textsf{Paper(id, title, abstract)} and \textsf{Citation(src, dst)}, where \textsf{src} and \textsf{dst} reference \textsf{Paper.id}. 
A tuple \textsf{Citation(src, dst)} denotes a directed citation from \textsf{src} to \textsf{dst} (\textsf{src}$\rightarrow$\textsf{dst}). 
Suppose we want to apply an LLM operator to triples of papers connected by two consecutive citations, $P_1\rightarrow P_2\rightarrow P_3$, for example, to extract technical terms common to all three papers. 
The query below joins the two citation relations with the corresponding paper records and then applies an LLM \texttt{map} operator to attributes in the resulting intermediate relation: 
\begin{lstlisting}[language=SQL]
	SELECT llm_map({'model': '...', 'prompt': '...', 
		'relevant_columns': ['P1.title','P1.abstract',
		'P2.title','P2.abstract','P3.title','P3.abstract']})
	FROM Citation AS C1
	JOIN Citation AS C2 ON C2.src = C1.dst
	JOIN Paper AS P1 ON P1.id = C1.src
	JOIN Paper AS P2 ON P2.id = C1.dst
	JOIN Paper AS P3 ON P3.id = C2.dst;
\end{lstlisting}

Semantic operators (\eg \texttt{llm\_map}) declare a set of relevant columns whose values are serialized into the prompt. 
To amortize prompt overhead, such as a shared task description and instructions, these operators batch multiple tuples into a single LLM call using a shared prompt template. 
In current systems, each batch is typically serialized as a flat table (Markdown, JSON, or XML) with one row per tuple. 
This layout mirrors the underlying flat-tuple intermediates and therefore inherits their redundancy under intermediate-result explosion.

\section{System Overview}
\label{sec:overview-and-design-principles}

We revisit the fundamental question of how to unify factorized and vectorized execution in a single engine. 
Our point of departure is list-based processing (LBP)~\cite{DBLP:journals/pvldb/GuptaMS21}, which adopts factorization in vectorized execution by aligning vectors with adjacency lists. 
While promising, as we showed in \sect{subsubsec:list-based-processing}, LBP underutilizes modern CPUs due to sparse vectors and interpretation overhead. It is also limited in the factorization layouts it can realize. 
At the same time, recent semantic DBMSs invoke LLMs on join outputs. Since these intermediates are flat tuples and are serialized as text for prompting, they inherit the redundancy of the relational model. 
This motivates a broader systems question that guides \textsc{FFX}: 
\emph{Can we preserve factorization in a fully vectorized pipeline, including as input to semantic operators?} 

\textsc{FFX} addresses these challenges with three core mechanisms:
(i) \emph{packed factorized vectors} that encode f-tree groupings while preserving fully packed, contiguous vectors~(\sect{sec:vector-representation});
(ii) a \emph{cascade update} operator that propagates tuple reductions across the grouping hierarchy within the vectorized execution model~(\sect{sec:cascade-update}); and
(iii) a \emph{structure-aware iterator and prompt serializer} that emits text derived from factorized intermediates for semantic operators without flattening~(\sect{sec:prompt-compression}). 
When factorization provides no benefit, \textsc{FFX} falls back to standard flat vectorized execution with no overhead.

\section{Packed Factorized Vectors}
\label{sec:vector-representation}

\textsc{FFX} introduces factorized vectors as an extension of the standard vector layout, also called \emph{data chunks}. 
By piggybacking on the existing vector abstraction for intermediates, we implement factorized vectors as an overloaded vector type, following the same approach used to support multiple vector types in DuckDB~\cite{DBLP:conf/sigmod/RaasveldtM19} and Kuzu~\cite{DBLP:conf/cidr/JinFCLS23}. 
This deliberately constrains factorized vectors to fixed-size, contiguous memory blocks, preserving cache locality and enabling vectorized execution. 

In LBP, parent-child groupings in the f-tree are maintained implicitly through the shared vector state. 
Join evaluation therefore binds each internal-node vector to one parent value at a time (\texttt{pos} $\ge 0$), while producing each child vector as a list of matches (\texttt{pos} $=-1$) conditioned on that current binding. 
This design induces two limitations: 
(i) child vectors are populated at the granularity of adjacency lists, \ie join matches, yielding sparse vectors and high interpretation overhead; and 
(ii) the implicit encoding restricts the supported f-trees to limited branching, \ie only one branch of a node can extend beyond height~1. 

The challenge is then twofold:
(i) how to support many-to-many join expansion without degenerating into LBP; and 
(ii) how to encode an arbitrary f-tree layout within a physically flat, packed vector. 
Addressing (i) requires preserving fully packed vectors at all times, whereas 
addressing (ii) requires capturing hierarchy explicitly within contiguous memory blocks. 
We refer to our solution as \emph{packed factorized vectors}, in contrast to the unpacked LBP approach, which cannot express arbitrary hierarchy or maintain packed vectors. 

\subsection{Vector Representation} 
\label{subsec:representation}

To support fully packed vectors, a join-key vector must remain as a list of values after a many-to-many join while preserving the parent-child groupings implied by the f-tree. 
Our solution is a packed, offset-based representation with a new \texttt{State}. 
A parent-child grouping is encoded by an offset array (\texttt{off}) that maps each parent entry to a range in the child vector. 
Concretely, the values associated with the parent at position $i$ occupy the slice $(\texttt{off}[i],\,\texttt{off}[i{+}1])$ in the child vector.

The \texttt{State} additionally stores the \texttt{start} and \texttt{end} positions to bound the active slice of a parent vector, as joins enumerate multiple child matches per parent value and are therefore iterated over in slices. 
Finally, we represent the selector as a bit array: it supports constant-size, in-place updates, which better suit our cascade operator (\sect{sec:cascade-update}). 
The slice length is $\texttt{end}-\texttt{start}$, while the number of active values is the number of set bits in the selector over that slice. 

The fields of the packed factorized vector \texttt{State} are as follows:

\lstdefinestyle{ffxcode}{
	basicstyle=\ttfamily\small,
	keywordstyle=\bfseries\color{blue},
	commentstyle=\itshape\color{gray},
	showstringspaces=false,
	columns=fullflexible,
	keepspaces=true,
	frame=single,
	framerule=0.3pt,
	aboveskip=0.6em,
	belowskip=0.6em,
	xleftmargin=0.5em,
	framexleftmargin=0.5em
}

\begin{lstlisting}[style=ffxcode,language=C++]
struct State {
	uint16_t start;
	uint16_t end;
	alignas(64) uint64_t sel[MAX_SIZE / 64]; // a single bit per position
	alignas(64) uint16_t off[MAX_SIZE];      // offset array
};
\end{lstlisting}

The vector packages the values, which are stored as a contiguous byte array, together with a pointer to a \texttt{State}. 
The byte array supports multiple data types. 
For variable-length data types, entries store the size, an inline prefix, and a pointer into an overflow block that carries the remaining bytes. 
Some scans and joins produce multiple vectors that share the same \texttt{State}, \eg dependent attributes of the same relationship that must be filtered or iterated over in lockstep, while other vectors carry their own state. 
We allocate \texttt{State} objects from an arena to ensure stable lifetimes and good locality; values and their associated state are co-located when possible, and shared-state vectors simply point to the same \texttt{State}. 

We order the \texttt{State} fields to match operator access patterns so that operators can scan contiguous bytes in tight loops without the overhead of virtual-function dispatch. 
Algorithm~\ref{alg:iterate-factorized-vector} sketches the common iteration pattern for scanning active positions using count trailing zeros (CTZ). 

\algtext*{EndIf}
\algtext*{EndFor}
\algtext*{EndWhile}
\algtext*{EndFunction}

\begin{algorithm}[h!]
	\caption{Iterating over a packed factorized vector}
	\label{alg:iterate-factorized-vector}
	\begin{algorithmic}[1]
		\Require $\text{vec}$ -- \emph{packed factorized vector}
		\State $sel \gets \text{vec.state.sel}$
		\State $start \gets \text{vec.state.start}$,\; $end \gets \text{vec.state.end}$
		\While{$\text{pos} \gets \text{next\_valid} (\text{sel}, \text{start}, \text{end})$}
		\State ... \textit{/* op-specific processing of $\text{vec.vals[pos]}$ */}
		\EndWhile
	\end{algorithmic}
\end{algorithm}

\subsubsection{Evaluation Example}
\label{subsec:evaluation-example}

We use the running example from \sect{subsec:background-f-representations}: $Q_{2H} = R(a_1,a_2)\Join R(a_2,a_3)$, evaluated with the pipeline 
\texttt{Scan[$a_2$]$\rightarrow$Join[$R(a_1,a_2)$]$\rightarrow$Join[$R(a_2,a_3)$]} under query attribute ordering $(a_2,a_1,a_3)$. 
After one call to \texttt{Scan} and the two subsequent \texttt{Join}s, the first output chunk is materialized in the output vectors of \fig{fig:ffx-examples}, omitting \texttt{sel} for readability. 

The \texttt{Scan} produces the output vector $a_2$ with values $\{2,3,4\}$, so it initializes \texttt{start}~$=0$ and \texttt{end}~$=3$. 
This vector represents the one-attribute relation over $a_2$, which is equivalent to an f-representation over a single-node f-tree rooted at $a_2$. 
We note that the offset array (\texttt{off}) is unused for the root of an f-tree and is therefore set to \texttt{NULL}, because the root has no grouping with a parent node to encode. 

The first \texttt{Join[$R(a_1,a_2)$]} produces $a_1$ matches for each $a_2$ value and encodes the grouping using the offset array. 
The resulting intermediate relation represented by the two vectors is the f-representation
$(\cup_{a_2}\{2\}\times \cup_{a_1}\{0,1,3\})\ \cup\
(\cup_{a_2}\{3\}\times \cup_{a_1}\{1\})\ \cup\
(\cup_{a_2}\{4\}\times \cup_{a_1}\{1,3\})$, 
where the offset array (\texttt{off}) identifies, for each $a_2$ entry, the contiguous slice of $a_1$ values that belongs to it. 

The second \texttt{Join[$R(a_2,a_3)$]} similarly produces the $a_3$ matches for each $a_2$ value and encodes the grouping using the offset array. 
The resulting output in the three vectors is the f-representation 
$(\cup_{a_2}\{2\} \times \cup_{a_1}\{0,1,3\} \times \cup_{a_3}\{5,6,7\})\ \cup\
(\cup_{a_2}\{3\} \times \cup_{a_1}\{1\} \times \cup_{a_3}\{2,4\})\ \cup\
(\cup_{a_2}\{4\} \times \cup_{a_1}\{1,3\} \times \cup_{a_3}\{8\})$. 
This representation is over the f-tree $\mathcal{T}_2$ in~\fig{fig:t-2}, where both $a_1$ and $a_3$ are grouped under $a_2$. 

\subsubsection{Packed Vectors}
\label{subsec:packed-vectors}

\begin{figure}[t!]
	\centering
	\includegraphics[scale=0.36]{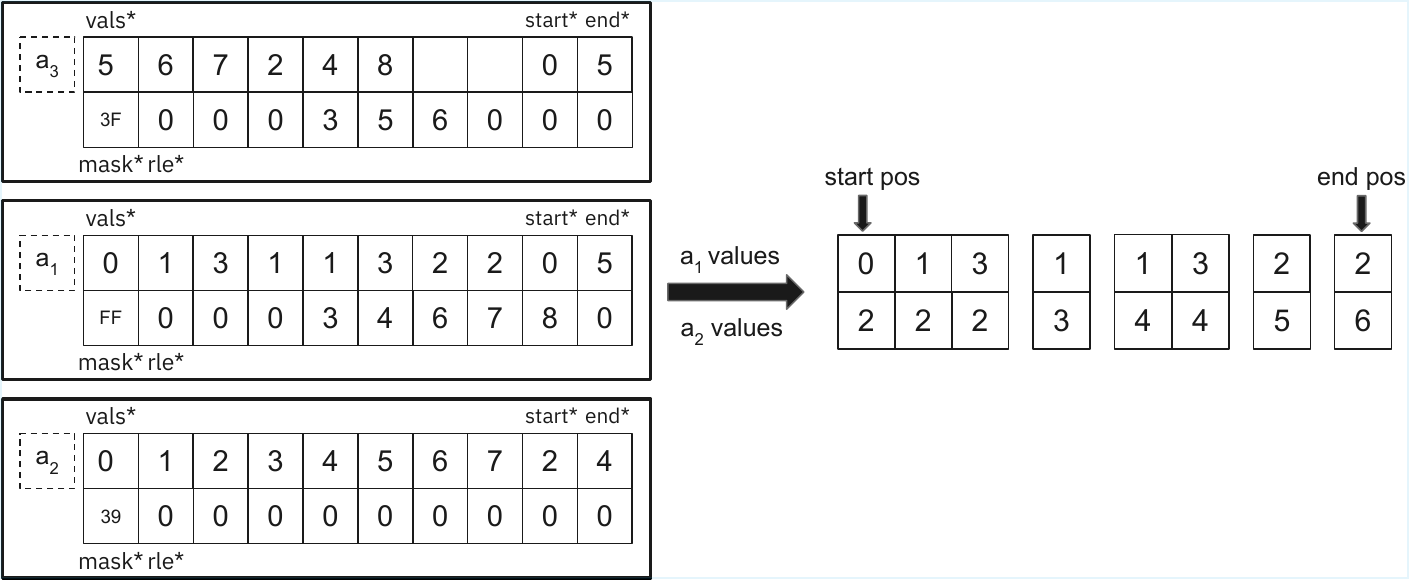}
	\caption{First output vectors produced by \textsc{FFX} for $R(a_1,a_2)\Join R(a_2,a_3)$ 
		and ordering $(a_2,a_1,a_3)$, evaluated on the base relation $R$ in~\fig{subfig:input-relation-r}.}
	\label{fig:ffx-examples}
\end{figure}

\fig{fig:ffx-examples} encodes the output as packed factorized vectors: each attribute $a_2$, $a_1$, and $a_3$ is stored in a fixed-size, contiguous vector, and the offset array (\texttt{off}) is used to recover the grouping required by $\mathcal{T}_2$. 
Unlike LBP, \textsc{FFX} does not bind a parent to a single value and materialize its matches as a separate child list; instead, it keeps both parent and child values packed and uses the offset array to delimit each parent's child slice. 
Operators then scan these contiguous ranges, with invalidations tracked by \texttt{sel}, while preserving the f-tree-defined associations. 

For the evaluation example (\sect{subsec:evaluation-example}), LBP would require backtracking to process the three $a_2$ bindings and therefore would require $3\times$ as many operator calls. 
On larger datasets, this repeated backtracking 
translates into orders of magnitude higher interpretation overhead.

\subsubsection{Arbitrary Factorizations}
\label{subsec:arbitrary-factorizations} 

\begin{figure}[t!]
	\centering
	\begin{subfigure}[b]{0.265\columnwidth}
		\includegraphics[width=\columnwidth]{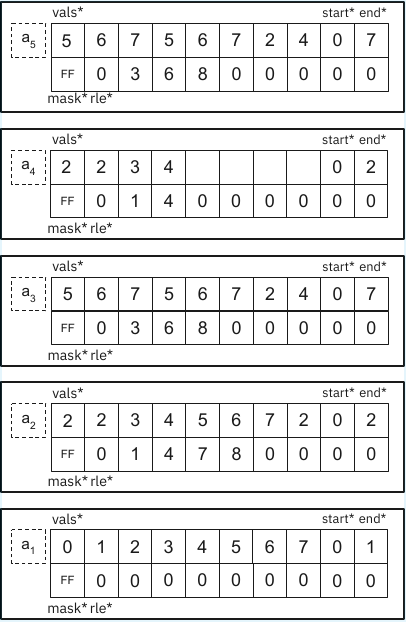}
		\caption{Packed factorized vectors.}
		\label{fig:packed-factorized-vectors}
	\end{subfigure}
	\begin{subfigure}[b]{0.48\columnwidth}
		\centering
		\captionsetup{justification=centering}
		\begin{tikzpicture}[every node/.style={circle,draw},
			level  1/.style={sibling distance=2.5cm,level distance=0.7cm},
			level  2/.style={sibling distance=2.0cm,level distance=0.7cm},
			level  3/.style={sibling distance=1.2cm,level distance=0.7cm},
			level  4/.style={sibling distance=0.6cm,level distance=0.7cm},
			level  7/.style={sibling distance=0.25cm,level distance=0.7cm}]
			\node [rectangle,draw=none] {$\cup_{a_1}$}
			child { node [rectangle,draw=none] {$0$} 
				child { node [rectangle,draw=none] {$\times$}
					child { node [rectangle,draw=none] {$\cup_{a_2}$}
						child { node [rectangle,draw=none] {$2$}
							child { node [rectangle,draw=none] {$\times$}
								child { node [rectangle,draw=none] {$\cup_{a_3}$}
									child { node [rectangle,draw=none] {$5$} }
									child { node [rectangle,draw=none] {$6$} }
									child { node [rectangle,draw=none] {$7$} }
								}
							}
						}
					}
					child { node [rectangle,draw=none] {$\cup_{a_4}$}
						child { node [rectangle,draw=none] {$2$}
							child { node [rectangle,draw=none] {$\times$}
								child { node [rectangle,draw=none] {$\cup_{a_5}$}
									child { node [rectangle,draw=none] {$5$} }
									child { node [rectangle,draw=none] {$6$} }
									child { node [rectangle,draw=none] {$7$} }
								}
							}
						}
					}
				}
			}
			child { node [rectangle,draw=none] {$1$} 
				child { node [rectangle,draw=none] {$\times$}
					child { node [rectangle,draw=none] {$\cup_{a_2}$}
						child { node [rectangle,draw=none] {$2$}
							child { node [rectangle,draw=none] {$\times$}
								child { node [rectangle,draw=none] {$\cup_{a_3}$}
									child { node [rectangle,draw=none] {$5$} }
									child { node [rectangle,draw=none] {$6$} }
									child { node [rectangle,draw=none] {$7$} }
								}
							}
						}
						child { node [rectangle,draw=none] {$3$}
							child { node [rectangle,draw=none] {$\times$}
								child { node [rectangle,draw=none] {$\cup_{a_3}$}
									child { node [rectangle,draw=none] {$2$} }
									child { node [rectangle,draw=none] {$4$} }
								}
							}
						}
					}
					child { node [rectangle,draw=none] {$\cup_{a_4}$}
						child { node [rectangle,draw=none] {$2$}
							child { node [rectangle,draw=none] {$\times$}
								child { node [rectangle,draw=none] {$\cup_{a_5}$}
									child { node [rectangle,draw=none] {$5$} }
									child { node [rectangle,draw=none] {$6$} }
									child { node [rectangle,draw=none] {$7$} }
								}
							}
						}
						child { node [rectangle,draw=none] {$3$}
							child { node [rectangle,draw=none] {$\times$}
								child { node [rectangle,draw=none] {$\cup_{a_5}$}
									child { node [rectangle,draw=none] {$2$} }
									child { node [rectangle,draw=none] {$4$} }
								}
							}
						}
					}
				}
			};
		\end{tikzpicture}
		\caption{F-representation.}
		\label{subfig:f-rep-in-packed-vectors}
	\end{subfigure}
	\caption{First output vectors produced by \textsc{FFX} for $R(a_1,a_2)\Join R(a_2,a_3)\Join R(a_1,a_4)\Join R(a_4,a_5)$ 
		and ordering $(a_1,a_2,a_3,a_4,a_5)$, evaluated on the base relation $R$ in~\fig{subfig:input-relation-r}.}
	\label{fig:4H}
\end{figure}

Packed factorized vectors also remove the branching restriction of LBP. 
Because parent-child groupings are stored explicitly in the offset array (\texttt{off}), rather than implicitly through \texttt{pos}, \textsc{FFX} can represent intermediates over any valid f-tree and propagate them across operators in the same packed layout. 
For example, \fig{fig:4H} shows the first output for $R(a_1,a_2)\Join R(a_2,a_3)\Join R(a_1,a_4)\Join R(a_4,a_5)$ under the ordering $(a_1,a_2,a_3,a_4,a_5)$, leading to a compact f-tree. 
The left subfigure shows the first output chunk as packed factorized vectors, while the right shows the equivalent logical f-representation. 
This is an illustration of branching layouts that \textsc{FFX} supports without sacrificing vector packing or requiring full materialization.

\section{Query Execution}
\label{subsec:query-processing}

\textsc{FFX} evaluates queries using vectorized operators over packed factorized vectors (\sect{subsec:representation}). 
Operators read and write fixed-size, contiguous vectors, yet the logical representations they consume and produce are f-representations, each defined over a particular f-tree. 
These intermediate f-trees are determined by the chosen query attribute ordering $(a_i,\dots,a_m)$ and evolve as evaluation extends the current binding prefix. 
The first attribute $a_i$ is scanned and yields a one-attribute f-representation, \ie an f-tree rooted at $a_i$. 
Each subsequent join extends the current intermediate by introducing the next attribute and selecting a \emph{grouping attribute} in the input f-tree under which the new values will be nested. 
We choose this grouping attribute as high as possible while respecting the root-to-leaf path constraint induced by the query and the chosen ordering, thereby yielding a valid f-tree for each ordering prefix and maximizing compactness under that ordering. 

When a binary join produces a new attribute $a_k$, the input join attribute is always an ancestor of $a_k$ in the output f-tree and is the parent of $a_k$ exactly when $a_k$ can be attached directly under it. 
Multi-way joins follow the same principle: all input join keys must lie on a single root-to-leaf path in the current f-tree because of the dependency. 
All input join keys are ancestors of $a_k$, and the highest node to which $a_k$ can be attached is the lowest input join key. 

We adapt join operators in \textsc{FFX} to our vector representation and introduce a \texttt{Cascade} \texttt{Update} operator that propagates tuple reductions across the grouping hierarchy. 

\subsection{Join Evaluation} 
\label{subsec:join-evaluation}

Each join consumes an input join-key vector (probe side) and produces one or more output-attribute vectors. 
Joins fall into two classes: 
(i) \emph{non-expanding} joins produce at most one match per active key value; and 
(ii) \emph{fanout-expanding} joins may produce multiple matches per key. 
\textsc{FFX} supports both \emph{binary} and \emph{multi-way} joins; a multi-way join is non-expanding if it matches at most one tuple across the joined relations. 
These classes differ in how they handle and update the vector \texttt{State}. 

\smallskip

\subsubsection{Non-expanding Join.} 
For 1-to-1 and many-to-1 joins, 
each active key yields at most one matching tuple, 
so the operator emits output values in lockstep with the input join-key vector. 
Implementation-wise, it does not write to the offset array (\texttt{off}); the join-key vector and the produced attribute vectors share the same \texttt{State}.
The operator updates only the selector (\texttt{sel}), invalidating keys that produce no matches. 

\subsubsection{Fanout-expanding Join.} 
For 1-to-many and many-to-many joins, 
each active key may yield multiple matching tuples. 
As described earlier, the join operator has a specific attribute chosen as a grouping attribute, possibly one of the input join keys or one of their descendant attributes in the input f-tree. 
Given the chosen grouping attribute, the operator appends produced matches contiguously and modifies the offset array so that each grouping entry at index $i$ maps to its child slice $(\texttt{off}[i],\,\texttt{off}[i{+}1])$ in the child vector. 
If the grouping attribute is the join key, this mapping is direct. 
If the grouping attribute is instead a descendant in the input f-tree, the join applies the produced matches to each value in the descendant slice; that is, it applies a Cartesian product.

Fanout expansion sets not only the offset array of the output vector but can also update the active slices of vectors: 
packed output may shrink the effective input slice (\texttt{start}, \texttt{end}), and keys with no matches are invalidated in the selector (\texttt{sel}). 
These reductions must propagate through the grouping hierarchy: if a slice becomes fully inactive, its ancestors may become inactive as well. 
\textsc{FFX} propagates active-slice (\texttt{start}, \texttt{end}) changes during join evaluation and uses \texttt{Cascade} \texttt{Update} (\sect{sec:cascade-update}) to propagate \texttt{selector} updates, which may also be triggered by filters. 

\subsubsection{Multi-way Joins.} 
\textsc{FFX} evaluates a multi-way join as a join-at-a-time pipeline that alternates between (i) \emph{extensions}, which produce candidate bindings for the next attribute via a binary join; and (ii) \emph{intersections}, which enforce the remaining relations by intersecting the corresponding key vectors.
This follows a worst-case-optimal join-at-a-time strategy~\cite{CiucanuOlteanu2016WCOJAT}, adapted to packed factorized vectors. 
As a practical optimization, we fix the extension and intersection order at compile time, 
rather than dynamically at run time in the order of the smallest lists, 
which enables reuse of intersections on shared ancestor attributes~\cite{DBLP:journals/pvldb/MhedhbiS19,DBLP:journals/tods/MhedhbiKS21} and cleanly supports filter operators interleaved between extension and intersection steps. 

\smallskip

Our execution model gracefully 
falls back to standard vectorized execution on non-expanding joins. 
The input and output vectors simply share the same \texttt{State}, and no offset array updates are required. 
In these cases, factorization offers no additional benefit. 

\subsection{Cascade Update}
\label{sec:cascade-update}

Reducing operators (filters and joins) can invalidate values within a packed factorized vector. 
Because vectors encode an f-representation, these invalidations are not local: if all values in a child slice become inactive, the corresponding parent binding must be invalidated. 
Once a parent binding becomes inactive, the invalidation can eliminate entire sibling slices and may propagate further upward and to siblings. 
Cascading these invalidations is required for correctness. 

\subsubsection{Operator Overview} 
\textsc{FFX} 
propagates reductions in place in two ways: 
(i) \emph{slice update}: shrink active slices, \ie \texttt{start} and \texttt{end} positions, common in fanout-expanding joins. As join evaluation descends the f-tree and enumerates larger numbers of child values, it binds the ancestor vectors to progressively smaller slices because each join contracts the active slice during iteration; and 
(ii)~\emph{selector update}: propagate \texttt{selector} invalidations across dependent groupings when a join or filter invalidates entries in the input vector. 

We implement a dedicated \texttt{Cascade} \texttt{Update} (\texttt{CU}) to propagate state changes induced by reducing operators.
It can be configured to update only the active slice (\texttt{start}, \texttt{end}), only the \texttt{selector}, or both.
It takes as input the modified \texttt{State} and applies the implied slice contraction or \texttt{selector} invalidations to all dependent vectors.
To do so, \texttt{CU} is initialized from a dependency tree derived from the f-tree of the current intermediate.
This tree specifies which ancestor and sibling \texttt{State}s must be updated, and in what order, when a given vector changes.
For correctness, the execution plan must ensure that every state modification produced by a reducing operator is propagated to the other dependent \texttt{State}s before downstream operators need to consume them. 

\begin{figure}[t!]
	\centering
	\begin{subfigure}[b]{0.18\textwidth}
		\centering
		\begin{tikzpicture}[
			level 1/.style={sibling distance=0.8cm,level distance=0.6cm},
			level 2/.style={sibling distance=1cm,level distance=0.6cm}]
			\node { $a_6$ }
			child { node { $a_3$ }
				child { node { $a_2$ }
					child { node { $a_1$ } }
				}
				child { node { $a_4$ }
					child { node { $a_5$ } }
				}
			};
		\end{tikzpicture}
		\caption{$\mathcal{T}_p$.}
		\label{fig:cascade-ftree}
	\end{subfigure}
	\begin{subfigure}[b]{0.26\textwidth}
		\centering
		\begin{tikzpicture}[
			level 1/.style={sibling distance=0.8cm,level distance=0.8cm},
			level 2/.style={sibling distance=0.8cm,level distance=0.8cm},
			level 3/.style={sibling distance=2cm,level distance=0.8cm}]
			\node { $a_5$ }
			child { node { $a_4$ (\texttt{AGG}) }
				child { node { $a_3$ (\texttt{AGG}) }
					child { node { $a_6$ (\texttt{AGG}) } }
					child { node { $a_2$ (\texttt{SCATTER}) }
						child { node { $a_1$ (\texttt{SCATTER}) } }
					}
				}
			};
		\end{tikzpicture}
		\caption{Dependency tree.}
		\label{fig:cascade-dag}
	\end{subfigure}
	\caption{Dependency Tree for $R(a_1,a_2)$ $\Join$ $R(a_2,a_3)$ $\Join$ $R(a_3,a_4)$ $\Join$ $R(a_4,a_5)$ $\Join$ $R(a_6,a_3)$, derived from the f-tree $\mathcal{T}_p$ and with $a_5$ as the source of the reduction to be propagated.}
	\label{fig:dependency-tree}
\end{figure}
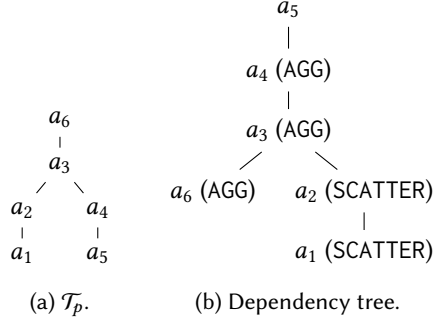

\subsubsection{Dependency Tree} 

\texttt{Cascade} \texttt{Update} propagates reductions by traversing a \emph{dependency tree} derived from the f-tree of the current intermediate. 
The tree is rooted at the f-tree node whose \texttt{State} changed, \eg the join key whose \texttt{selector} was updated. 
We construct the dependency tree by walking from this node up to the f-tree root and, at each step, adding exactly those other attributes whose validity may change as a consequence. 
We label dependency-tree nodes by their type of propagation: 
\texttt{AGG} nodes rely on a child slice to update an ancestor binding, while \texttt{SCATTER} nodes push newly invalidated bindings into dependent subtrees. 
For each visited f-tree node $v$: 

\begin{enumerate}
	\item \emph{Add the parent (\texttt{AGG}).} If $v$ has a parent in $\mathcal{T}$, we add $\mathrm{parent}(v)$ as a child of $v$ in the dependency tree and label that node \texttt{AGG}.
	\item \emph{Add side descendants (\texttt{SCATTER}).} We add nodes for all descendants of $v$ that are not on the path from $v$ to the f-tree root as children of $v$ and label each such node \texttt{SCATTER}. 
	\item Set $v \leftarrow \mathrm{parent}(v)$ and repeat until reaching the root.
\end{enumerate}

An \texttt{AGG} node denotes an ancestor update: \texttt{Cascade} \texttt{Update} checks the parent-child grouping and invalidates a parent binding whenever all values in the corresponding child slice become inactive. 
A \texttt{SCATTER} node denotes a descendant update: \texttt{Cascade} \texttt{Update} pushes newly introduced parent invalidations into the dependent subtree, pruning all slices grouped by the invalidated bindings. 

\noindent\textbf{Example.}
\fig{fig:cascade-ftree} shows an f-tree rooted at $a_6$ with child $a_3$ and two branches under $a_3$: one rooted at $a_4$ with child $a_5$, and one rooted at $a_2$ with child $a_1$. 
Suppose a filter or join updates the \texttt{selector} of $a_5$; \fig{fig:dependency-tree} shows the corresponding dependency tree rooted at $a_5$. 

\texttt{Cascade} \texttt{Update} first propagates \texttt{AGG}: logically, it checks whether each $a_4$ binding has any active values in its $a_5$ slice and invalidates the binding if the slice is empty, then repeats the same check from $a_4$ to $a_3$ and from $a_3$ to $a_6$. 
If some $a_3$ bindings are invalidated, it then propagates \texttt{SCATTER} from $a_3$ into the other branch, pruning the slices under $a_2$, and consequently under $a_1$, that are grouped by those newly invalidated $a_3$ bindings.

\subsubsection{Operator Evaluation}

\texttt{Cascade} \texttt{Update} propagates updates along the dependency tree using a preorder traversal. 
In effect, it first applies all \texttt{AGG} steps toward the root and then applies \texttt{SCATTER} to the affected subtrees. 
At each visited node, it updates the corresponding vector \texttt{State} with tight loops, avoiding recursion and tuple-at-a-time control flow. 
The traversal supports early stopping: if a node produces no new invalidations, propagation terminates for that subtree because no descendant can be affected. 
\alg{alg:set-propagate} shows the \texttt{selector} propagation.

\begin{algorithm}[t!]
	\caption{Selector propagation from a dependency-tree parent node (\texttt{p\_node}) to a child node (\texttt{c\_node}) based on the parent's newly invalidated positions (\texttt{p\_inv\_pos})}
	\label{alg:set-propagate}
	\begin{algorithmic}[1]
	\Function{selector\_propagate}{p\_node, c\_node, p\_inv\_pos}
		\State $p\_state \gets p\_node.state$;\;\; $c\_state \gets c\_node.state$;\;\; 
		\State $new\_inv\_pos \gets \emptyset$ \Comment{Set to accumulate new invalidations due to propagation.}
		\If{$c\_node.\text{label} = \texttt{AGG}$}
			\State \Comment{For AGG, an f-tree child node propagates invalidations to its parent node. Note that the node relationship in the dependency tree is reversed.}
			\While{$i \gets \text{next\_valid}(c\_state.\text{sel},\, c\_state.\text{start},\, c\_state.\text{end})$}\label{alg:set-propagate-7}
				\If{$\text{no\_set\_bits}(p\_state.\text{sel},\, p\_state.\text{off}[i],\, p\_state.\text{off}[i{+}1])$}\label{alg:set-propagate-8}
					\State $\text{CLEAR\_BIT}(c\_state.\text{sel}, i)$
					\State $new\_inv\_pos \gets new\_inv\_pos \cup \{i\}$
				\EndIf
			\EndWhile
		\Else \Comment{\texttt{SCATTER}: For each child slice, add active pos as new invalidations $\rightarrow$ clear the slice.}
		\For{\textbf{each} $i$ \textbf{in} $p\_inv\_pos$}
			\While{$j \gets \text{next\_valid}(c\_state.\text{sel},\, c\_state.\text{off}[i],\, c\_state.\text{off}[i{+}1])$}
				\State $new\_inv\_pos \gets new\_inv\_pos \cup \{j\}$
			\EndWhile
			\State $\text{CLEAR\_BITS}(c\_state.\text{sel},\, c\_state.\text{off}[i],\, c\_state.\text{off}[i{+}1])$
		\EndFor
		\EndIf
		\State \Call{trim\_slice}{$c\_state$} \Comment{Trim leading and trailing inactive positions $\rightarrow$ update \texttt{start}/\texttt{end}.}
		\If{$new\_inv\_pos \neq \emptyset$}\label{alg:set-propagate-not-empty}
			\For{\textbf{each} $child$ \textbf{in} $c\_node.\text{children}$}
				\State \Call{selector\_propagate}{c\_node, $child$, $new\_inv\_pos$}
			\EndFor
		\EndIf
	\EndFunction
	\end{algorithmic}
\end{algorithm}

\subsubsection{Optimizations}
\label{subsec:cu-optimizations}

\texttt{Cascade} \texttt{Update} (\texttt{CU}) represents additional overhead introduced by factorized execution. 
To minimize this overhead, we apply three complementary optimizations. 

\smallskip

\noindent\textbf{Placement.} 
While correctness allows inserting a \texttt{CU} after every reducing operator, \textsc{FFX}'s planner inserts \texttt{CU} operators lazily to amortize repeated propagation. 
Specifically, we delay \texttt{selector} propagation until it is required: (i) before a sink operator; or (ii) before a vector is consumed as input when its \texttt{State} may be stale. 
Moreover, we avoid inserting \texttt{selector} cascades along frontier-only extensions, where reductions cannot yet affect sibling subtrees. 

\smallskip

\noindent\textbf{Operator fusion.} 
\texttt{CU} can be inserted as a standalone operator or fused into the preceding reducing operator to reduce call overhead. 

\smallskip

\noindent\textbf{Delta-driven \texttt{AGG}.} 
A naïve \texttt{AGG} scans all active parent bindings and checks whether each child slice is empty, even if only a few child positions have changed. 
We instead drive \texttt{AGG} from the delta: newly invalidated child positions identify their invalidated parent bindings via the offset array (\texttt{off}), forming a smaller set of parent bindings that may need to be invalidated. This is an optimization for \alg{alg:set-propagate}, lines \ref{alg:set-propagate-7}--\ref{alg:set-propagate-8}. 
\texttt{AGG} then tests slice emptiness only for parents in this workset, making the cost scale with the number of new invalidations. 

\smallskip

Together, these optimizations make the cost of the \texttt{Cascade} \texttt{Update} (\texttt{CU}) operator largely proportional to the number of newly introduced invalidations while still linear in the size of the input f-representation: 
\textsc{FFX}'s planner amortizes their insertion, fusion reduces per-invocation overhead, and the refinements to \texttt{AGG} avoid scanning unchanged parent slices unless a child slice actually becomes empty. 
As a result, \texttt{CU} remains lightweight on low-selectivity joins yet still prunes correctly when reductions are selective and cascade.

\section{Factorized Semantic Processing}
\label{sec:prompt-compression}

\textsc{FFX} extends semantic operators to execute directly over factorized intermediates. 
The main challenge is that LLM-powered operators consume text representations of flat tuples, whereas \textsc{FFX} represents intermediates as packed factorized vectors over an f-tree. 
A straightforward implementation would flatten the intermediate into tuples and then serialize those tuples into prompts, but doing so would forfeit the compactness benefits of factorization. 

Instead, \textsc{FFX} preserves factorization throughout prompt construction. 
It traverses the relevant portion of the input using windowed iteration, serializes the corresponding f-representation in a structure-aware format, and instructs the LLM to perform the implied Cartesian expansion while producing one prediction per logical tuple. 
As we show in this section, this design reduces redundancy in serialized prompts with little to no loss in quality. 

\subsection{Semantic Operator Model}
\label{subsec:semantic-operator-model}

We focus on the \texttt{llm\_map} primitive, which we implement as a dedicated operator. 
Other semantic primitives, such as \texttt{llm\_reduce}, \texttt{llm\_filter}, and \texttt{llm\_rerank}, are left to future work. 

The \texttt{llm\_map} operator takes as input a factorized intermediate together with a set of relevant attributes and logically produces one prediction per logical input tuple. 
When the relevant attributes lie on a single root-to-leaf path of the current f-tree, the operator serializes that path directly in hierarchical form and produces one prediction for each resulting logical tuple. 
When they span multiple branches, \textsc{FFX} avoids flattening the input: it serializes the factorized structure and instructs the model to apply the implied Cartesian product and return one prediction per resulting tuple. 
This preserves the compactness benefits of factorization on the input side, although the logical output cardinality remains unchanged. 
Empirically, and somewhat surprisingly, our evaluation in \sect{subsec:prompt-compression-benefits} shows that even non-reasoning models can often perform this Cartesian expansion accurately while still carrying out the semantic task. 
Errors become noticeable mainly at large batch sizes, where flat representations already exceed the context-window limit. 

Because each prediction depends on all relevant attributes, the predicted attribute value and all relevant attributes must lie on the same root-to-leaf path in the output f-representation. 
\textsc{FFX} reconstructs the output without full materialization: it inserts each tuple enumerated by the model, together with its prediction, at the appropriate position in the output factorized vectors. 
The resulting output is then passed to the next operator in the pipeline.

\subsection{Windowing Over Factorized Intermediates}
\label{subsec:prompt-windowing}

Prompt construction begins with a deterministic iterator over the relevant portion of the current f-tree. 
For each root-to-leaf path containing relevant attributes, the iterator maintains a \emph{window} at the corresponding relevant leaf, \ie a relevant attribute with no relevant descendant on that path. 
Each window tracks a bounded set of bindings for that leaf together with the shared ancestor prefix above it. 
One iterator step returns a factorized fragment formed by the current windows across all relevant paths. 
This fragment corresponds to a bounded set of logical tuples without flattening the full intermediate.

Window sizes control the trade-off between compactness and flatness. 
If all window sizes are $1$, the iterator degenerates to ordinary flat-tuple iteration. 
Larger windows preserve common prefixes and expose multiple logical tuples together, so repeated context is emitted once rather than duplicated across rows. 
During iteration, windows advance in a fixed order and only one window moves at a time while earlier windows remain fixed. 
As a result, consecutive prompts share long prefixes, which improves prefix reuse, controls context size, and stabilizes output quality.

\subsection{Structure-Aware Prompt Serialization}
\label{subsec:prompt-serialization}

Each factorized fragment emitted by the iterator is serialized from the underlying packed factorized vectors into a format that follows the f-tree rather than a flat row-per-tuple layout. 
\textsc{FFX} supports JSON and XML, but the particular surface syntax is orthogonal to the core idea. 
In all cases, shared ancestor bindings are emitted once, and descendant bindings are nested under their corresponding parent, so the prompt directly mirrors the factorized structure of the input. 
The serializer projects only the relevant attributes and embeds the result inside a meta-prompt that specifies the task, the expected output schema, the f-tree structure, and the required Cartesian expansion. 

The model is instructed to return output tuples in a fixed enumeration order. 
\textsc{FFX} uses the same order for its enumeration and aligns each returned prediction with its corresponding tuple. 
When predictions are missing for some tuples due to an LLM failure with the Cartesian expansion, \textsc{FFX} assigns \texttt{NULL}; alternatively, one could retry the LLM invocation. 
Thus, \texttt{llm\_map} consumes a factorized input, avoids flattening for prompt construction, and flattens after LLM invocation because the relevant input attributes and the prediction are dependent and must therefore lie on the same root-to-leaf path in the output f-representation. 
The remaining attributes remain factorized. 

\section{Evaluation}
\label{sec:evaluation}

We evaluate the key design decisions underlying \textsc{FFX}. 
Our empirical study addresses three questions: 
(i)~How does our vector representation compare with prior state-of-the-art approaches such as LBP~\cite{DBLP:journals/pvldb/GuptaMS21}? 
(ii)~What overhead, if any, does our vector representation introduce when factorization provides no benefit?
(iii)~How effective is our factorized semantic processing at reducing input token usage, and what is the impact on accuracy?

We also compare \textsc{FFX} against \textsc{DuckDB}~\cite{DBLP:conf/sigmod/RaasveldtM19}, a relational vectorized baseline, and \textsc{Kuzu}~\cite{DBLP:conf/cidr/JinFCLS23}, an LBP-based system, and we evaluate multi-core scalability on large datasets. 

\subsection{Setup}

\subsubsection{Hardware}
\label{subsubsec:hardware} 
Unless otherwise stated, we run all experiments on a 
Google Cloud \texttt{c4-standard-8} instance with 8 vCPUs, 30\,GB of main memory, and an \texttt{INTEL(R)} \texttt{XEON(R)} \texttt{PLATINUM} \texttt{8581C} \texttt{CPU@2.30GHz}. The machine runs Debian v6.1.148.

\subsubsection{Compiler} 
All binaries are compiled with \texttt{clang v15.0.6} using the \texttt{-O3} flag. 
Build files are generated using \texttt{CMake} \texttt{v3.25.1}.

\subsubsection{Datasets}
The datasets used in our evaluation are listed in \tab{tab:datasets-used}. 
They are drawn from the Open Graph Benchmark (OGB)~\cite{hu2021opengraphbenchmarkdatasets}, 
a 2010 Twitter crawl~\cite{Kwak10www}, and the SNAP collection~\cite{snapnets}. 
Each dataset is represented as a base relation $R(\textsf{src},\textsf{dst})$. 
In \tab{tab:datasets-used}, \textit{Edges} denotes the number of tuples in $R$, and \textit{Nodes} denotes the number of distinct values across \textsf{src} and \textsf{dst}. 
These datasets span diverse domains, including citation, social, web, and product networks, and exhibit varying structural characteristics such as scale, neighbourhood degree distributions, and skew. 

\begin{table}[t!]
	\centering
	\begin{tabular}{lllllllll}\toprule
		Domain    & Name             & Nodes & Edges  \\\midrule
		Social    & Epinions   &   76K &  509K  \\
		& LiveJournal &  4.8M &   69M  \\
		& Twitter  & 41.6M & 1.46B  \\
		Web       & Google  &  876K &  5.1M  \\
		Product   & Amazon     &  403K &  3.5M  \\
		Citation  & ArXiv      &   169K    &  1.17B     \\\bottomrule
	\end{tabular}
	\caption{Datasets used.}
	\label{tab:datasets-used}
\end{table}

\subsubsection{Queries}
\label{subsubsec:queries}

The queries used are listed in \tab{tab:queries} in Datalog-style conjunctive form. 
Each query consists exclusively of equi-joins and self-joins over the same base relation $R$, where an equi-join is expressed by reusing the same query variable across multiple atoms, \eg $R(a,b) \Join R(b,c)$. 

We use queries with varying join-graph shapes, including path, star, and tree queries, to cover a range of factorized representations and execution challenges, from deep chain groupings in paths to high-branching layouts in stars and trees. 
Our main focus is on acyclic join graphs, where factorization yields asymptotic size reductions and where query performance is most sensitive to intermediate-result explosion. 
To complement these workloads, we also evaluate \textsc{FFX} on the 
\texttt{LSQB}~\cite{DBLP:conf/sigmod/MhedhbiLKWS21} and \texttt{JOB}~\cite{DBLP:journals/pvldb/LeisGMBK015} benchmarks 
in our end-to-end evaluation, which stress the system along additional dimensions, including cyclic queries and string predicate evaluation. 
These choices are consistent with those from prior work~\cite{DBLP:conf/icde/KaluminD25,DBLP:journals/pvldb/MhedhbiS19,DBLP:conf/sigmod/AbergerTOR16,DBLP:journals/pvldb/BirlerKN24}.

\begin{table}[t!]
	\centering
	\begin{tabular}{>{\raggedright\arraybackslash}p{0.4cm} l}
		\toprule
		& \textbf{Query} \\
		\midrule
		\multirow{2}{*}{Path}
		& \texttt{Q1} $=$ $R(a$$,$$b)$ $\Join$ $R(b$$,$$c)$ $\Join$ $R(c$$,$$d)$ $\Join$ $R(d$$,$$e)$ \\
		& \texttt{Q2} $=$ $R(a$$,$$b)$ $\Join$ $R(b$$,$$c)$ $\Join$ $R(c$$,$$d)$ $\Join$ $R(d$$,$$e)$ $\Join$ $R(e$$,$$f)$ \\
		\midrule
		\multirow{3}{*}{Star} 
		& \texttt{Q3} $=$ $R(a$$,$$b)$ $\Join$ $R(a$$,$$c)$ $\Join$ $R(a$$,$$d)$ $\Join$ $R(a$$,$$e)$ \\
		& \texttt{Q4} $=$ $R(a$$,$$b)$ $\Join$ $R(a$$,$$c)$ $\Join$ $R(a$$,$$d)$ $\Join$ $R(a$$,$$e)$ $\Join$ $R(a$$,$$f)$ \\
		& \texttt{Q5} $=$ $R(a$$,$$b)$ $\Join$ $R(a$$,$$c)$ $\Join$ $R(a$$,$$d)$ $\Join$ $R(a$$,$$e)$ $\Join$ $R(a$$,$$f)$ $\Join$ $R(a$$,$$g)$\\ 
		\midrule
		\multirow{4}{*}{Tree}
		& \texttt{Q6} $=$ $R(a$$,$$b)$ $\Join$ $R(a$$,$$c)$ $\Join$ $R(b$$,$$d)$ $\Join$ $R(c$$,$$e)$ \\
		& \texttt{Q7} $=$ $R(a$$,$$b)$ $\Join$ $R(a$$,$$c)$ $\Join$ $R(a$$,$$d)$ $\Join$ $R(b$$,$$e)$ $\Join$ $R(c$$,$$f)$ \\
		& \texttt{Q8} $=$ $R(a$$,$$b)$ $\Join$ $R(b$$,$$c)$ $\Join$ $R(b$$,$$d)$ $\Join$ $R(c$$,$$e)$ $\Join$ $R(d$$,$$f)$ \\
		& \texttt{Q9} $=$ $R(a$$,$$b)$ $\Join$ $R(a$$,$$c)$ $\Join$ $R(b$$,$$d)$ $\Join$ $R(e$$,$$d)$ \\
		\bottomrule
	\end{tabular}
	\caption{Queries used, grouped by join query graph structure.}
	\label{tab:queries}
\end{table}

\subsubsection{Query Execution} 
We run each query three times and report the median unless otherwise stated. 
After each query, the cache is manually cleared to minimize bias from execution order 
using the following command: \texttt{sudo sync \&\& echo 3 | sudo tee /proc/sys/vm/drop\_caches}

\subsection{Factorization and Vectorization Benefits}
\label{subsec:factorization-and-vectorization-benefits}

We compare our packed factorized vector representation with that of list-based processing (LBP)~\cite{DBLP:journals/pvldb/GuptaMS21}. 
To ensure a fair comparison, both representations are implemented in \textsc{FFX}. 
We denote ours by \textsc{P} and LBP's by \textsc{UP}, standing for packed and unpacked, respectively. 
We evaluate the nine queries in \tab{tab:queries} using \textsc{P} and \textsc{UP} across three datasets: 
\texttt{Amazon}, \texttt{Google}, and \texttt{Epinions}. 
Each query is evaluated over all possible plans, \ie all query attribute orderings. 
Some plans are equivalent to processing flat tuples since their intermediates are encoded over a root-to-leaf f-tree; 
for example, plan $(a,b,c,d,e)$ for \texttt{Q1} produces intermediates over $a-b-c-d-e$ rooted at $a$. 
Other plans yield more compact representations with intermediates over f-trees with branches and smaller heights; 
for example, plan $(c,b,a,d,e)$ for \texttt{Q1} produces intermediates over $c-b-a$ and $c-d-e$, rooted at $c$. 

\begin{figure}[t!] 
	\centering
	\begin{subfigure}[b]{0.15\columnwidth}
		\centering
		\includegraphics[width=\textwidth]{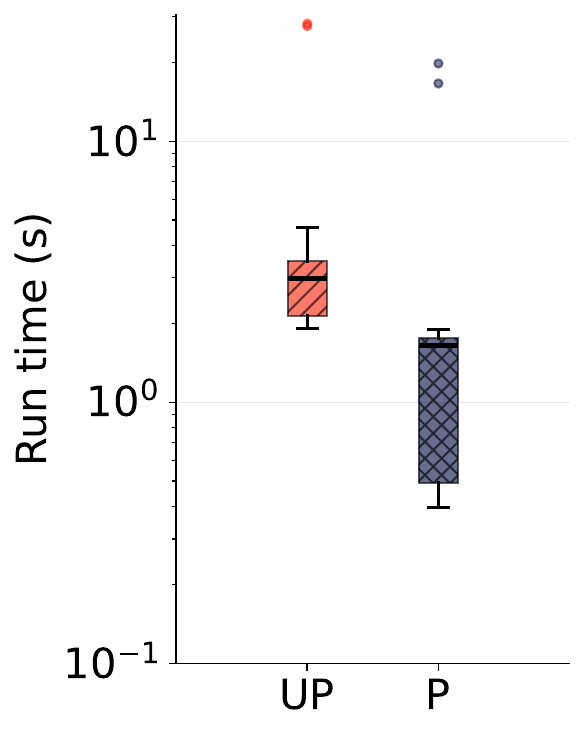}
		\captionsetup{font=tiny,skip=0.5pt}
		\caption{Q1 ($n = 16$)}
		\label{fig:wg_q2}
	\end{subfigure}
	\hfill
	\begin{subfigure}[b]{0.15\columnwidth}
		\centering
		\includegraphics[width=\textwidth]{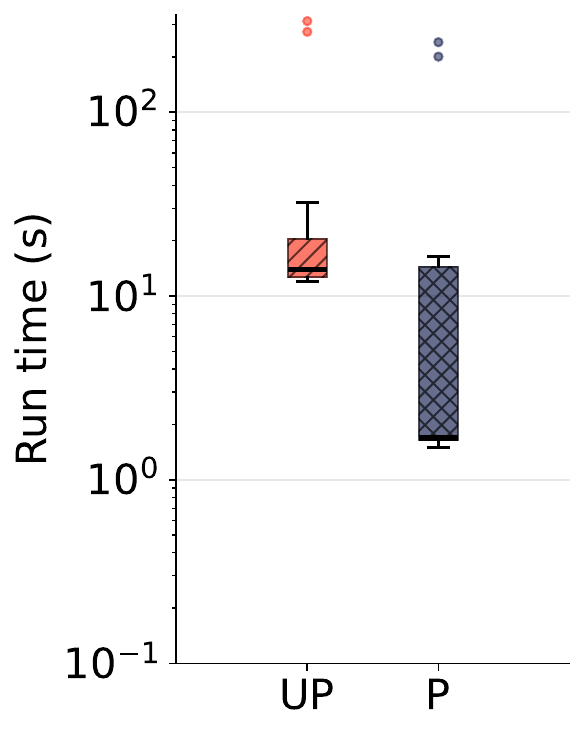}
		\captionsetup{font=tiny,skip=1pt}
		\caption{Q2 ($n = 31$)}
		\label{fig:wg_q3}
	\end{subfigure}
	\hfill
	\begin{subfigure}[b]{0.15\columnwidth}
		\centering
		\includegraphics[width=\textwidth]{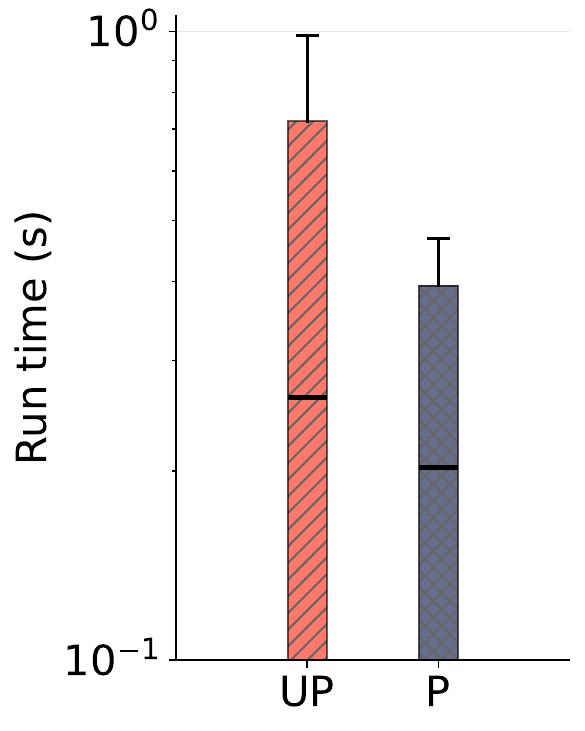}
		\captionsetup{font=tiny,skip=1pt}
		\caption{Q3 ($n = 48$)}
		\label{fig:wg_q4}
	\end{subfigure}
	\hfill
	\begin{subfigure}[b]{0.15\columnwidth}
		\centering
		\includegraphics[width=\textwidth]{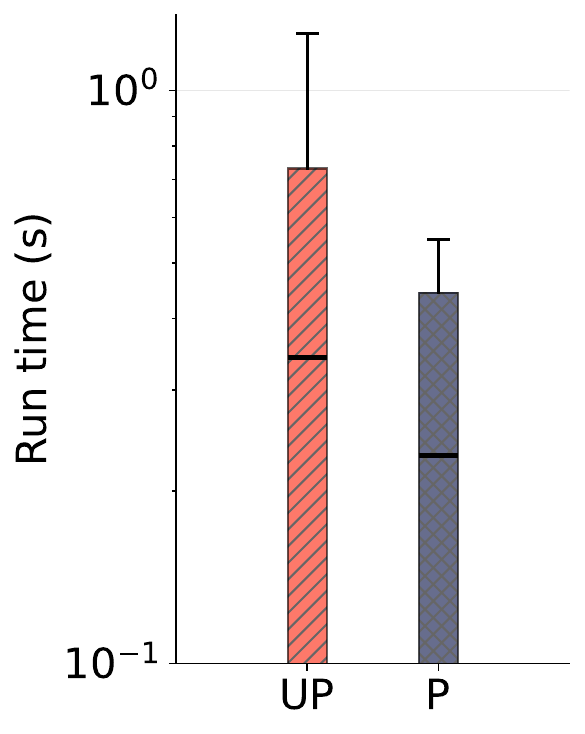}
		\captionsetup{font=tiny,skip=1pt}
		\caption{Q4 ($n = 240$)}
		\label{fig:wg_q5}
	\end{subfigure}
	\hfill
	\begin{subfigure}[b]{0.15\columnwidth}
		\centering
		\includegraphics[width=\textwidth]{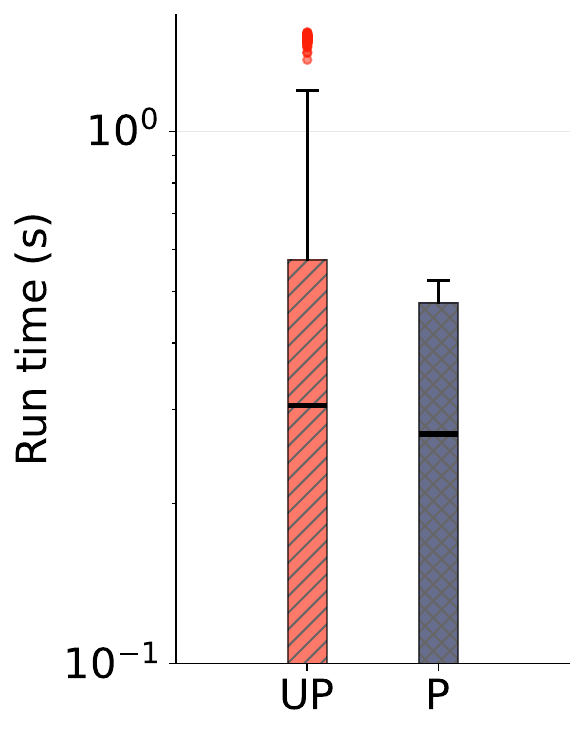}
		\captionsetup{font=tiny,skip=1pt}
		\caption{Q5 ($n = 1440$)}
		\label{fig:wg_q6}
	\end{subfigure}
	
	\begin{subfigure}[b]{0.15\columnwidth}
		\centering
		\includegraphics[width=\textwidth]{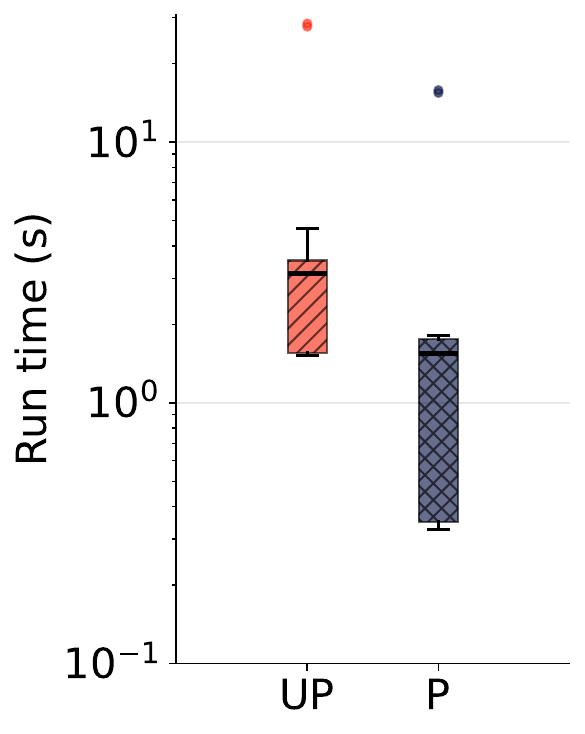}
		\captionsetup{font=tiny,skip=1pt}
		\caption{Q6 ($n = 16$)}
		\label{fig:wg_q7}
	\end{subfigure}
	\hspace*{2em}
	\begin{subfigure}[b]{0.15\columnwidth}
		\centering
		\includegraphics[width=\textwidth]{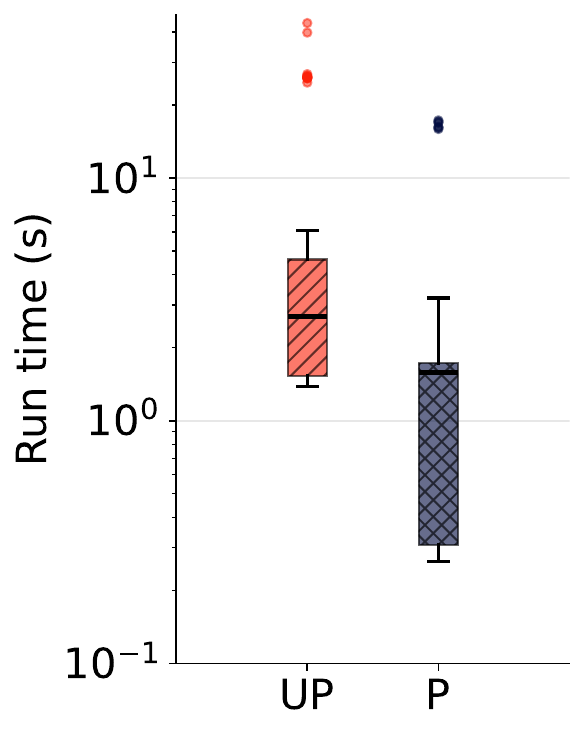}
		\captionsetup{font=tiny,skip=1pt}
		\caption{Q7 ($n = 72$)}
		\label{fig:wg_q8}
	\end{subfigure}
	\hspace*{2em}
	\begin{subfigure}[b]{0.15\columnwidth}
		\centering
		\includegraphics[width=\textwidth]{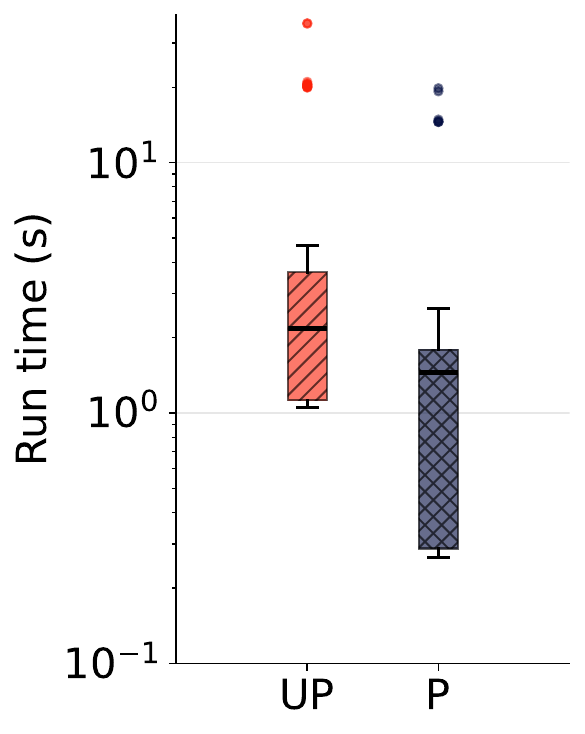}
		\captionsetup{font=tiny,skip=1pt}
		\caption{Q8 ($n = 72$)}
		\label{fig:wg_q9}
	\end{subfigure}
	\hspace*{2em}
	\begin{subfigure}[b]{0.15\columnwidth}
		\centering
		\includegraphics[width=\textwidth]{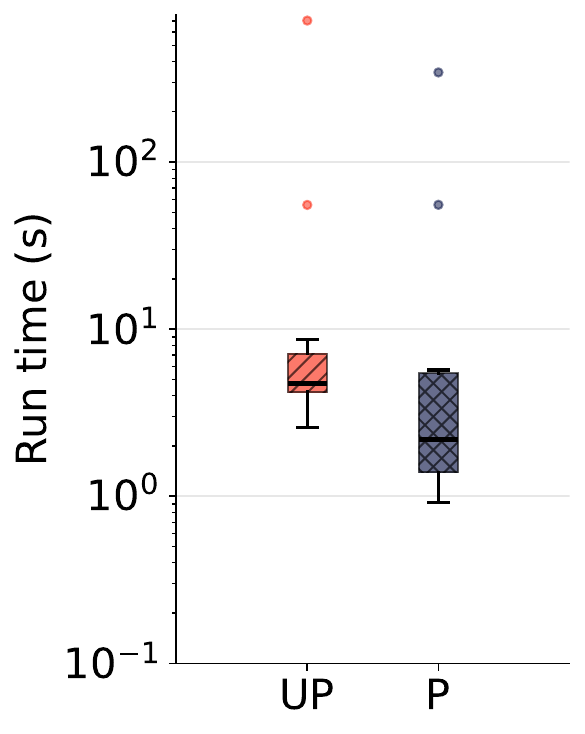}
		\captionsetup{font=tiny,skip=1pt}
		\caption{Q9 ($n = 16$)}
		\label{fig:wg_q10}
	\end{subfigure}
	\caption{Performance comparison on \texttt{Google} for Q1--Q9 using 
		(i) LBP's unpacked vectors (UP) and (ii) packed factorized vectors (P), both implemented within \textsc{FFX}. 
		The figure reports run time in seconds (log scale) across all plans for each query, where $n$ denotes the number of plans.}
	\label{fig:web-Google_plots}
\end{figure}

\fig{fig:web-Google_plots} shows the results on \texttt{Google}; we observe similar trends on \texttt{Epinions} and \texttt{Amazon}. 
On \texttt{Google}, \textsc{P} outperforms \textsc{UP} on $1883$ plans ($96.53\%$) and underperforms on only $68$ plans ($3.47\%$). 
The largest observed speedup occurs on \texttt{Q2} for plan $(d,c,e,f,b,a)$, where run time drops from $16.5$ to $1.76$~seconds ($9.39\times$). 
Across all queries, the three highest average speedups are for \texttt{Q2}, \texttt{Q7}, and \texttt{Q8}, at $5.72\times$, $4.11\times$, and $3.32\times$, respectively. 
Overall, the mean speedup on \texttt{Google} across all plans and queries is $2.08\times$, and the $68$ slowdowns all correspond to \texttt{Q5} plans with run times below $100$~ms. 

We attribute these improvements to two factors: (i) vectorization, which lowers interpretation cost, with, on average, $78.77\times$ fewer virtual function calls, and improves microarchitectural efficiency, as reflected in reductions in cycles, instruction count, branch-misprediction rate, and cache-miss rate; and (ii) factorization, which leads to more compact representations, as measured by the number of processed tuples, and hence less redundant work. 
These results highlight that factorization and vectorization offer complementary benefits. 
We next isolate and assess the contribution of each.

\subsubsection{Vectorization Benefits Analysis}

To isolate the contribution of vectorization and quantify the overhead of our packed factorized vectors, we need queries and plans for which factorization provides no benefit. 
To this end, we focus on plans for path queries \texttt{Q1} and \texttt{Q2}, whose intermediates are over root-to-leaf f-trees and are thus equivalent to processing flat tuples. 
For these queries, plans $P_{Q1,1}=(a,b,c,d,e)$ and $P_{Q2,1}=(a,b,c,d,e,f)$ yield no asymptotic factorization benefits, with intermediates over flat representations:
$a-b-c-d-e$ and $a-b-c-d-e-f$, both rooted at $a$.

\tab{tab:experiment2-linear-queries-vectorization} reports results on the \texttt{Google} dataset, where packed execution consistently outperforms the unpacked baseline. 
Using packed factorized vectors, \textsc{FFX} achieves $1.66\times$ and $1.46\times$ speedups on $P_{Q1,1}$ for \texttt{Q1} and on $P_{Q2,1}$ for \texttt{Q2}, while reducing operator function calls by up to $108.4\times$. 
Microarchitecturally, packed execution reduces total CPU cycles by $1.35$--$3.59\times$ and improves IPC by $1.02$--$1.58\times$. 
These gains arise from both reduced work, with $1.47$--$12.84\times$ fewer instructions and $1.17$--$4.91\times$ fewer branch mispredictions, and higher per-cycle throughput, which compound to drive end-to-end speedups. 
In contrast, LBP's representation cannot realize these benefits, since it continues to pass unpacked vectors even when factorization offers no benefit.

Overall, our analysis shows that even when packed factorized vectors offer no compactness benefit, 
\textsc{FFX} retains substantial speedups over LBP. 
The packed layout enables cache-efficient processing and allows the compiler to exploit autovectorization more effectively when possible. 

\begin{table*}[t!]
	\centering
	\small
	\begin{tabular}{llrrrrrrrrr}
		\toprule
		\textbf{Plan} & & \textbf{RT (s)} & \textbf{\# Calls} & \textbf{\# I} & \textbf{\# C} & \textbf{IPC} & \textbf{\# BM} & \textbf{\# CM} & \textbf{\# Proc. T} \\
		\midrule
		\multirow{3}{*}{$P_{Q1,1}$} & UP & 27663 & 913.0M & 131.41B & {86.74B} & {1.52} & {869.4M} & {1.46B} & {15.89B} \\
		& P & 16671 & 8.00M & 89.14B & {55.23B} & {1.61} & {591.8M} & {1.29B} & {945.54M} \\
		& & \textbf{(1.66$\times$)} & \textbf{(114.13$\times$)} & \textbf{(1.47$\times$)} & \textbf{(1.57$\times$)} & \textbf{(1.06$\times$)} & \textbf{(1.47$\times$)} & \textbf{(1.13$\times$)} & \textbf{(16.80$\times$)} \\
		\midrule
		\multirow{3}{*}{$P_{Q2,1}$} & UP & 294089 & 15.38B & 2.20T & {905.72B} & {2.18} & {7.24B} & {12.37B} & {282.28B} \\
		& P & 200793 & 141.84M & 1.46T & {668.47B} & {2.43} & {6.17B} & 11.75B & 15.89B \\
		& & \textbf{(1.46$\times$)} & \textbf{(108.43$\times$)} & \textbf{(1.52$\times$)} & \textbf{(1.35$\times$)} & \textbf{(1.11$\times$)} & \textbf{(1.17$\times$)} & \textbf{(1.05$\times$)} & \textbf{(17.76$\times$)} \\
		\midrule
		\multirow{3}{*}{$P_{Q1,2}$} & UP & 2348 & 56.91M & 9.45B & {8.11B} & {1.15} & {29.62M} & {133.17M} & {11.48B} \\
		& P & 441 & 82.76K & 2.60B & {2.26B} & {1.17} & {15.86M} & {73.90M} & {141.79M} \\
		& & \textbf{(5.32$\times$)} & \textbf{(687.69$\times$)} & \textbf{(3.63$\times$)} & \textbf{(3.59$\times$)} & \textbf{(1.02$\times$)} & \textbf{(1.87$\times$)} & \textbf{(1.80$\times$)} & \textbf{(80.96$\times$)} \\
		\midrule
		\multirow{3}{*}{$P_{Q2,2}$} & UP & 16502 & 812.1M & 113.1B & {54.20B} & {1.32} & {415.6M} & {608.55M} & {138.11B} \\
		& P & 1757 & 0.59M & 8.81B & 6.66B & 1.32 & 84.57M & 275.59M & 1.12B \\
		& & \textbf{(9.39$\times$)} & \textbf{(1378.84$\times$)} & \textbf{(12.84$\times$)} & \textbf{(8.14$\times$)} & \textbf{(1.58$\times$)} & \textbf{(4.91$\times$)} & \textbf{(2.21$\times$)} & \textbf{(122.60$\times$)} \\
		\bottomrule
	\end{tabular}
	\caption{Comparison of FFX packed (P) and LBP unpacked (UP) representations on \texttt{Google} for queries \texttt{Q1} and \texttt{Q2} in~\tab{tab:queries}.
		We use two plans for each: $P_{Q1,1}=(a,b,c,d,e)$, $P_{Q1,2}=(c,d,b,a,e)$, $P_{Q2,1}=(a,b,c,d,e,f)$, and $P_{Q2,2}=(d,c,e,f,b,a)$. Metrics include run time (RT) in seconds and the number of operator function calls (\#~Calls), instructions (\# I), CPU cycles (\# C), instructions per cycle (IPC), branch misses (\# BM), cache misses (\#~CM), and number of processed tuples (\# Proc. T). Improvements are shown in parentheses.}
	\label{tab:experiment2-linear-queries-vectorization}
\end{table*}

\subsubsection{Factorization Benefits Analysis}
\label{subsubsec:fact-benefits}

Our goal here is to isolate the contribution of factorization, \ie of compactness, on top of vectorization. 
We therefore focus on plans whose intermediates are over branched f-trees rather than root-to-leaf paths. 
For \texttt{Q1} and \texttt{Q2}, examples of such plans are 
$P_{Q1,2}=(c,d,b,a,e)$ and $P_{Q2,2}=(d,c,e,f,b,a)$. 
Their factorized representations contain branching and are over the f-trees 
$c-d-e,\, c-b-a$, rooted at $c$, and $d-c-b-a,\, d-e-f$, rooted at $d$. 
These branched representations eliminate heavily repeated enumerations. 
They are the most compact representations for \texttt{Q1} and \texttt{Q2} under f-representations~\cite{DBLP:conf/icdt/OlteanuZ12}. 

\tab{tab:experiment2-linear-queries-vectorization} reports the results on the \texttt{Google} dataset. 
Packed factorized execution (\textsc{P}) delivers large gains over the unpacked baseline (\textsc{UP}) on these plans: 
$5.32\times$ for \texttt{Q1} and $9.39\times$ for \texttt{Q2}. 
These improvements are due to the substantial elimination of redundant computation, as reflected in the 
$3.63\times$ and $12.84\times$ reductions in executed instructions. 
This elimination yields much larger benefits than those obtained from vectorization alone. 

Branching reduces f-tree height, leading to greater compactness. 
By propagating packed factorized vectors, \textsc{FFX} can exploit these benefits. 
In contrast, \textsc{Kuzu}'s unpacked, list-based representation cannot achieve the same reduction in computation and effectively linearizes these branches, yielding the f-trees 
$c-d,\, c-b-a-e$ 
for $P_{Q1,2}$ 
and  
$d-c,\, d-e-f-b-a$ 
for $P_{Q2,2}$. 
This increases the effective f-tree height and leads to redundant computation. 

Across the nine queries, 22\% (2/9: \texttt{Q1}, \texttt{Q9}) have the same best ordering under packed and unpacked execution, yet packed execution is still faster by $4.94\times$ on average (Q1: $(c,b,d,a,e)$, 397 vs.\ 2106 ms, $5.30\times$; Q9: $(b,d,a,c,e)$, 921 vs.\ 4216 ms, $4.58\times$). 
For the remaining 78\% (7/9: \texttt{Q2}--\texttt{Q8}), the optimal orderings differ slightly, and packed execution achieves a $5.47\times$ average speedup when comparing the best plan under each representation. 
The largest gap is on \texttt{Q2}, where the packed best ordering $(c,b,d,e,a,f)$ is $7.98\times$ faster than the unpacked best ordering $(c,b,d,e,f,a)$ (1.50 vs.\ 11.99 seconds); the smallest is on \texttt{Q4}, where packed execution under $(a,e,b,c,f,d)$ improves by $1.22\times$ over unpacked execution under $(a,b,f,c,d,e)$ (49 vs.\ 60 ms). 

Optimal plans diverge because the same ordering induces different factorization schemes under packed and unpacked execution. 
Unpacked execution is restricted to a limited class of f-trees (Figure~\ref{fig:ftrees-supported}), which can force less compact layouts and lead to additional redundant work. 
By supporting arbitrary f-trees, \textsc{FFX} can realize the compact layout implied by the chosen ordering, reducing computation and shifting the plan trade-offs. 
This effect is most pronounced for deep path queries ($5.30\times$--$7.98\times$), whereas shallow star queries see smaller gains ($1.22\times$--$1.26\times$) because flatter f-trees incur lower per-tuple overhead. 

In summary, when branching enables compact factorized intermediates, \textsc{FFX} preserves vectorization and exploits compactness, reducing redundant computation.
In contrast to prior approaches, vectors remain fully packed, interpretation overhead drops, microarchitectural efficiency increases, and \textsc{FFX} obtains the benefits of both compact representations and CPU-efficient execution. 

\subsection{End-to-End Join Benchmark Evaluation}
\label{subsec:end-to-end-system-evaluation}

We compare \textsc{FFX} against two state-of-the-art systems: 
\textsc{DuckDB}~\cite{DBLP:conf/sigmod/RaasveldtM19} and 
\textsc{Kuzu}~\cite{DBLP:conf/cidr/JinFCLS23}. 
\textsc{DuckDB} is a high-performance analytical RDBMS with a vectorized execution engine, 
while \textsc{Kuzu} is a graph DBMS with factorized execution based on a list-based representation. 

While both \textsc{DuckDB} and \textsc{Kuzu} are mature, feature-rich production systems, 
\textsc{FFX} is a research prototype optimized for the core problem studied here. 
Hence, the absolute numbers are not directly comparable, but they are indicative of the potential efficiency gains achievable by tightly integrating factorization and vectorization.

\subsubsection{System Setup} 
We ensure that table statistics are computed so that the DBMS optimizers can select join orders based on cost. 
We run all systems single-threaded in in-memory mode, with disk spilling disabled and sufficient memory provisioned. 
We perform one warm-up run, which is discarded from measurement. 
We then report the median of three timed runs and measure execution time only, excluding compilation time.

\subsubsection{Workload}

We evaluate the three systems on the \texttt{LSQB}~\cite{DBLP:conf/sigmod/MhedhbiLKWS21} and \texttt{JOB}~\cite{DBLP:journals/pvldb/LeisGMBK015} benchmarks, as well as on the nine queries in \tab{tab:queries} over the \texttt{Amazon} dataset. 

\subsubsection{LSQB Results} 
\tab{tab:lsqb-comparison} reports single-threaded run times at scale factor~10 for Q1--Q6. We omit Q7--Q9 because \textsc{FFX} does not support outer and anti-joins. 
Overall, \textsc{FFX} is consistently competitive and often substantially faster than both \textsc{Kuzu} and \textsc{DuckDB}, with the largest gains on join patterns that admit compact factorization.

For \textsc{FFX}, we pick orderings to maximize factorization, favouring short and branching f-trees; otherwise, we pick plans arbitrarily among equally compact f-trees. 
This strategy is most effective for path queries such as \texttt{Q1} and \texttt{Q6}, where a compact hierarchy substantially reduces redundant work. 
In contrast, cyclic join queries lead to taller f-trees and reduce compactness. 
In these cases (\eg \texttt{Q2}), we evaluate \textsc{FFX} under \textsc{DuckDB}'s chosen ordering.  

\subsubsection{JOB Results} 
\begin{figure*}[t!]
	\centering
	\begin{subfigure}[b]{0.16\columnwidth}
		\centering
		\includegraphics[width=\textwidth]{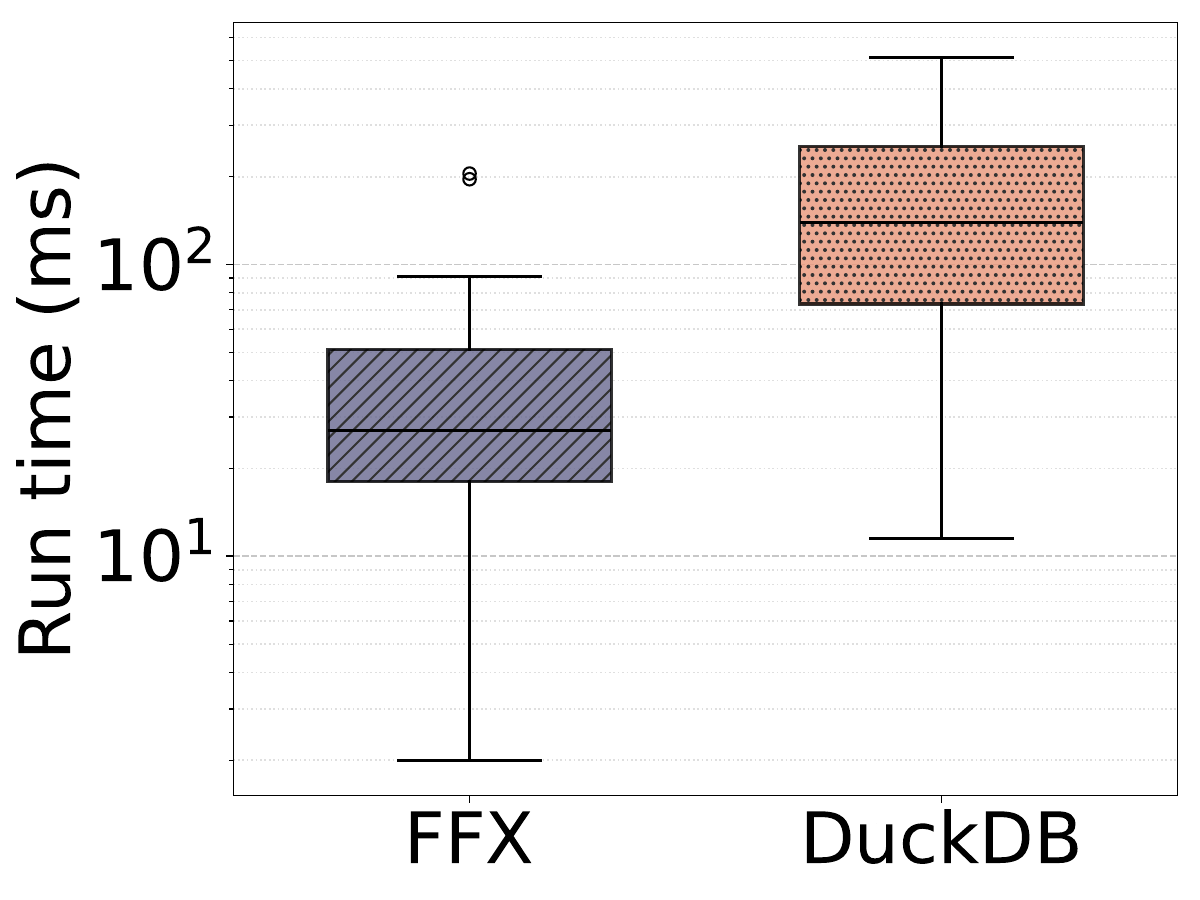}
		\caption{J.a run time}
		\label{fig:ffx_duckdb_runtime_comp_a}
	\end{subfigure}
	\hfill
	\begin{subfigure}[b]{0.16\columnwidth}
		\centering
		\includegraphics[width=\textwidth]{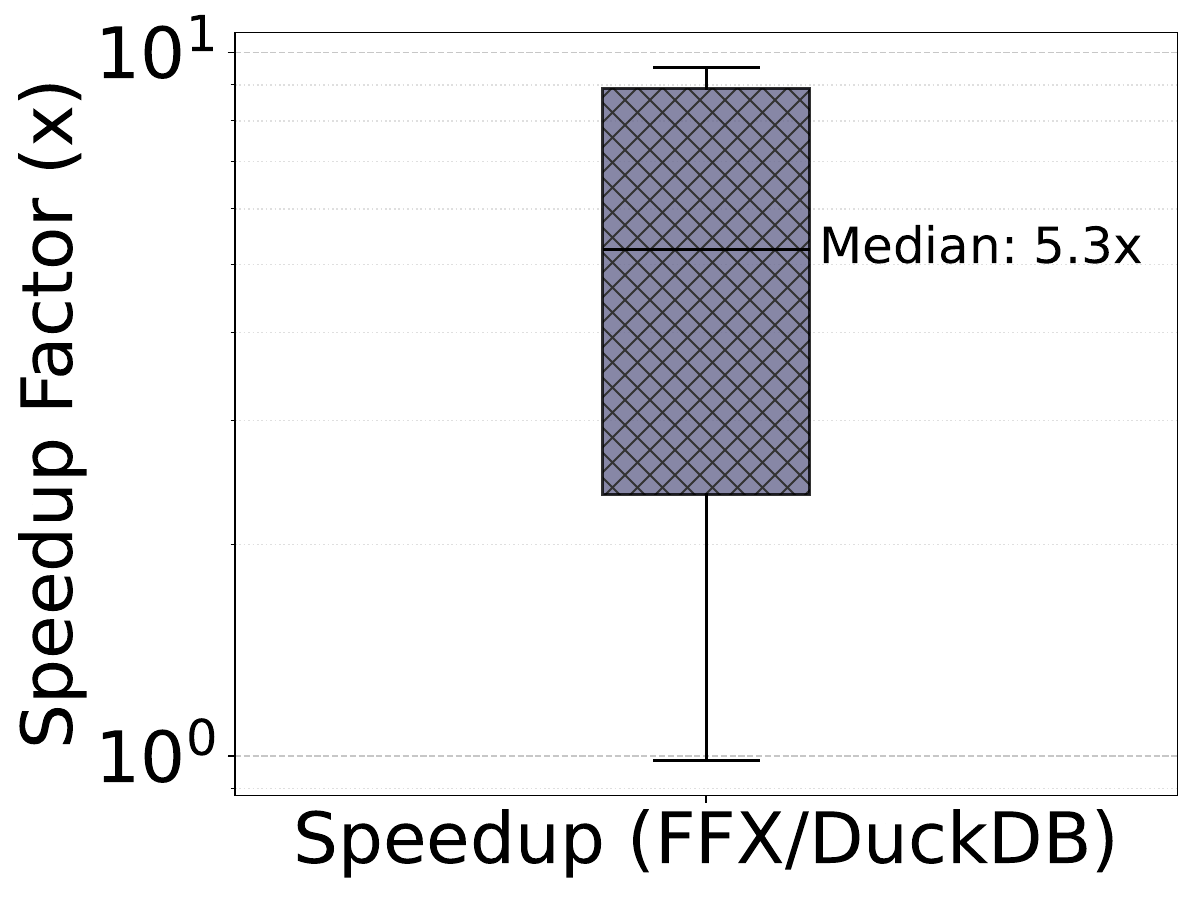}
		\caption{J.a speedup}
		\label{fig:ffx_duckdb_speedup_comp_a}
	\end{subfigure}
	\hfill
	\begin{subfigure}[b]{0.16\columnwidth}
		\centering
		\includegraphics[width=\textwidth]{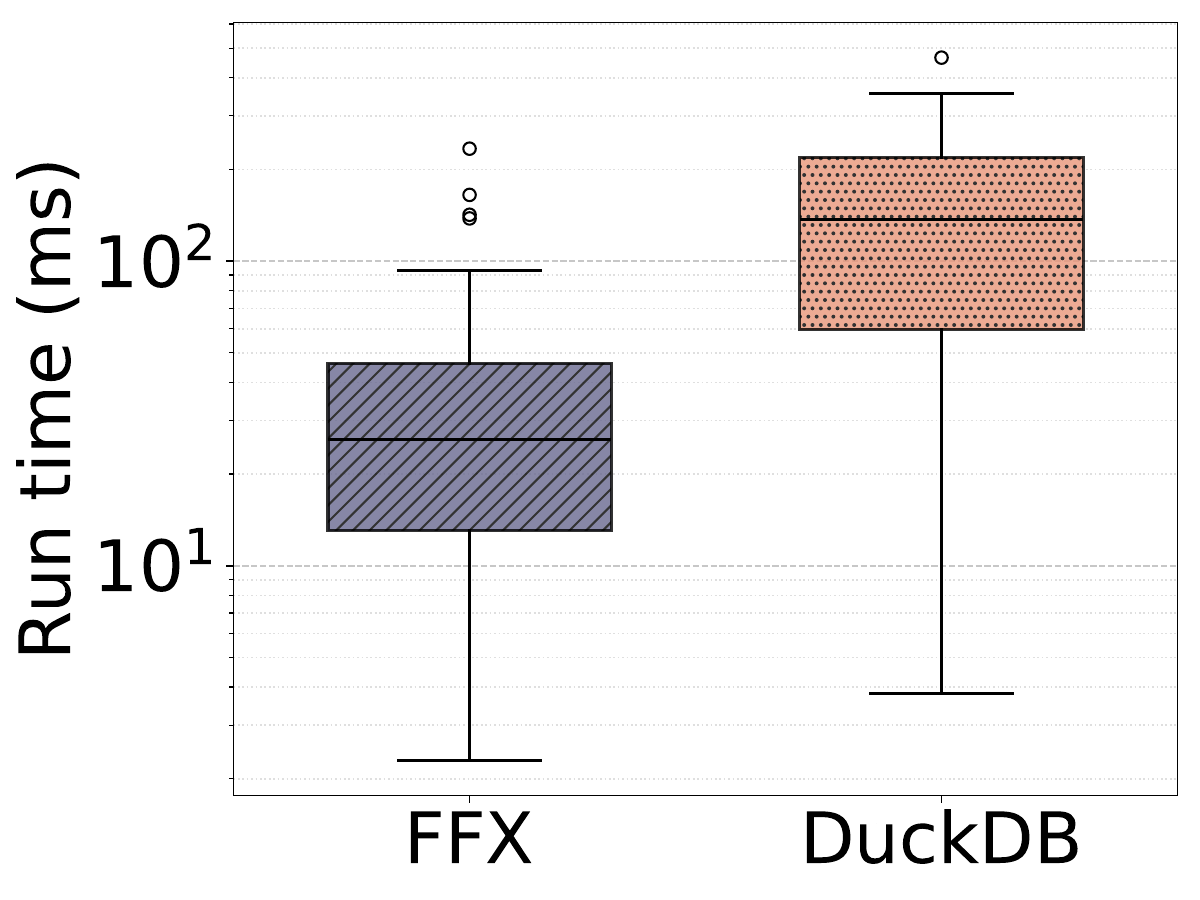}
		\caption{J.b run time}
		\label{fig:ffx_duckdb_runtime_comp_b}
	\end{subfigure}
	\hfill
	\begin{subfigure}[b]{0.16\columnwidth}
		\centering
		\includegraphics[width=\textwidth]{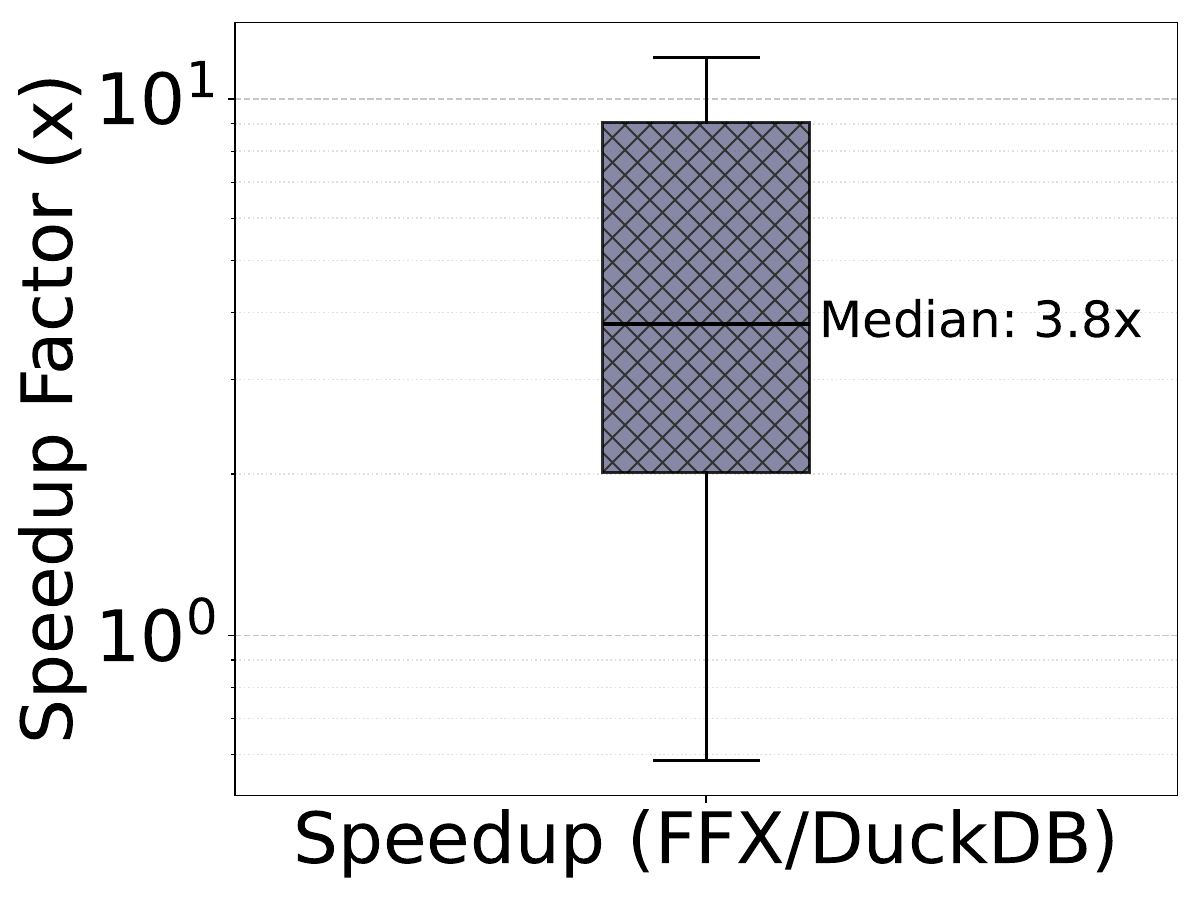}
		\caption{J.b speedup}
		\label{fig:ffx_duckdb_speedup_comp_b}
	\end{subfigure}
	\hfill
	\begin{subfigure}[b]{0.16\columnwidth}
		\centering
		\includegraphics[width=\textwidth]{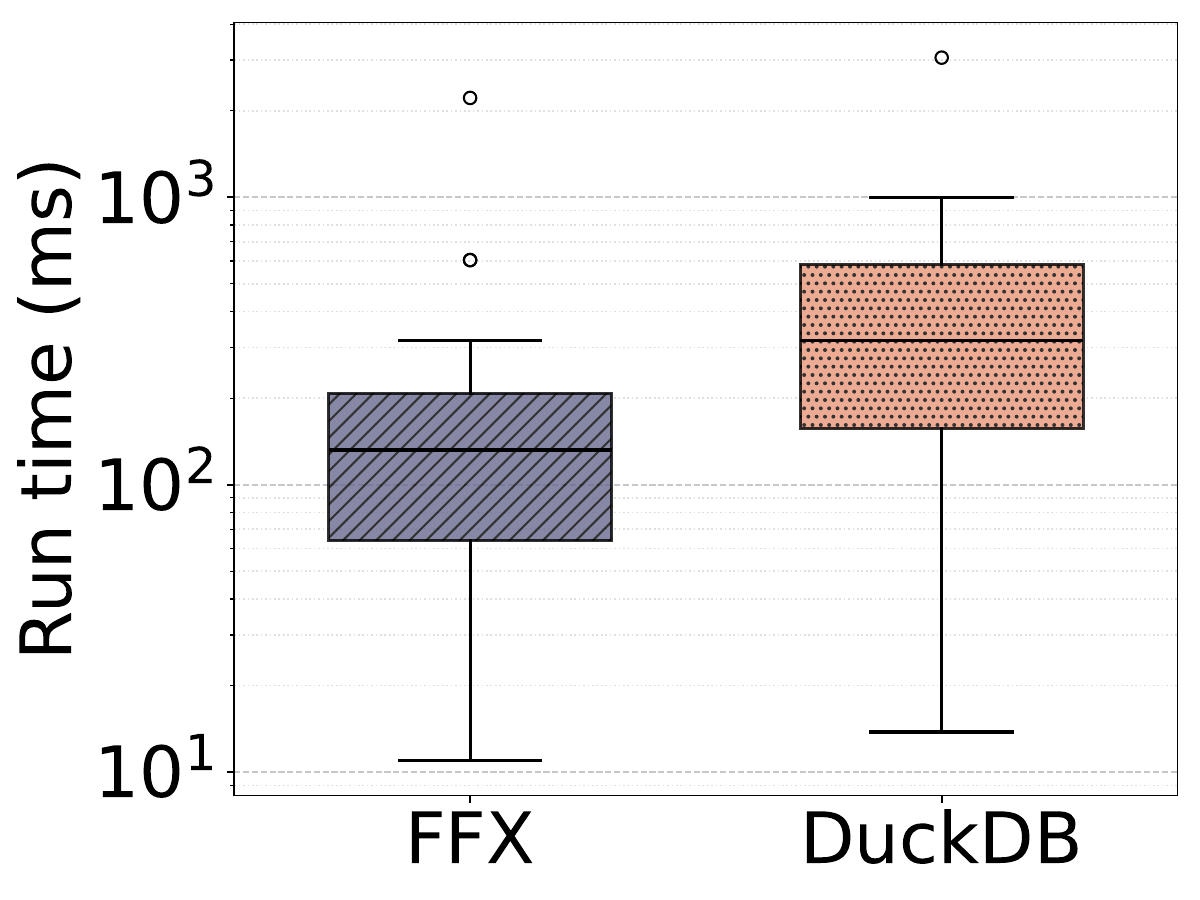}
		\caption{J.c run time}
		\label{fig:ffx_duckdb_runtime_comp_c}
	\end{subfigure}
	\hfill
	\begin{subfigure}[b]{0.16\columnwidth}
		\centering
		\includegraphics[width=\textwidth]{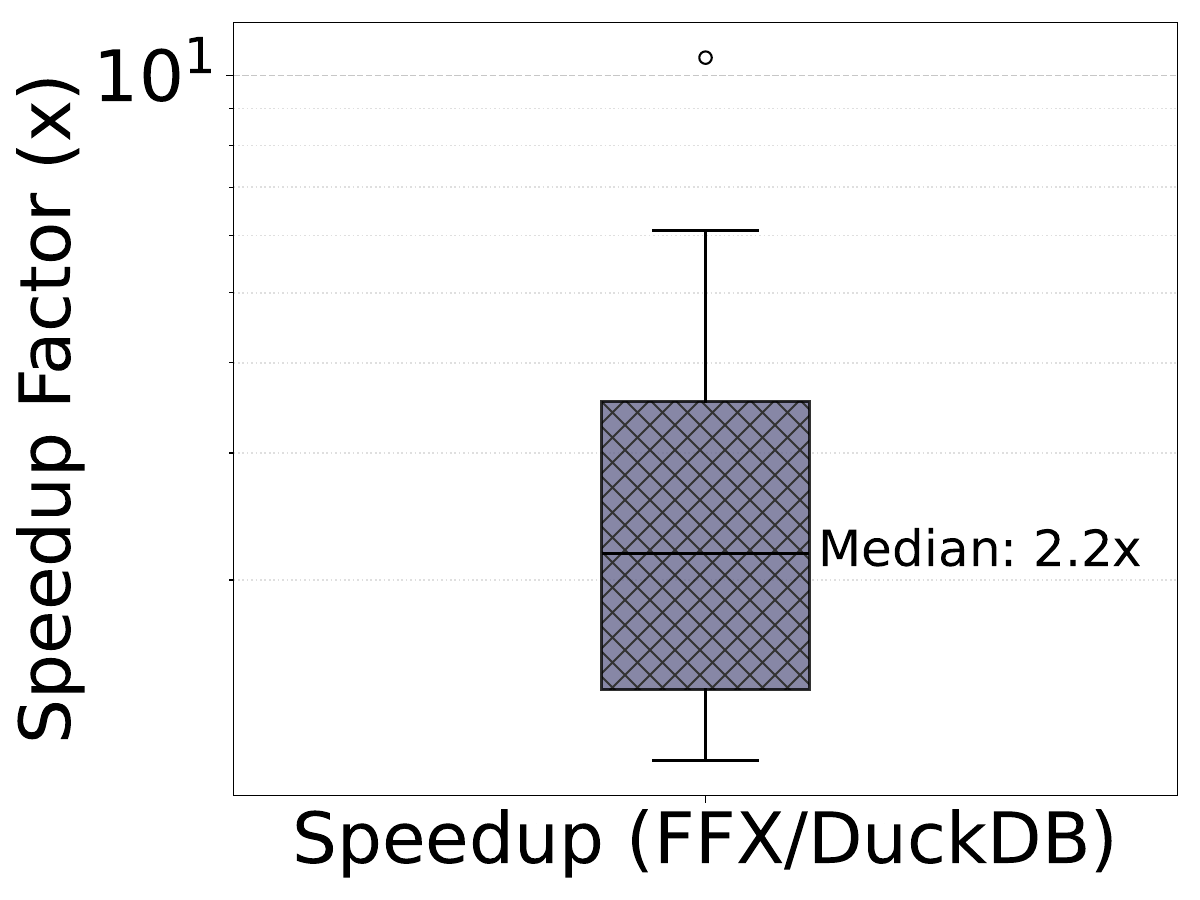}
		\caption{J.c speedup}
		\label{fig:ffx_duckdb_speedup_comp_c}
	\end{subfigure}
	\caption{Run time (ms) and speedup of \textsc{FFX} over \textsc{DuckDB} on the JOB~\cite{DBLP:journals/pvldb/LeisGMBK015} (J) variants a, b, and c.}
	\label{fig:ffx_duckdb_JOB_characteristics}
\end{figure*}

\fig{fig:ffx_duckdb_JOB_characteristics} compares \textsc{FFX} and \textsc{DuckDB} on the \texttt{JOB} benchmark variants \texttt{a}, \texttt{b}, and \texttt{c}. 
Across all variants and queries, \textsc{FFX} is fastest when the ordering yields succinct f-trees, often star-like patterns, allowing it to keep intermediates factorized and avoid enumerating a large number of flat tuples. 

We highlight variant \texttt{c} because it has less selective predicates, which increase intermediate sizes. 
As a result, speedups are smaller because of increased computation during join processing: 
lower selectivity causes more index nested-loop probes (\texttt{FFX}'s core join operator) and more irregular join-index lookups, which in turn raise cache-miss rates. 
For example, from \texttt{Q18b} to \texttt{Q18c}, the cache-miss rate increases from 2.19\% to 3.89\%. 

\begin{table}[t!]
	\centering
	\small
	\begin{tabular}{@{}ccrr@{}}
		\toprule
		\textbf{Query} & \textbf{FFX} & \textbf{Kuzu} & \textbf{DuckDB} \\
		\midrule
		Q1 & 1.28 & 591.10 \textbf{(545.5$\times$)} & 18.43 \textbf{(17.0$\times$)} \\
		Q2 & 2.71 & 15.16 \textbf{(5.6$\times$)} & 2.72 \textbf{(1.0$\times$)} \\
		Q3 & 3.32 & 105.61 \textbf{(31.8$\times$)} & 6.14 \textbf{(1.9$\times$)} \\
		Q4 & 1.26 & 7.48 \textbf{(5.9$\times$)} & 9.86 \textbf{(7.8$\times$)} \\
		Q5 & 5.56 & 10.08 \textbf{(1.8$\times$)} & 6.94 \textbf{(1.2$\times$)} \\
		Q6 & 5.84 & 49.42 \textbf{(8.6$\times$)} & 153.39 \textbf{(26.3$\times$)} \\
		\bottomrule
	\end{tabular}
	\caption{Run time comparison in seconds for queries Q1--Q6 (single-threaded execution) on \texttt{LSQB}. Improvement factors relative to Kuzu and DuckDB are shown in parentheses.}
	\label{tab:lsqb-comparison}
\end{table}

\subsubsection{Nine Queries Results} Following the \texttt{JOB} benchmark methodology, for the nine queries in~\tab{tab:queries}, we also project the \texttt{MIN} value for all attributes to isolate join execution performance and avoid heavy tuple copying or enumeration in the sink operator. 

\begin{figure}[t!]
	\centering
	\begin{subfigure}[b]{0.17\columnwidth}
		\centering
		\includegraphics[width=\textwidth]{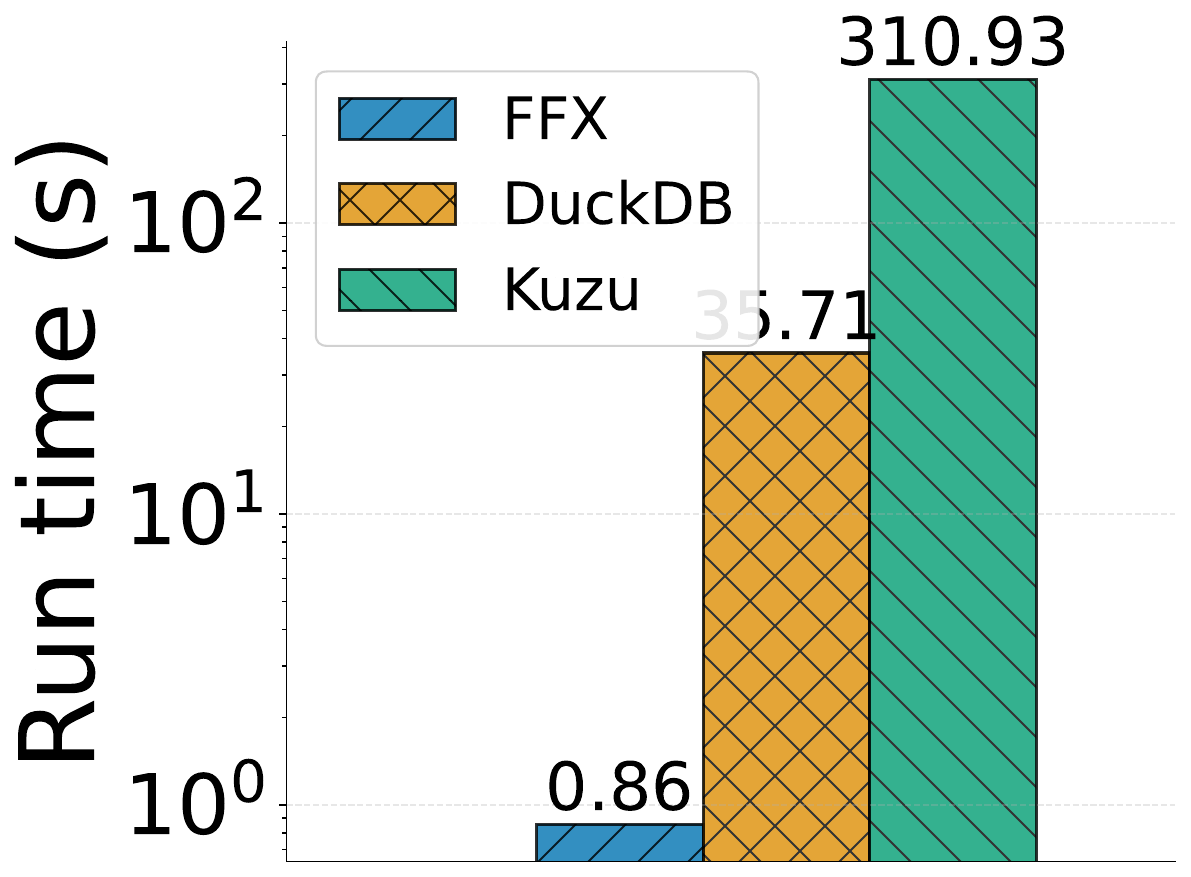}
		\captionsetup{font=tiny,skip=1pt}
		\caption{Q1}
		\label{fig:exp4_q1}
	\end{subfigure}
	\hspace*{1em}
	\begin{subfigure}[b]{0.17\columnwidth}
		\centering
		\includegraphics[width=\textwidth]{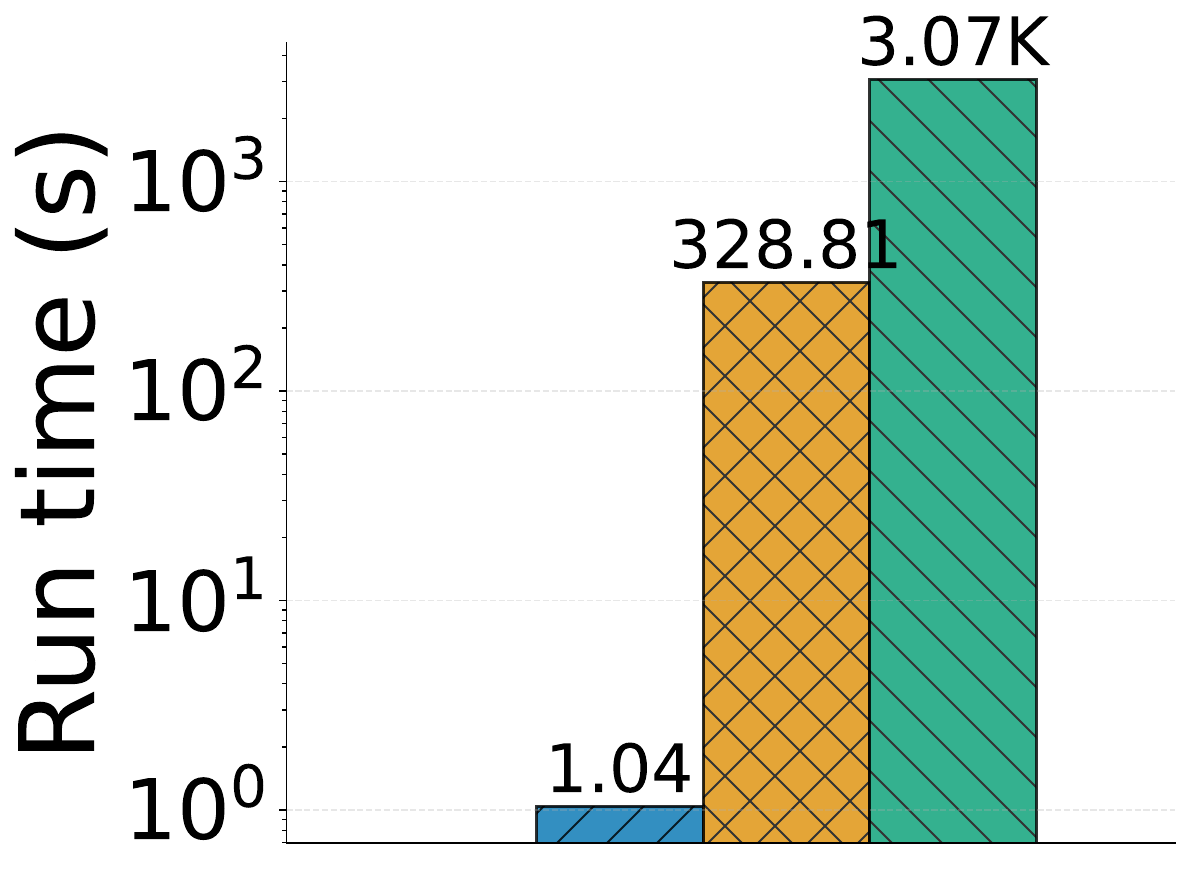}
		\captionsetup{font=tiny,skip=1pt}
		\caption{Q2}
		\label{fig:exp4_q2}
	\end{subfigure}
	\hspace*{1em}
	\begin{subfigure}[b]{0.17\columnwidth}
		\centering
		\includegraphics[width=\textwidth]{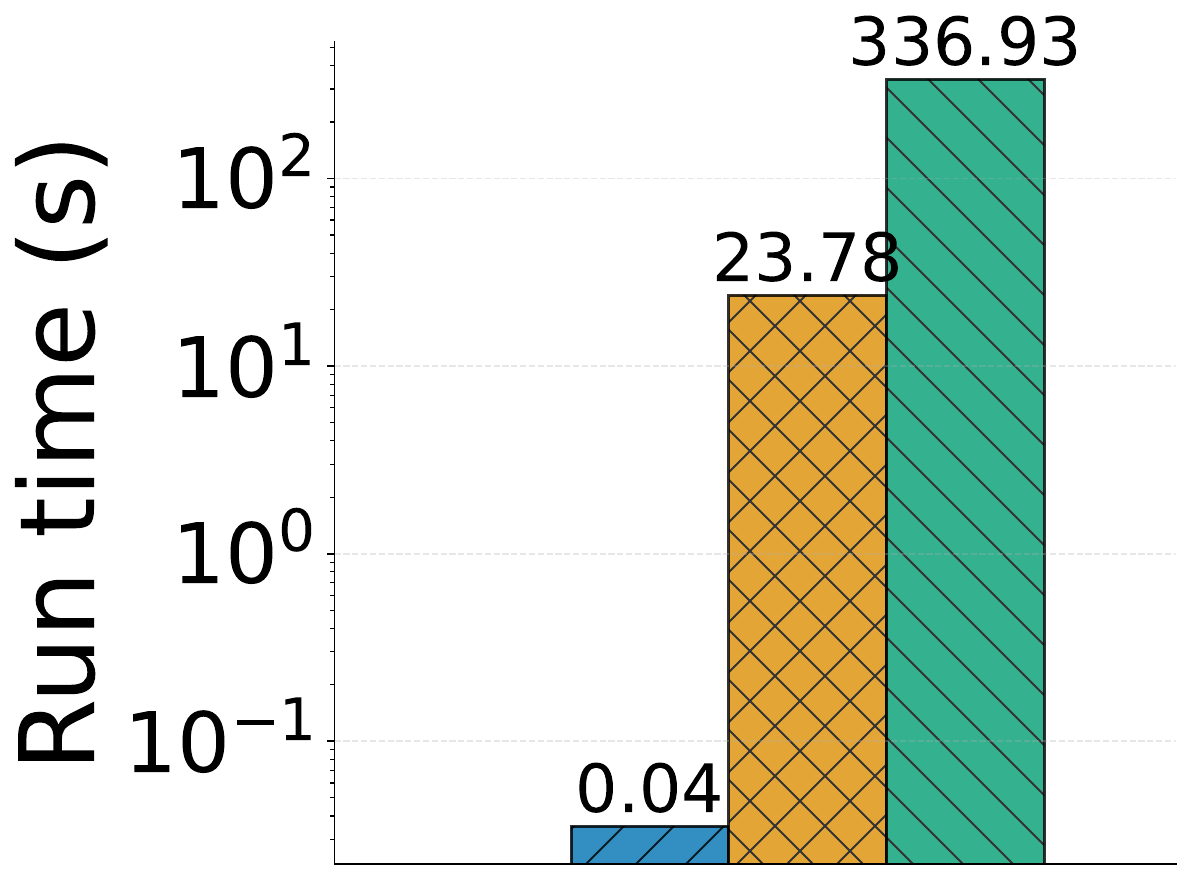}
		\captionsetup{font=tiny,skip=1pt}
		\caption{Q3}
		\label{fig:exp4_q3}
	\end{subfigure}
	\hspace*{1em}
	\begin{subfigure}[b]{0.17\columnwidth}
		\centering
		\includegraphics[width=\textwidth]{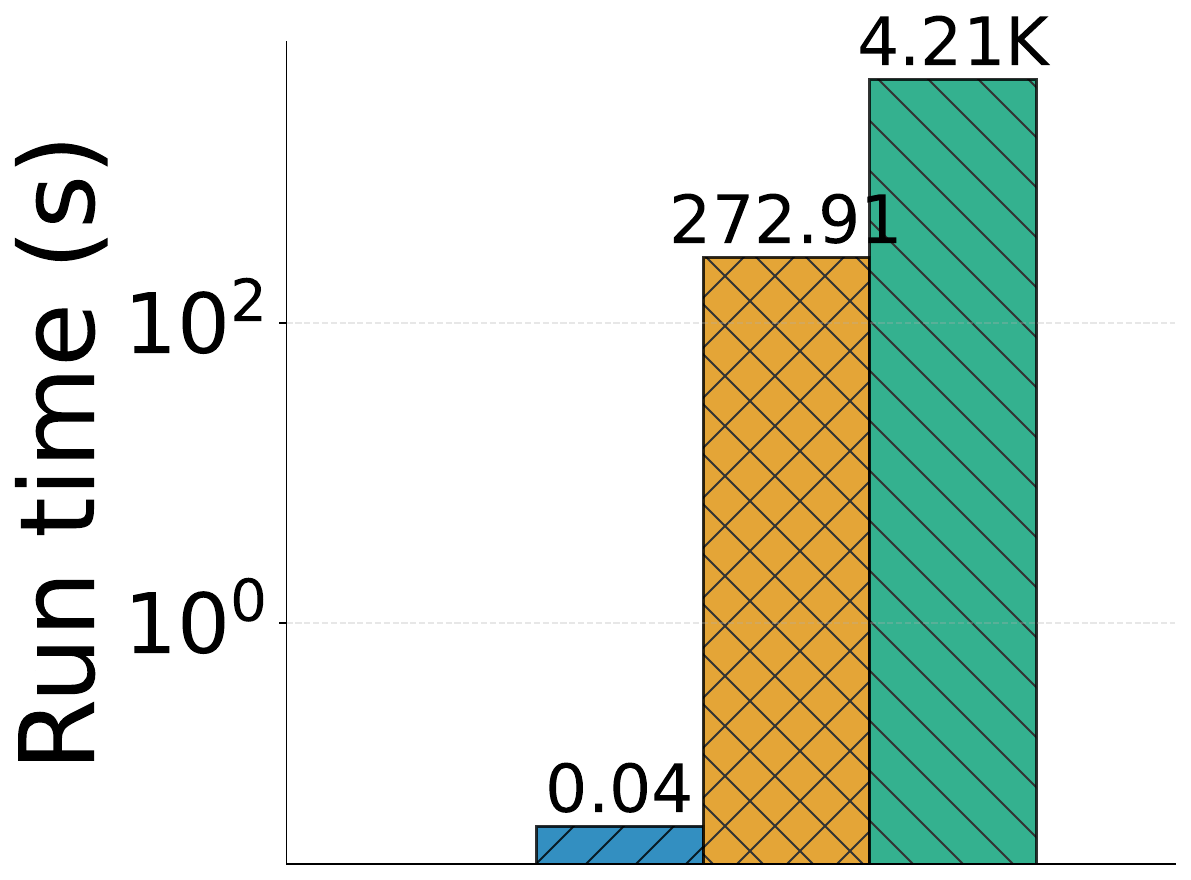}
		\captionsetup{font=tiny,skip=1pt}
		\caption{Q4}
		\label{fig:exp4_q4}
	\end{subfigure}
	\hspace*{1em}
	\begin{subfigure}[b]{0.17\columnwidth}
		\centering
		\includegraphics[width=\textwidth]{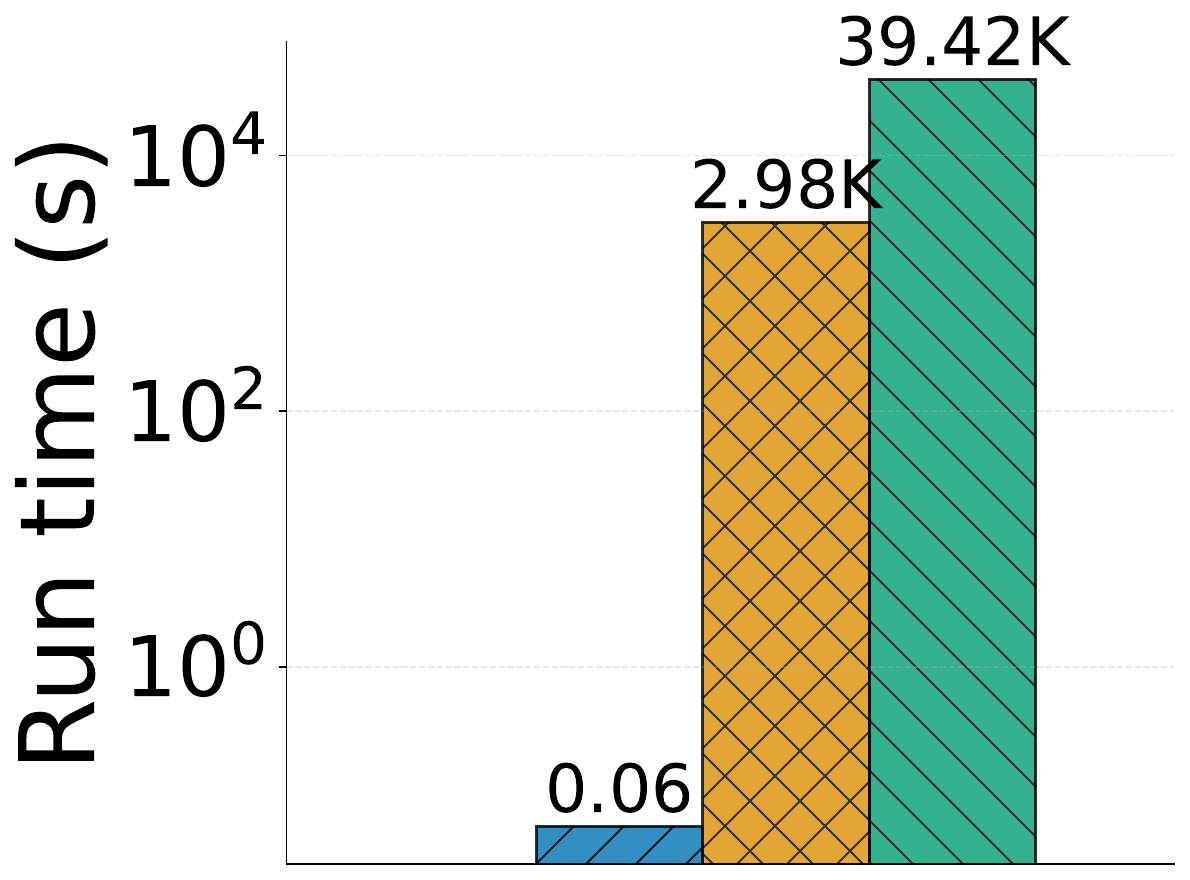}
		\captionsetup{font=tiny,skip=1pt}
		\caption{Q5}
		\label{fig:exp4_q5}
	\end{subfigure}\vspace*{1em}
	
	\begin{subfigure}[b]{0.17\columnwidth}
		\centering
		\includegraphics[width=\textwidth]{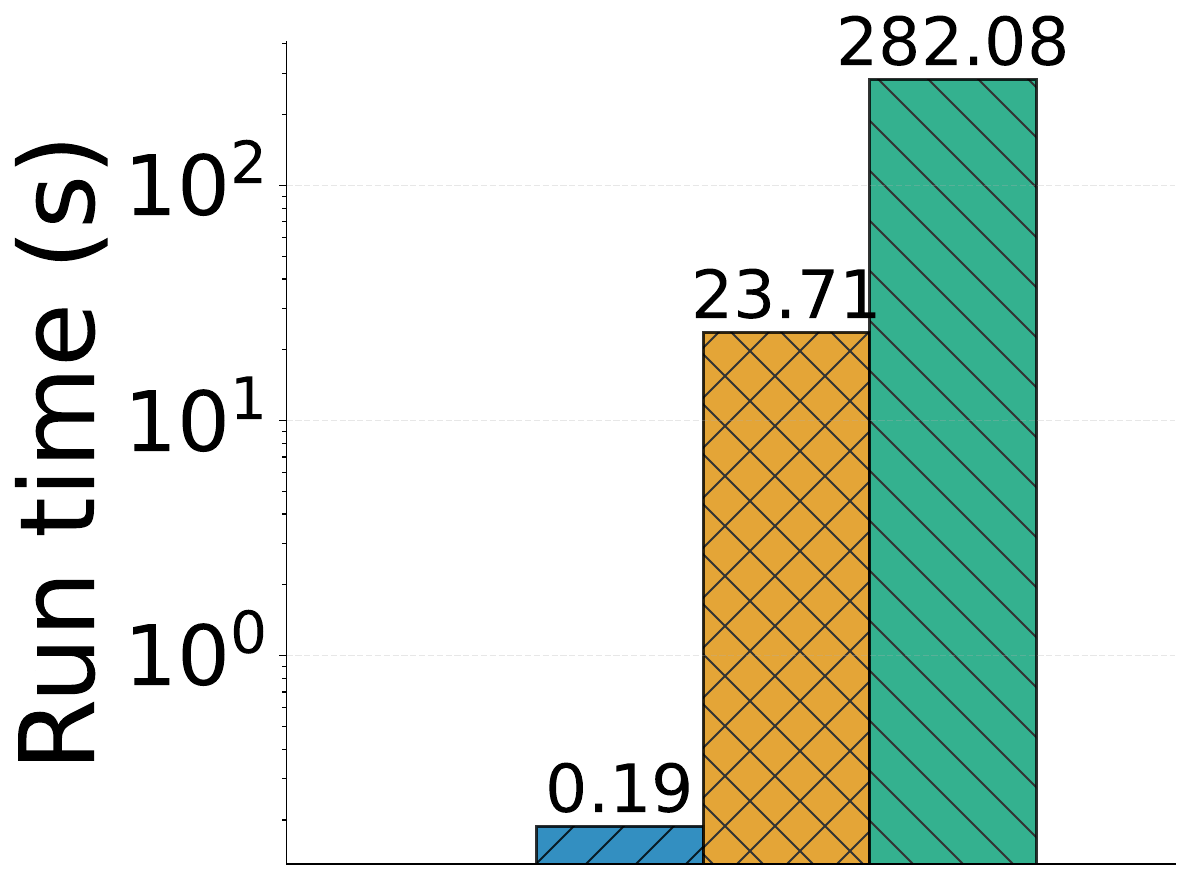}
		\captionsetup{font=tiny,skip=1pt}
		\caption{Q6}
		\label{fig:exp4_q6}
	\end{subfigure}
	\hspace*{2em}
	\begin{subfigure}[b]{0.17\columnwidth}
		\centering
		\includegraphics[width=\textwidth]{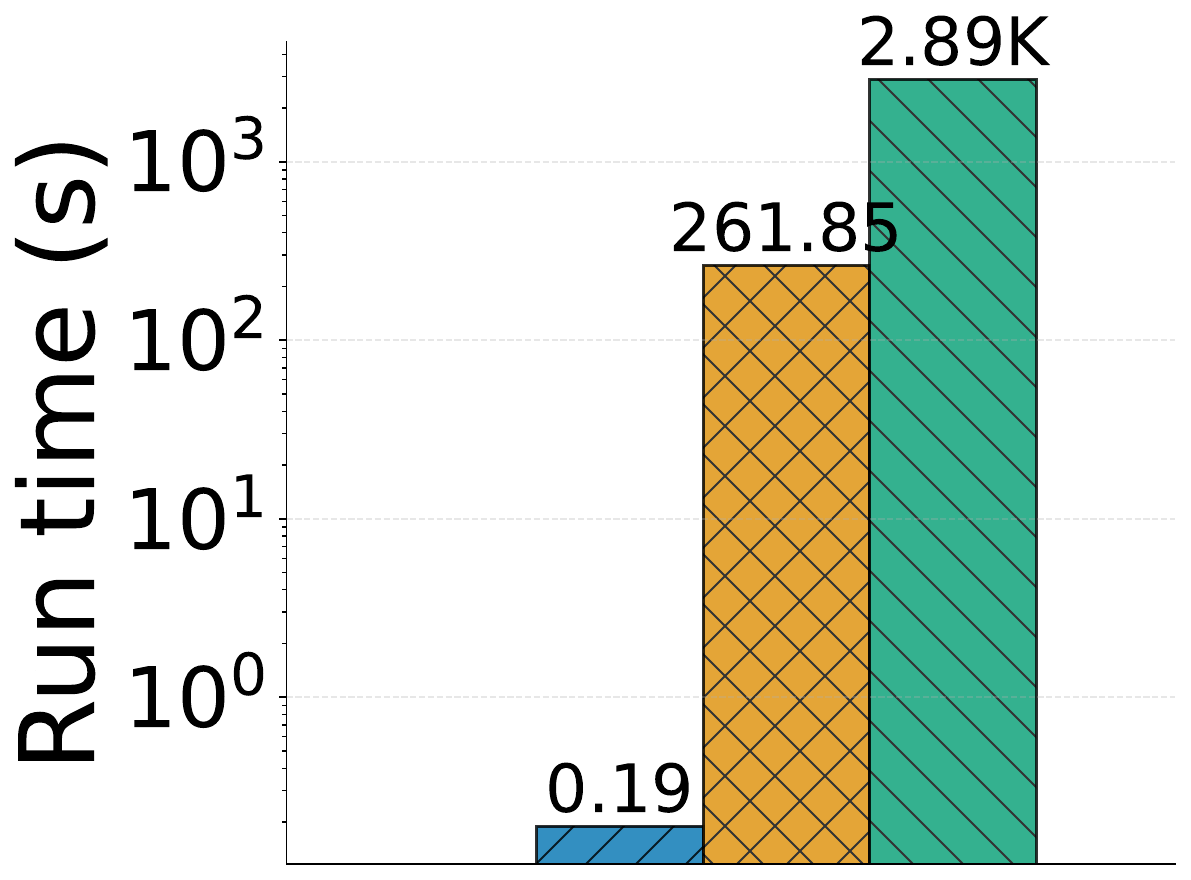}
		\captionsetup{font=tiny,skip=1pt}
		\caption{Q7}
		\label{fig:exp4_q7}
	\end{subfigure}
	\hspace*{2em}
	\begin{subfigure}[b]{0.17\columnwidth}
		\centering
		\includegraphics[width=\textwidth]{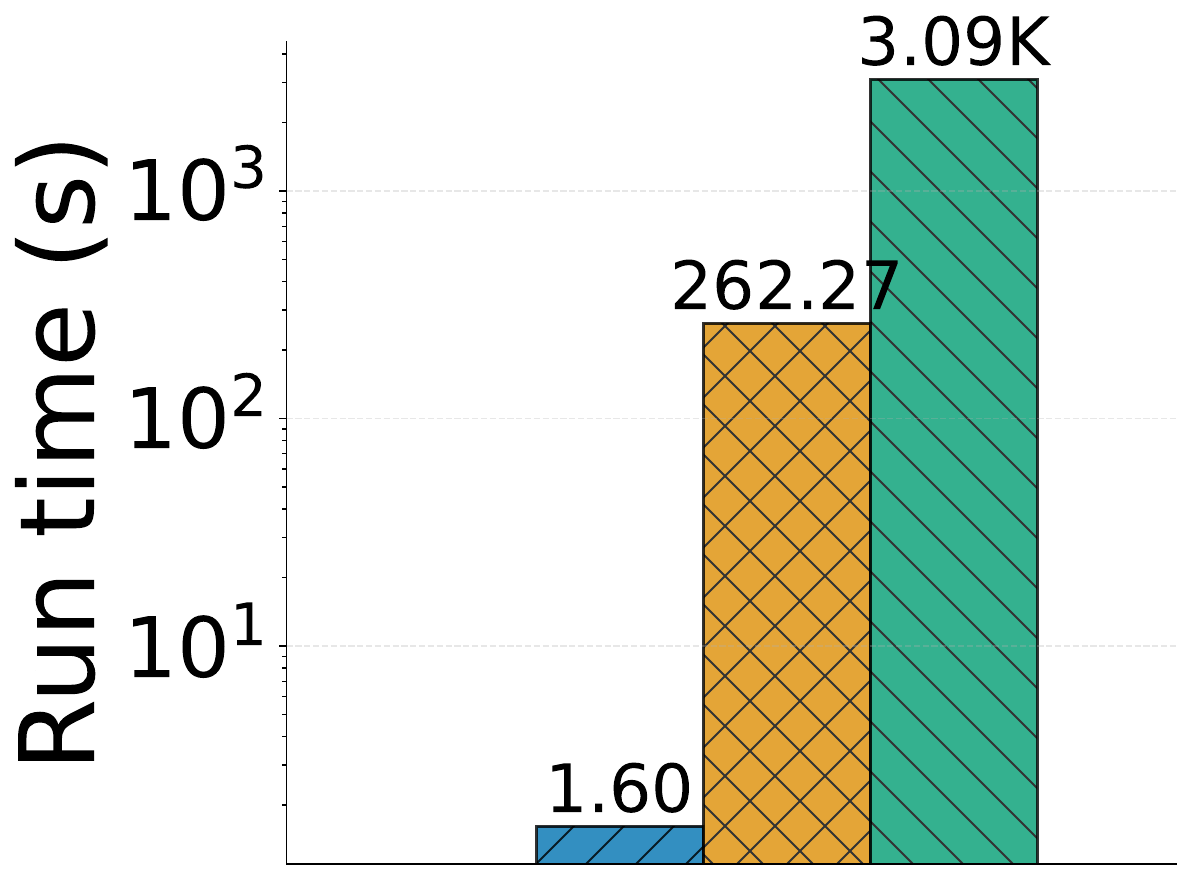}
		\captionsetup{font=tiny,skip=1pt}
		\caption{Q8}
		\label{fig:exp4_q8}
	\end{subfigure}
	\hspace*{2em}
	\begin{subfigure}[b]{0.17\columnwidth}
		\centering
		\includegraphics[width=\textwidth]{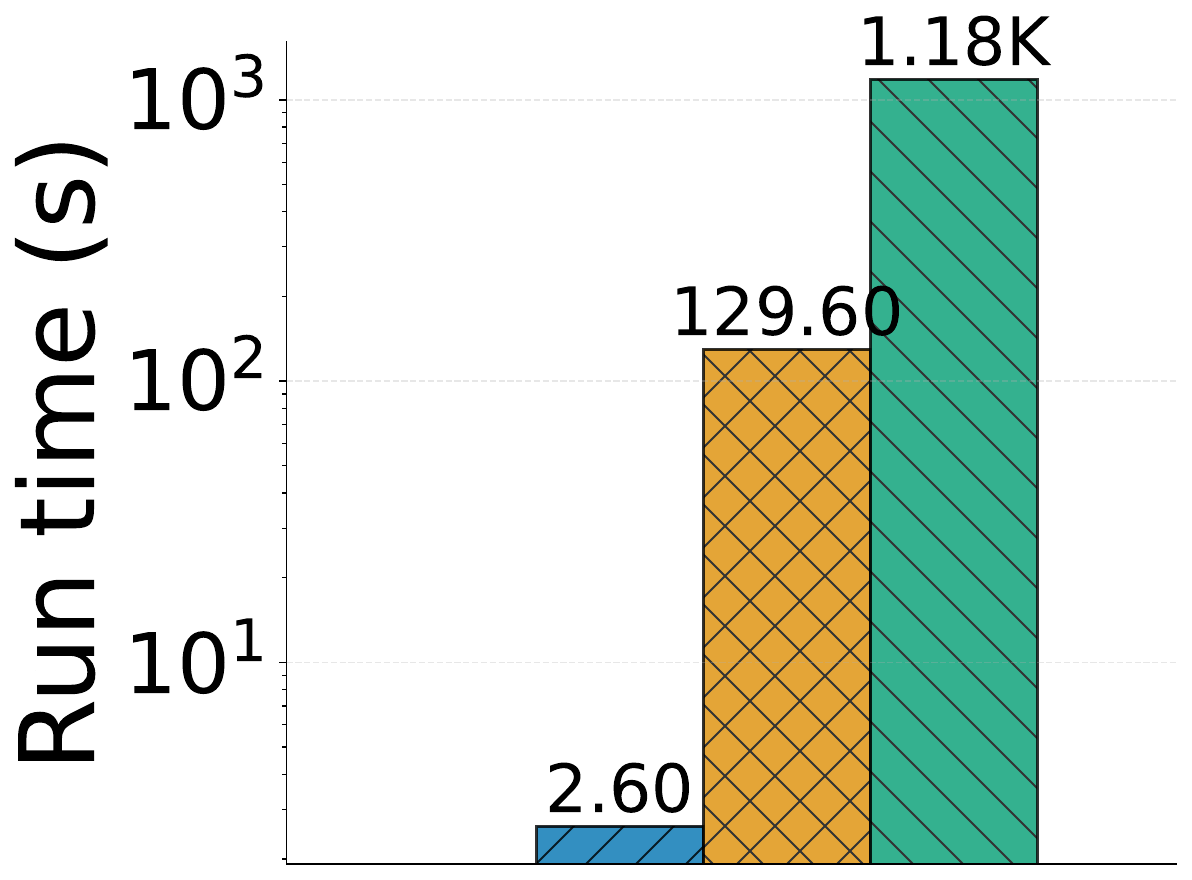}
		\captionsetup{font=tiny,skip=1pt}
		\caption{Q9}
		\label{fig:exp4_q9}
	\end{subfigure}
	\caption{Comparison of FFX, DuckDB, and Kuzu on \texttt{Amazon} for all nine 
		queries, from \texttt{Q1} to \texttt{Q9}, with \texttt{MIN} aggregation. Each system is run in single-threaded mode.
		FFX achieves speedups: $10^2$--$10^5\times$ across queries.}
	\label{fig:exp4_scalability}
\end{figure}

\fig{fig:exp4_scalability} reports run time in seconds across all queries in~\tab{tab:queries}. 
\textsc{FFX} achieves substantial speedups, ranging from $10^2\times$ to $10^5\times$ over both \textsc{DuckDB} and \textsc{Kuzu}. 
The advantage is particularly pronounced for star-shaped queries such as \texttt{Q5}, where \textsc{FFX} is $10^5\times$ faster than \textsc{Kuzu}. 
In such queries, data associated with the central join key (\texttt{a}) is repeatedly reused across multiple joins, and our factorized vectors allow \textsc{FFX} to eliminate potential redundancy by processing shared join keys once. 

To better understand the performance gap, we inspect the query plans using \texttt{EXPLAIN}-like utilities and compare them with \textsc{FFX}'s execution strategy. 
We find that \textsc{FFX}'s main advantage arises when it chooses a join order that yields the most compact intermediate representation and therefore minimizes redundant computation. 
For example, on \texttt{Q1}, \textsc{DuckDB} and \textsc{Kuzu} evaluate the logical plan
$(R(a,b) \Join R(b,c)) \Join (R(c,d) \Join R(d,e))$. 
Executing this plan requires materializing $2.98$B tuples before the final projection applies \texttt{MIN} to all attributes. 
In contrast, \textsc{FFX} uses its factorized representation to avoid generating this large result. 
Instead, it pushes projection and aggregation down to smaller intermediates corresponding to the subqueries
$R(a,b) \Join R(b,c)$ and $R(c,d) \Join R(d,e)$, 
which together contain only $63.2$M tuples.

\subsection{Multi-core Scalability}
\label{sec:multi-core-scalability} 

To evaluate \textsc{FFX}'s scalability, we execute the query $R(a,b) \Join R(a,c)$ on the \texttt{LiveJournal} and \texttt{Twitter} datasets.
Because \textsc{FFX} operates fully in memory, without disk I/O, query execution can be embarrassingly parallel. 
We compare \textsc{FFX} against \textsc{DuckDB} and \textsc{Kuzu} on \texttt{LiveJournal}; comparisons on \texttt{Twitter} are omitted because of its larger scale and frequent timeouts in the baselines. 
Both \textsc{DuckDB} and \textsc{Kuzu} are run in in-memory mode. 

Since the \texttt{Twitter} dataset does not fit on the hardware described in \sect{subsubsec:hardware}, 
we use a machine equipped with 64~vCPUs, 64~GB of main memory, and an \texttt{AMD EPYC 9655 (Zen 5) @ 2.7 GHz} CPU. 
The query is evaluated using up to 32~cores. 

\begin{figure}[t!]
	\centering
	\begin{subfigure}[b]{0.3\columnwidth}
		\centering
		\includegraphics[width=\textwidth]{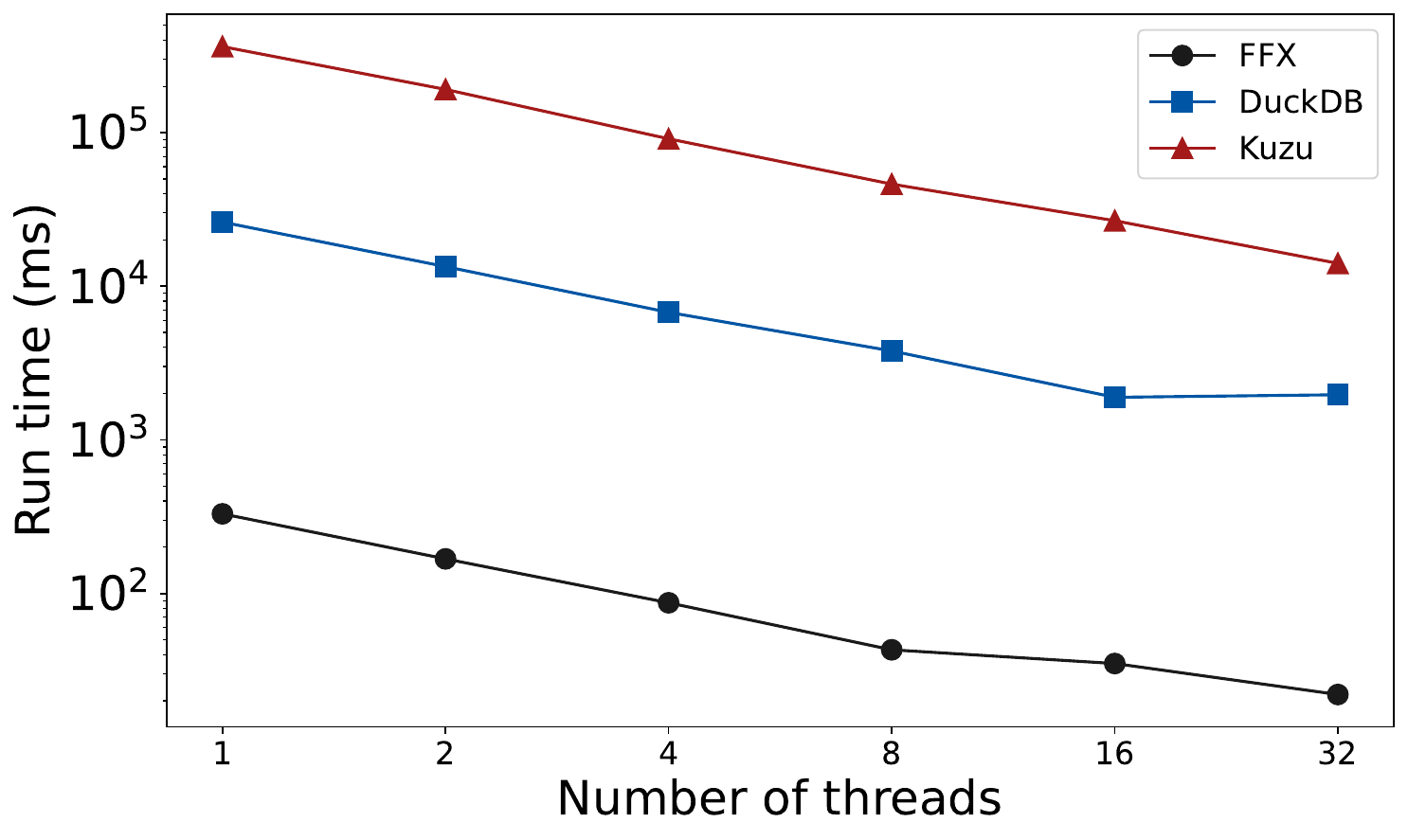}
		\caption{\texttt{LiveJournal}}
		\label{fig:multithreaded-LJ}
	\end{subfigure}
	\hspace*{4em}
	\begin{subfigure}[b]{0.3\columnwidth}
		\centering
		\includegraphics[width=\textwidth]{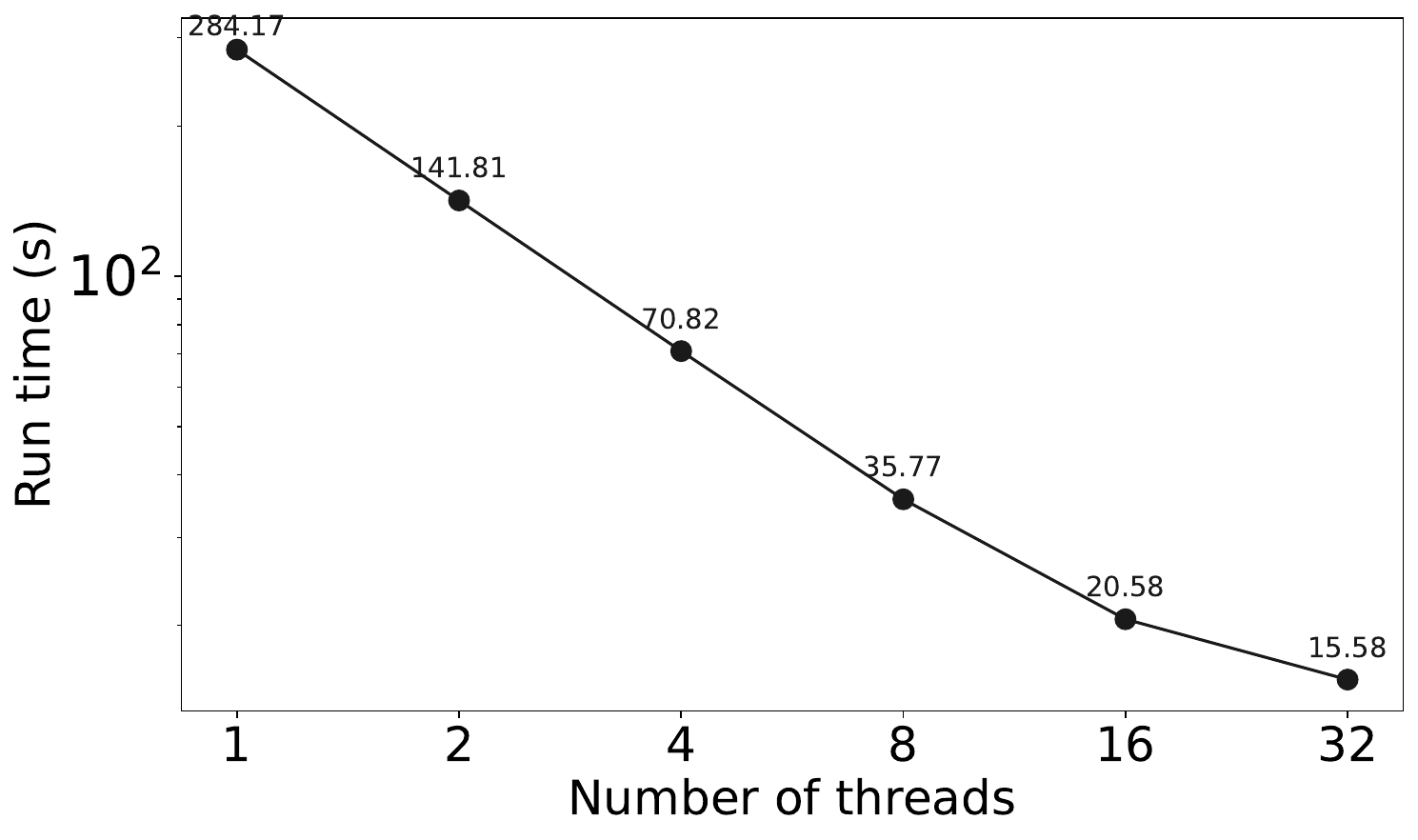}
		\caption{\texttt{Twitter}}
		\label{fig:multithreaded-Twitter}
	\end{subfigure}
	\caption{Multi-threaded performance comparison for query $R(a,b)\Join R(a,c)$ on \texttt{LiveJournal} and \texttt{Twitter} datasets. All systems ran in in-memory mode.}
	\label{fig:multithreaded-scalability}
\end{figure}

\fig{fig:multithreaded-scalability} presents the results. 
All systems exhibit near-linear speedup. 
\textsc{FFX} remains faster than both \textsc{Kuzu} and \textsc{DuckDB} because of its combined factorized and vectorized execution. 
On \texttt{LiveJournal} and \texttt{Twitter}, \textsc{FFX} scales by up to $13.2\times$ and $13.8\times$, respectively. 
We note that as the number of cores increases beyond 16, fewer scalability benefits are observed on this specific workload.

\subsection{Evaluation of Factorized Semantic Processing}
\label{subsec:prompt-compression-benefits}

\subsubsection{Setup}
We evaluate the benefits of factorized semantic processing on the \texttt{llm\_map} operator. 
All experiments run on the citation \texttt{ArXiv} dataset in \tab{tab:datasets-used} and apply \texttt{llm\_map} to the output of a many-to-many join query, allowing us to compare different semantic operator implementations with different serialized data representations. 

\subsubsection{Dataset}
We construct a dataset of academic-paper triplets from 2-hop citations by evaluating:\vspace*{-0.1em}
$$Q_p = Citation(a,b) \Join Citation(b,c) \Join Paper(a, -, abs_a) \Join Paper(b, -, abs_b) \Join Paper(c, -, abs_c)$$ 
where $a$ cites $b$ and $b$ cites $c$, and \texttt{Citation} and \texttt{Paper} are base tables defined in \sect{subsec:semantic-llm-operators}. 
To keep inference costs manageable while enabling sweeps over batch sizes, operator implementations, and models, we sample the query result by randomly selecting six distinct $b$ values. 
This yields $2028$ $(a,b,c)$ triplets spanning $232$ unique papers, including $153$ distinct $a$ papers and $77$ distinct $c$ papers. 
We use one separately selected value of $b$, not among the six, to create a training set to iterate on and improve our prompts. We report the results on the $2028$ triplets as a dev set. 

\subsubsection{Task}\label{subsubsec:task}
For each output tuple of $Q_p$, 
we apply \texttt{llm\_map} to generate five representative keywords from the abstracts to summarize the 2-hop citation chain $(a,b,c)$. 

\subsubsection{Metrics}\label{subsubsec:metric} 
Because there is no ground-truth keyword set, we evaluate keyword generation through retrieval. 
We use the concatenated keywords as a query and attempt to retrieve the corresponding triplet from the full set of $2028$ abstract triplets using a hybrid search method that includes BM25. 
For each query, we report Recall@5 and Recall@10, that is, whether the correct triplet is retrieved in the top 5 or top 10 results, respectively, and we aggregate the results. 

\begin{table*}[t!]
	\centering
	\begin{tabular}{c c c c c c r r c}
		\toprule
		\textbf{\# T} & \textbf{System} & \textbf{\# MT} & \textbf{MR} & \textbf{R@5} & \textbf{R@10} &
		\textbf{In. T} & \textbf{Out. T} & \textbf{Run time (s)} \\
		\midrule
		1   & \texttt{FFX}-Flat & 0 & 0.000 & 0.7170 & 0.8023 & 2.41M & 108.44K & 88.69 \\
		\midrule
		4   & \texttt{FFX}-Flat & 0 & 0.000& 0.7416 & 0.8333 & 1.65M & 103.77K & 58.14 \\
		     & \texttt{FFX}-Fact & 0 & 0.000& 0.7569 & 0.8333 & 0.97M (\textbf{1.70$\times$}) & 110.51K & 63.01 \\
		\midrule
		8   & \texttt{FFX}-Flat & 0 & 0.000 & 0.7801 & 0.8688 & 1.53M & 102.50K & 59.67 \\
		     & \texttt{FFX}-Fact & 0 & 0.000 & 0.7554 & 0.8388 & 0.61M (\textbf{2.51$\times$}) & 107.60K & 61.40 \\
		\midrule
		16  & \texttt{FFX}-Flat & 0 & 0.000 & 0.8057 & 0.8861 & 1.46M & 100.53K & 63.45 \\
		      & \texttt{FFX}-Fact & 0 & 0.000 & 0.7643 & 0.8521 & 0.39M (\textbf{3.74$\times$}) & 103.62K & 71.89 \\
		\midrule
		32  & \texttt{FFX}-Flat & 2 & 0.001 & 0.8215 & 0.8856 & 1.43M & 97.59K & 75.16 \\
		      & \texttt{FFX}-Fact & 4 & 0.002 & 0.7505 & 0.8185 & 0.27M (\textbf{5.29$\times$}) & 102.57K & 98.69 \\
		\midrule
		64  & \texttt{FFX}-Flat & 33 & 0.016 & 0.8210 & 0.8861 & 1.42M & 95.07K & 105.48 \\
		       & \texttt{FFX}-Fact & 0 & 0.000 & 0.7525 & 0.8304 & 0.18M (\textbf{7.88$\times$}) & 97.79K & 112.40 \\
		\midrule
		128 & \texttt{FFX}-Flat & 347 & 0.171 & 0.6800 & 0.7446 & 1.41M & 79.93K & 148.04 \\
		       & \texttt{FFX}-Fact & 75 & 0.037 & 0.6805 & 0.7608  & 0.12M (\textbf{11.75$\times$}) & 95.56K & 174.06 \\
		\midrule
		256 & \texttt{FFX}-Flat & 1199 & 0.591 & 0.3319 & 0.3669 & 1.41M & 39.07K & 165.38 \\
		        & \texttt{FFX}-Fact & 517 & 0.255 & 0.5439 & 0.6154 & 0.09M (\textbf{15.67$\times$}) & 70.86K & 213.80 \\
		\midrule
		512  & \texttt{FFX}-Fact & 976 & 0.481 & 0.4004 & 0.4329 & 70.14K (\textbf{20.10$\times$}) & 52.60K & 303.78 \\
		2048 & \texttt{FFX}-Fact & 1112 & 0.548 & 0.3319 & 0.3782 & 62.56K (\textbf{22.54$\times$}) & 40.29K & 164.16 \\
		\midrule
		-- & \texttt{Lotus} & 0 & 0.000 & 0.6824 & 0.7978 & 2.26M & 95.95K & 92.68 \\
		\bottomrule
	\end{tabular}
		\caption{Keyword generation with \texttt{GPT-5.2}. 
			Comparison of factorized semantic processing and semantic processing over flat tuples (\texttt{FFX}-Flat) for the task in \sect{subsubsec:task}. 
			Results are grouped by the number of tuples in each batch (\# T) processed across successive LLM invocations. 
			For each batch, we report the number of missing output tuples (\# MT) caused by errors when the LLM applies Cartesian products, and their ratio (MR) relative to the total expected output of $2{,}028$, which provides an upper bound on recall. 
			We also report Recall@5 (R@5) and Recall@10 (R@10), the total number of input tokens (In.~T), output tokens (Out.~T), and run time in seconds under asynchronous execution. 
			Values in parentheses report the achieved input-token reduction. For batch sizes larger than 256 tuples, flat representations lead to errors because the prompts exceed the context window; therefore, reductions are computed relative to \texttt{FFX}-Flat at batch size 256. We note that beyond batch sizes of 128, both approaches lead to degradation.}
	\label{tab:results-GPT52}
\end{table*}

\subsubsection{Baselines}
Our primary external baseline is Lotus~\cite{DBLP:journals/corr/abs-2407-11418}, a recent system for LLM-powered data processing. 
For an apples-to-apples comparison, we also compare against an \texttt{llm\_map} implementation within \texttt{FFX} that serializes data as flat tuples for inclusion in prompts (\texttt{FFX-Flat}). 
The operator takes factorized vectors as input, fully enumerates the flat tuples by setting all window sizes to $1$, and serializes each row's relevant attributes independently. 
This baseline reflects the standard way LLM operators are applied to relational intermediates, but it exposes the model to substantial redundancy since many output tuples in $Q_p$ share the same value of $b$. 

\subsubsection{Models} 
We ran our evaluation using \texttt{GPT-5.2}, \texttt{GPT-4o}, and \texttt{GPT-4o-mini}, and observed the same overall trends across all three models. 
For brevity, we report only the results for \texttt{GPT-5.2}. 
We also experimented with smaller LLMs such as \texttt{Qwen3-8B}, which we found to be less reliable at simultaneously applying the Cartesian expansion and performing the prediction, and therefore we omit them.

\subsubsection{Analysis} 
We compare our approach from \sect{sec:prompt-compression}, denoted here as \texttt{FFX-Fact}, against the baselines. 
\texttt{FFX-Fact} serializes prompts directly from packed factorized vectors without flattening the intermediates and adds instructions that describe the f-tree structure and the required Cartesian expansion. 
In all executions, we use the attribute ordering $(b,a,c)$, which yields the f-tree $b-a,b-c$, \ie the f-tree rooted at $b$ with children $a$ and $c$. 
Under this ordering, each attribute of paper $b$ appears once, while the attributes of the associated papers $a$ and $c$ are emitted as separate lists beneath it rather than repeated across flattened tuples. 
We use \emph{JSON} serialization. 

\tab{tab:results-GPT52} reports the results, grouped by the number of tuples in each batch (\# T) processed across successive LLM invocations. 
For each batch, we report the number of missing output tuples (\# MT), caused by failures to produce the full Cartesian-product output, and the corresponding ratio (MR) relative to the expected total of $2028$ triplets, which provides an upper bound on recall. 
We also report Recall@5 (R@5), Recall@10 (R@10), the total number of input tokens (In.~T), output tokens (Out.~T), and run time in seconds under asynchronous execution. 

For the same batch size, \texttt{FFX-Fact} consistently reduces input tokens relative to \texttt{FFX-Flat}, often substantially, while remaining competitive in retrieval accuracy. 
Across batch sizes for which both representations produce results, \texttt{FFX-Fact} reduces input tokens by $1.70\times$ to $15.67\times$ relative to \texttt{FFX-Flat}; for example, at batch size $256$, it uses $0.09$M input tokens versus $1.41$M for \texttt{FFX-Flat}. 

At moderate batch sizes ($8$ to $64$), flat prompting achieves higher recall, with gaps of up to about $0.07$ for both Recall@5 and Recall@10. 
However, at larger batch sizes, \texttt{FFX-Flat} degrades sharply: Recall@5 and Recall@10 fall from $0.8210$ and $0.8861$ at batch size $64$ to $0.3319$ and $0.3669$ at batch size $256$. 
This drop is driven largely by missing output tuples, which arise when the LLM fails to return the full output tuples for larger inputs. 
\texttt{FFX-Fact} also degrades beyond batch size $128$, but more gradually, reaching $0.5439$ Recall@5 and $0.6154$ Recall@10 at batch size $256$. 

\texttt{FFX-Fact} also scales beyond the largest batch supported by \texttt{FFX-Flat}. 
While \texttt{FFX-Flat} cannot be evaluated beyond batch size $256$ because the prompts exceed the context window, \texttt{FFX-Fact} continues to operate at batch sizes up to $2048$ while keeping prompt sizes low: $70.14$K input tokens at batch size $512$ and $62.56$K at batch size $2048$. 
That said, at these larger batch sizes, the number of missing output tuples also rises substantially, indicating that output generation under large Cartesian expansions remains challenging. 
Thus, \texttt{FFX-Fact} significantly extends the usable batch-size range, but both approaches degrade beyond batch size $128$, with factorized prompting degrading more gracefully. 

Overall, these results show that, by serializing shared structure only once, factorized semantic processing achieves substantial token savings and avoids the much sharper quality collapse observed with flat prompting at large batch sizes. 
The main remaining bottleneck as input size grows is the reliable generation of the full Cartesian-product output. 

\medskip

\noindent\emph{Comparison to Lotus.} 
Under the same \texttt{GPT-5.2} setup, \texttt{FFX-Fact} at batch size $128$ achieves essentially the same Recall@5 as \texttt{Lotus} and lower Recall@10 ($0.6805$ vs.\ $0.6824$ for Recall@5 and $0.7608$ vs.\ $0.7978$ for Recall@10), while using substantially fewer input tokens ($0.12$M vs.\ $2.26$M, an $18.83\times$ reduction). 
At batch size $64$, \texttt{FFX-Fact} achieves higher recall than \texttt{Lotus} ($0.7525$ vs.\ $0.6824$ for Recall@5 and $0.8304$ vs.\ $0.7978$ for Recall@10) while using $0.18$M input tokens, which is about $12.55\times$ fewer than \texttt{Lotus}. 
At no evaluated batch size does \texttt{FFX-Fact} use more input tokens. 

\section{Conclusion} 
\label{sec:conclusion}

This work tackles a core systems challenge: how to unify factorized and vectorized execution in a single engine. 
We presented \textsc{FFX} to address this challenge through three main ideas: 
\emph{packed factorized vectors} that encode arbitrary f-trees while preserving fully packed, cache-friendly layouts; 
a \emph{cascade update} operator that propagates tuple reductions across dependent bindings without reconstruction or pointer chasing; and 
factorized semantic processing, which enables semantic operators to serialize factorized intermediates and predict over their implied Cartesian products, producing flat output tuples and predictions without first serializing flat tuples. 

Our results show that factorization and vectorization are complementary design choices. 
Across join-heavy workloads, \textsc{FFX} preserves tight inner loops, reduces interpretation overhead, and achieves substantial speedups over prior systems. When factorization offers little or no benefit, \textsc{FFX} gracefully falls back to standard vectorized execution without additional overhead. 
For semantic query processing, \textsc{FFX} reduces prompt sizes substantially by up to $18.83\times$ compared to \texttt{Lotus} with little to no degradation in output quality. 
Overall, \textsc{FFX} demonstrates that factorized representations can be propagated end-to-end to optimize join-heavy analytical queries, including those augmented with LLM-powered semantic operators.

\begin{acks}
This work was supported by a research grant from Pometry. We thank Benjamin Steer for valuable technical discussions and for support with the experimental infrastructure. 
\end{acks}

\bibliographystyle{ACM-Reference-Format}
\bibliography{references}

@article{DBLP:journals/pvldb/DorbaniYLM25,
  author       = {Anas Dorbani and
                  Sunny Yasser and
                  Jimmy Lin and
                  Amine Mhedhbi},
  title        = {Beyond Quacking: Deep Integration of Language Models and {RAG} into
                  DuckDB},
  journal      = {PVLDB},
  volume       = {18},
  number       = {12},
  pages        = {5415--5418},
  year         = {2025},
}

@article{DBLP:conf/sigir/LiuY0ZGZLSG25,
	author       = {Jiahao Liu and
	Xueshuo Yan and
	Dongsheng Li and
	Guangping Zhang and
	Hansu Gu and
	Peng Zhang and
	Tun Lu and
	Li Shang and
	Ning Gu},
	title        = {Improving LLM-powered Recommendations with Personalized Information},
	booktitle    = {SIGIR},
	year         = {2025},
}

@article{DBLP:conf/nsdi/WangALZGMWLZDJW24,
	author       = {Haopei Wang and
	Anubhavnidhi Abhashkumar and
	Changyu Lin and
	Tianrong Zhang and
	Xiaoming Gu and
	Ning Ma and
	Chang Wu and
	Songlin Liu and
	Wei Zhou and
	Yongbin Dong and
	Weirong Jiang and
	Yi Wang},
	title        = {NetAssistant: Dialogue Based Network Diagnosis in Data Center Networks},
	journal    = {NSDI},
	year         = {2024},
}

@article{DBLP:journals/vldb/SahuMSLO20,
	author       = {Siddhartha Sahu and
	Amine Mhedhbi and
	Semih Salihoglu and
	Jimmy Lin and
	M. Tamer {\"{O}}zsu},
	title        = {The ubiquity of large graphs and surprising challenges of graph processing: extended survey},
	journal      = {VLDBJ},
	year         = {2020},
}

@article{DBLP:conf/nsdi/NarayanaTRW16,
	author       = {Srinivas Narayana and
	Mina Tahmasbi and
	Jennifer Rexford and
	David Walker},
	title        = {Compiling Path Queries},
	journal    = {NSDI},
	year         = {2016},
}

@article{DBLP:conf/www/GuptaGLSWZ13,
	author       = {Pankaj Gupta and
	Ashish Goel and
	Jimmy Lin and
	Aneesh Sharma and
	Dong Wang and
	Reza Zadeh},
	title        = {{WTF:} the who to follow service at Twitter},
	journal    = {WWW},
	year         = {2013},
}

@inproceedings{DBLP:conf/icde/KaluminD25,
	author       = {Hasara Kalumin and
	Amol Deshpande},
	title        = {Optimizing Queries with Many-to-Many Joins},
	booktitle    = {ICDE},
	year         = {2025},
}

@article{DBLP:journals/sigmod/ZhaoSYYKZ25,
	author = {Zhao, Junyi and Su, Kai and Yang, Yifei and Yu, Xiangyao and Koutris, Paraschos and Zhang, Huanchen},
	title = {Debunking the Myth of Join Ordering: Toward Robust SQL Analytics},
	year = {2025},
	journal = {SIGMOD},
}

@article{DBLP:journals/pvldb/StoianZKNDKK25,
	author       = {Mihail Stoian and
	Andreas Zimmerer and
	Skander Krid and
	Amadou Ngom and
	Jialin Ding and
	Tim Kraska and
	Andreas Kipf},
	title        = {Parachute: Single-Pass Bi-Directional Information Passing},
	journal      = {PVLDB.},
	year         = {2025},
}

@article{DBLP:journals/ftdb/MhedhbiDS24,
	author       = {Amine Mhedhbi and
	Amol Deshpande and
	Semih Salihoglu},
	title        = {Modern Techniques For Querying Graph-structured Databases},
	journal      = {FnT DB},
	year         = {2024},
}

@article{DBLP:journals/pvldb/BirlerKN24,
	author       = {Altan Birler and
	Alfons Kemper and
	Thomas Neumann},
	title        = {Robust Join Processing with Diamond Hardened Joins},
	journal      = {PVLDB},
	year         = {2024},
}

@article{DBLP:conf/cidr/YangZYK24,
	author       = {Yifei Yang and
	Hangdong Zhao and
	Xiangyao Yu and
	Paraschos Koutris},
	title        = {Predicate Transfer: Efficient Pre-Filtering on Multi-Join Queries},
	journal    = {CIDR},
	publisher    = {www.cidrdb.org},
	year         = {2024},
}

@article{DBLP:conf/sigmod/LambSHCKHS24,
	author       = {Andrew Lamb and
	Yijie Shen and
	Dani{\"{e}}l Heres and
	Jayjeet Chakraborty and
	Mehmet Ozan Kabak and
	Liang{-}Chi Hsieh and
	Chao Sun},
	title        = {Apache Arrow DataFusion: {A} Fast, Embeddable, Modular Analytic Query Engine},
	journal    = {SIGMOD},
	year         = {2024},
}

@article{DBLP:journals/pvldb/WoldeSB23,
	author       = {Daniel ten Wolde and
	G{\'{a}}bor Sz{\'{a}}rnyas and
	Peter A. Boncz},
	title        = {DuckPGQ: Bringing {SQL/PGQ} to DuckDB},
	journal      = {PVLDB},
	year         = {2023},
}

@article{DBLP:conf/cidr/WoldeSSB23,
	author       = {Daniel ten Wolde and
	Tavneet Singh and
	G{\'{a}}bor Sz{\'{a}}rnyas and
	Peter A. Boncz},
	title        = {DuckPGQ: Efficient Property Graph Queries in an analytical {RDBMS}},
	journal    = {CIDR},
	year         = {2023},
}

@article{DBLP:conf/cidr/JinFCLS23,
	author       = {Guodong Jin and 
	Xiyang Feng and 
	Ziyi Chen and 
	Chang Liu and 
	Semih Salihoglu},
	title        = {K{\`{U}}ZU Graph Database Management System},
	journal    = {CIDR},
	year         = {2023},
}

@article{DBLP:journals/pvldb/HuangSL023,
	author       = {Zezhou Huang and
	Rathijit Sen and
	Jiaxiang Liu and
	Eugene Wu},
	title        = {JoinBoost: Grow Trees Over Normalized Data Using Only {SQL}},
	journal      = {PVLDB},
	year         = {2023},
}

@article{DBLP:journals/pvldb/PedreiraEBWSPHC22,
	author       = {Pedro Pedreira and
	Orri Erling and
	Maria Basmanova and
	Kevin Wilfong and
	Laith Sakka and
	Krishna Pai and
	Wei He and
	Biswapesh Chattopadhyay},
	title        = {Velox: Meta's Unified Execution Engine},
	journal      = {PVLDB},
	year         = {2022},
}

@inproceedings{DBLP:conf/sigmod/BehmPAACDGHJKLL22,
	author       = {Alexander Behm and
	Shoumik Palkar and
	Utkarsh Agarwal and
	Timothy Armstrong and
	David Cashman and
	Ankur Dave and
	Todd Greenstein and
	Shant Hovsepian and
	Ryan Johnson and
	Arvind Sai Krishnan and
	Paul Leventis and
	Ala Luszczak and
	Prashanth Menon and
	Mostafa Mokhtar and
	Gene Pang and
	Sameer Paranjpye and
	Greg Rahn and
	Bart Samwel and
	Tom van Bussel and
	Herman Van Hovell and
	Maryann Xue and
	Reynold Xin and
	Matei Zaharia},
	title        = {Photon: {A} Fast Query Engine for Lakehouse Systems},
	booktitle    = {SIGMOD},
	year         = {2022},
}

@article{DBLP:journals/tods/MhedhbiKS21,
	author       = {Amine Mhedhbi and
	Chathura Kankanamge and
	Semih Salihoglu},
	title        = {Optimizing One-time and Continuous Subgraph Queries using Worst-case Optimal Joins},
	journal      = {TODS},
	year         = {2021},
}

@article{DBLP:journals/pvldb/GuptaMS21,
	author       = {Pranjal Gupta and
	Amine Mhedhbi and
	Semih Salihoglu},
	title        = {Columnar Storage and List-based Processing for Graph Database Management
	Systems},
	journal      = {PVLDB},
	year         = {2021},
}

@article{DBLP:journals/pvldb/SchleichO20,
	author       = {Maximilian Schleich and
	Dan Olteanu},
	title        = {{LMFAO:} An Engine for Batches of Group-By Aggregates},
	journal      = {PVLDB},
	year         = {2020},
}

@article{DBLP:conf/sigmod/RaasveldtM19,
	author       = {Mark Raasveldt and Hannes M{\"{u}}hleisen},
	title        = {DuckDB: an Embeddable Analytical Database},
	journal    = {SIGMOD},
	year         = {2019},
}

@article{DBLP:journals/pvldb/MhedhbiS19,
	author       = {Amine Mhedhbi and 
	Semih Salihoglu},
	title        = {Optimizing Subgraph Queries by Combining Binary and Worst-Case Optimal Joins},
	journal      = {PVLDB.},
	year         = {2019},
}

@article{DBLP:journals/pvldb/KerstenLKNPB18,
	author       = {Timo Kersten and
	Viktor Leis and
	Alfons Kemper and
	Thomas Neumann and
	Andrew Pavlo and
	Peter A. Boncz},
	title        = {Everything You Always Wanted to Know About Compiled and Vectorized
	Queries But Were Afraid to Ask},
	journal      = {PVLDB},
	year         = {2018},
}

@inproceedings{DBLP:conf/sigmod/KankanamgeSMCS17,
	author       = {Chathura Kankanamge and
	Siddhartha Sahu and
	Amine Mhedhbi and
	Jeremy Chen and
	Semih Salihoglu},
	title        = {Graphflow: An Active Graph Database},
	booktitle    = {SIGMOD},
	year         = {2017},
}

@article{DBLP:conf/sigmod/AbergerTOR16,
	author       = {Christopher R. Aberger and 
	Susan Tu and 
	Kunle Olukotun and 
	Christopher R{\'{e}}},
	title        = {EmptyHeaded: {A} Relational Engine for Graph Processing},
	journal    = {SIGMOD},
	year         = {2016},
}

@inproceedings{DBLP:conf/sigmod/KumarNPZ16,
	author       = {Arun Kumar and
	Jeffrey F. Naughton and
	Jignesh M. Patel and
	Xiaojin Zhu},
	title        = {To Join or Not to Join?: Thinking Twice about Joins before Feature
	Selection},
	booktitle    = {SIGMOD},
	year         = {2016},
}

@techreport{CiucanuOlteanu2016WCOJAT,
	title       = {Worst-Case Optimal Join at a Time},
	author      = {Ciucanu, Radu and Olteanu, Dan},
	institution = {Department of Computer Science, University of Oxford},
	type        = {Technical Report},
	year        = {2016},
	month       = mar,
	note        = {First version: November 2015. Latest version: March 2016.},
	url         = {https://www.cs.ox.ac.uk/dan.olteanu/papers/co-tr16.pdf}
}

@article{DBLP:journals/tods/OlteanuZ15,
	author       = {Dan Olteanu and
	Jakub Z{\'{a}}vodn{\'{y}}},
	title        = {Size Bounds for Factorised Representations of Query Results},
	journal      = {TODS},
	year         = {2015},
}

@article{DBLP:journals/pvldb/BakibayevKOZ13,
	author       = {Nurzhan Bakibayev and
	Tom{\'{a}}s Kocisk{\'{y}} and
	Dan Olteanu and
	Jakub Zavodny},
	title        = {Aggregation and Ordering in Factorised Databases},
	journal      = {PVLDB},
	year         = {2013},
}

@article{DBLP:journals/pvldb/RamanABCKKLLLLMMPSSSSZ13,
	author       = {Vijayshankar Raman and
	Gopi K. Attaluri and
	Ronald Barber and
	Naresh Chainani and
	David Kalmuk and
	Vincent KulandaiSamy and
	Jens Leenstra and
	Sam Lightstone and
	Shaorong Liu and
	Guy M. Lohman and
	Tim Malkemus and
	Ren{\'{e}} M{\"{u}}ller and
	Ippokratis Pandis and
	Berni Schiefer and
	David Sharpe and
	Richard Sidle and
	Adam J. Storm and
	Liping Zhang},
	title        = {{DB2} with {BLU} Acceleration: So Much More than Just a Column Store},
	journal      = {PVLDB},
	year         = {2013},
}

@article{DBLP:journals/pvldb/BakibayevOZ12,
	author       = {Nurzhan Bakibayev and
	Dan Olteanu and
	Jakub Zavodny},
	title        = {{FDB:} {A} Query Engine for Factorised Relational Databases},
	journal      = {PVLDB},
	year         = {2012},
}

@article{DBLP:conf/icdt/OlteanuZ12,
	author       = {Dan Olteanu and
	Jakub Zavodny},
	title        = {Factorised representations of query results: size bounds and readability},
	booktitle    = {ICDT},
	year         = {2012},
}

@inproceedings{DBLP:conf/icde/ZukowskiWB12,
	author       = {Marcin Zukowski and
	Mark van de Wiel and
	Peter A. Boncz},
	title        = {Vectorwise: {A} Vectorized Analytical {DBMS}},
	booktitle    = {ICDE},
	year         = {2012},
}

@article{DBLP:conf/focs/AtseriasGM08,
	author       = {Albert Atserias and 
	Martin Grohe and 
	D{\'{a}}niel Marx},
	title        = {Size Bounds and Query Plans for Relational Joins},
	journal    = {FOCS},
	year         = {2008},
}

@article{DBLP:conf/cidr/BonczZN05,
	author       = {Peter A. Boncz and
	Marcin Zukowski and
	Niels Nes},
	title        = {MonetDB/X100: Hyper-Pipelining Query Execution},
	booktitle    = {CIDR},
	year         = {2005},
}

@article{monetdb-thesis,
	author = {Boncz, Peter},
	year = {2002},
	title = {Monet; a next-Generation DBMS Kernel For Query-Intensive Applications},
	journal = {PhD Thesis, Universiteit van Amsterdam, Amsterdam},
}

@article{DBLP:journals/tkde/Graefe94,
	author       = {Goetz Graefe},
	title        = {Volcano - An Extensible and Parallel Query Evaluation System},
	journal      = {TKDE},
	year         = {1994},
}

@article{DBLP:conf/pvldb/Yannakakis81,
	author = {Yannakakis, Mihalis},
	title = {Algorithms for acyclic database schemes},
	journal = {PVLDB},
	year = {1981},
}

@article{DBLP:conf/sigmod/SatrianiVSRVP25,
	author = {Satriani, Dario and Veltri, Enzo and Santoro, Donatello and Rosato, Sara and Varriale, Simone and Papotti, Paolo},
	title = {Logical and Physical Optimizations for SQL Query Execution over Large Language Models},
	journal    = {SIGMOD},
	year = {2025},
}

@article{DBLP:conf/cidr/BiswalPJKLGGZ25,
	author       = {Asim Biswal et al},
	title        = {Text2SQL is Not Enough: Unifying AI and Databases with TAG},
	journal    = {CIDR},
	year         = {2025},
}

@article{DBLP:conf/cidr/LiuRCCCCFKMSG25,
	author       = {Chunwei Liu et al},
	title        = {Palimpzest: Optimizing AI-Powered Analytics with Declarative Query Processing},
	journal    = {CIDR},
	year         = {2025},
}

@article{DBLP:journals/corr/abs-2407-11418,
	author       = {Liana Patel et al},
	title        = {{LOTUS:} Enabling Semantic Queries with LLMs Over Tables of Unstructured and Structured Data},
	journal      = {CoRR},
	volume       = {abs/2407.11418},
	year         = {2024}
}

@article{DBLP:conf/sigmod/JoT24,
	author = {Jo, Saehan and Trummer, Immanuel},
	title = {ThalamusDB: Approximate Query Processing on Multi-Modal Data},
	journal = {SIGMOD},
	year = {2024},
}

@mastersthesis{Lombardo2016StorageLayerFactorized,
	title        = {Storage layer for factorized databases},
	author       = {Lombardo, Antonio},
	school       = {University of Oxford},
	year         = {2016},
	type         = {MSc thesis},
	note         = {New College},
	url          = {https://fdbresearch.github.io/papers/AntonioLombardo_Thesis.pdf},
	urldate      = {2026-01-27}
}

@article{hu2021opengraphbenchmarkdatasets,
	author={Weihua Hu and Matthias Fey and Marinka Zitnik and Yuxiao Dong and Hongyu Ren and Bowen Liu and Michele Catasta and Jure Leskovec},
	title={Open Graph Benchmark: Datasets for Machine Learning on Graphs}, 
	journal    = {CoRR},
	volume     = {abs/2005.00687},
	year={2021},
}

@misc{snapnets,
	author       = {Jure Leskovec and Andrej Krevl},
	title        = {{SNAP Datasets}: {Stanford} Large Network Dataset Collection},
	howpublished = {\url{http://snap.stanford.edu/data}},
	year         = {2014}
}

@inproceedings{DBLP:conf/icde/MhedhbiGKS21,
	author       = {Amine Mhedhbi and
	Pranjal Gupta and
	Shahid Khaliq and
	Semih Salihoglu},
	title        = {{A+} Indexes: Tunable and Space-Efficient Adjacency Lists in Graph
	Database Management Systems},
	booktitle    = {ICDE},
	year         = {2021},
}

@article{Kwak10www,
	author = {Kwak, Haewoon and Lee, Changhyun and Park, Hosung and Moon, Sue},
	title = "{W}hat is {T}witter, a social network or a news media?",
	journal = {WWW},
	year = {2010},
}

@article{DBLP:conf/sigmod/MhedhbiLKWS21,
	author       = {Amine Mhedhbi and
	Matteo Lissandrini and
	Laurens Kuiper and
	Jack Waudby and
	G{\'{a}}bor Sz{\'{a}}rnyas},
	title        = {{LSQB:} a large-scale subgraph query benchmark},
	journal    = {GRADES-NDA},
	year         = {2021},
}

@article{DBLP:journals/pvldb/LeisGMBK015,
	author       = {Viktor Leis and
	Andrey Gubichev and
	Atanas Mirchev and
	Peter A. Boncz and
	Alfons Kemper and
	Thomas Neumann},
	title        = {How Good Are Query Optimizers, Really?},
	journal      = {PVLDB},
	year         = {2015},
}

\end{document}